\documentclass[11pt,cite,epsfig,psfrag]{article}
\usepackage[utf8]{inputenc}
\usepackage{placeins}
\usepackage{geometry}
 \usepackage{graphicx}
 \usepackage{subfig}
  \usepackage{amsmath} 
  \graphicspath{{Images_matrix/}}
  \usepackage[usenames, dvipsnames]{color}
\usepackage{wrapfig}
\usepackage{boxedminipage}
\usepackage{multirow}

\usepackage{fullpage}
\usepackage{amsmath}
\usepackage{amssymb}
\usepackage{graphicx}
\usepackage{mathrsfs}
\usepackage{wrapfig}
\usepackage{boxedminipage}
\usepackage{epsfig}
\usepackage{setspace}
\usepackage{float}
\usepackage{hyperref}
\usepackage{enumerate}
\usepackage{cite}
\usepackage[font=footnotesize,labelfont=rm]{caption}

\usepackage[numbers,sort&compress]{natbib}

\def\be{\begin{equation}}
\def\ee{\end{equation}}
\def\bea{\begin{eqnarray}}
\def\eea{\end{eqnarray}}

\numberwithin{equation}{section}

 \newcommand{\RN}[1]{%
   \textup{\uppercase\expandafter{\romannumeral#1}}%
 }
\begin{document}

\thispagestyle{empty}

\vskip 2cm

\begin{center}
{\Large \bf Topology of critical points in boundary matrix duals}
\end{center}

\vskip .2cm

\vskip 1.2cm

\centerline{ \bf   Pavan Kumar Yerra \footnote{pk11@iitbbs.ac.in}, Chandrasekhar Bhamidipati\footnote{chandrasekhar@iitbbs.ac.in} and Sudipta Mukherji\footnote{mukherji@iopb.res.in}
}

\vskip 7mm 
\begin{center}{ $^{1},^{3}$ Institute of Physics, Sachivalaya Marg, \\ Bhubaneswar, Odisha, 751005, India \\ and\\Homi Bhabha National Institute, \\Training School Complex, \\Anushakti Nagar, Mumbai, 400085, India}	
\end{center}

\begin{center}{ $^{2}$ School of Basic Sciences\\ 
Indian Institute of Technology Bhubaneswar \\ Bhubaneswar, Odisha, 752050, India}
\end{center}

\vskip 1.2cm
\vskip 1.2cm
\centerline{\bf Abstract}
\vskip 0.5cm
\noindent
Computation of topological charges of the Schwarzschild and charged black holes in AdS in canonical and grand canonical ensembles allows for a classification of the phase transition points via the 
Bragg-Williams off-shell free energy. We attempt a topological classification of the critical points and 
the equilibrium phases of the dual gauge theory via a phenomenological matrix model, 
which captures the features of the ${\cal{N}}=4$, $SU(N)$ Super Yang-Mills theory on $S^3$ at finite temperature at large $N$. 
With minimal modification of parameters, critical points of the matrix model 
at finite chemical potential can be classified as well. The topological charges of locally stable and unstable dynamical phases of the system turn out to be opposite to each other, totalling to zero, and this matches the analysis in the bulk.

\vskip 0.5cm
\noindent

\newpage
\setcounter{footnote}{0}
\noindent

\baselineskip 15pt
\section{Introduction}

Phase transitions and critical phenomena in thermodynamic and statistical systems 
are quite appealing, especially when the context involves 
black holes~\cite{Bekenstein:1973ur,Bardeen:1973gs,Hawking:1975vcx}. 
Thermodynamics of black holes continues to furnish insights into the microscopic 
features of quantum gravity through the computation of entropy of black holes close 
to extremal limit~\cite{Sen:2007qy}, along with an understanding of strong gravitational phenomena~\cite{LIGO2017}. 
Phase structure of black holes in theories with a negative cosmological constant is better 
scrutinised, due to their relevance in discerning finite 
temperature aspects of dual strongly coupled field theories via the AdS/CFT correspondence. 
A particular example in this regard is the 
Hawking-Page (HP) transition~\cite{Hawking:1982dh} in five space-time dimensions which is understood as the 
deconfining 
transition in 
the ${\cal{N}} = 4$  $SU(N)$ supersymmetric Yang-Mills (SYM) theory at large 
$N$~\cite{,Maldacena:1997re,Gubser:1998bc,Witten:1998qj}. This deconfining transition is reflected on the boundary as a
jump in the free energy from order $N^0$ to $N^2$.
Subsequently, based on this 
indentification, 
in~\cite{Alvarez-Gaume:2005dvb}, a phenomenological matrix model was constructed. This model, build  out
of the Wilson-Polyakov loop variable, was proposed to be in the same universality class as that of
${\cal{N}} = 4$ $SU(N)$ SYM theory on $S^3$ at infinite 't Hooft coupling ($\lambda$). 
A further example involves the $R$-charged black hole in AdS that undergoes a continuous 
transition~\cite{Chamblin:1999tk,Chamblin:1999hg} apart from showing other characteristics. 
The corresponding boundary matrix model, 
extended in~\cite{Basu:2005pj} to include appropriate dependence on the charge, is found to 
capture qualitatively the behaviour emanating from the bulk black hole. Since these models 
typically contain 
two parameters $a$ and $b$, which could depend on various intensive variables of the boundary theory
and $\lambda$, in the following we will refer it as $(a,b)$ model. In this work, our aim is 
to make a comparative study of certain topological features in the bulk and in the boundary
matrix dual which we describe below.\\

\noindent
Recently, a novel technique has cropped up which allows inquiries into the 
phase structure of general thermodynamic systems, from the point of view of topology. The method relies on 
working out the topological charges of various critical points or the stable/unstable phases 
of the system, eventually leading to a classification. This topological approach has come 
to light from the study of light-rings and ultra compact objects 
 (see e.g.,~\cite{Cunha:2017qtt,Cunha:2020azh,Guo:2020qwk,Wei:2020rbh,Junior:2021svb}), and has been 
further customised to be useful for thermodynamics. Topological classification of second order critical 
points ~\cite{Wei:2021vdx}, with particular applications to extended 
thermodynamics~\cite{Caldarelli:1999xj,Kastor:2009wy,Cvetic:2010jb,Dolan:2011xt,Karch:2015rpa,Kubiznak:2012wp,Kubiznak:2016qmn,Wei:2019uqg} has been done. 
To wit, one starts by writing the temperature of the black hole 
as a function of entropy and other thermodynamic variables. Knowing that the critical points of the system can 
be obtained from the points of inflection of the temperature curve with respect to entropy, a potential is 
constructed by eliminating one of the thermodynamic variables, such that the vector field following from the potential has certain zeros which give rise to the critical points. The procedure is now well studied~\cite{Duan:1984ws,Duan:2018rbd,Wei:2021vdx} and leads to a conserved current as well as the notion of a topological charge which together govern the nature of the critical points. Of course when there are several critical points or situations where novel phases are created or destroyed, a more prudent analysis is required~\cite{Yerra:2022alz}. While initial developments focussed on investigating topological behaviour of second order critical points~\cite{Wei:2021vdx,Yerra:2022alz,Ahmed:2022kyv,Yerra:2022eov,Wei:2022dzw} in the extended thermodynamic set up, the methods are applicable more generally, and in particular in the case of first order phase transitions as well~\cite{Yerra:2022coh}.  More recently, with a slight modification of the definition of vector field, it has also been shown that a topological charge can be assigned not just to the phase transition points, but to the black holes themselves. This leads to a remarkable interpretation of the black hole as being a topological defect in the space-time\cite{Wei:2022dzw}. The order parameter for phase transitions can be quite general, such as charge, angular momentum or the horizon radius, and the topological charge in several of these cases is now computed~\cite{Ahmed:2022kyv,Wei:2022mzv,Bai:2022klw,Gogoi:2023xzy,Ahmed:2023snm,Liu:2022aqt,Fan:2022bsq,Wu:2023sue,Ye:2023gmk,Fang:2022rsb,Wu:2022whe,Du:2023nkr,Sharqui:2023mbx,Zhang:2023uay,Gogoi:2023xzy}.\\

\noindent  
Our aim concerns the computation of topological behaviour of critical points in the context of AdS/CFT 
correspondence, where the same quantity may be studied both in the bulk and on the boundary. There are diverse 
contexts where geometrical ideas  are being advanced for gaining holographic understanding of critical points. One recent example involves quantum error correction~\cite{Bao:2022agm,Almheiri:2014lwa}. In the theory of quantum computation there are certain fault-tolerance thresholds below which the quantum errors are corrected quite rapidly. It is believed with increasing evidence that confinement-deconfiment transition is one such natural threshold, similar to the behaviour of  the topological codes, crossing which leads to a disturbance of the fidelity of quantum computation~\cite{Bao:2022agm}. Geometrical arguments show that this threshold has intricate connection to the HP transition. A more detailed analysis however requires a precise model where the phase transition points of black holes can be tracked in the boundary gauge theories, at least in certain limits. In this spirit, a preliminary computation was set up in~\cite{Yerra:2022coh}, where an off-shell free energy of the black holes in the bulk was used to compute the topological charge of the HP transition point~\cite{Hawking:1982dh}  as well charges of the equilibrium phases of the system~\cite{Banerjee:2010ve}. On the boundary, a similar computation was then set up using an effective potential in the gauge theory, which can be derived systematically from the bulk free energy. The topological classification in the bulk and on the boundary were shown to match~\cite{Yerra:2022coh}, even though the order parameters turn out to be very different, namely horizon radius in the case of the HP transition and the charge parameter, which tracks the confinement-deconfinement transition in the gauge theory.\\

\noindent
 These computations referred to in the previous paragraph, required using an off-shell free energy in terms of the order parameter, with various phases of the system appearing from its equilibrium points~\cite{cha95,Banerjee:2010ng,Banerjee:2010ve,nayak2008bragg,Dey:2007vt,Dey:2006ds}. The key idea in constructing such a free energy is to use a Bragg-Williams approach to phase transitions, as opposed to standard mean field techniques such as the Landau's theory. 
The later method is more useful when the order parameter is small where a series expansion of free energy can 
be performed around the critical point~\cite{cha95}. However, while studying first order phase transitions, 
such as the HP transition, the discontinuous changes in order parameter make the traditional methods 
unreliable. Bragg-Williams method ~\cite{bw1, bw2} continues to be helpful in such situations as it has been tested in various condensed matter systems to black holes~\cite{cha95,kubo,Banerjee:2010ve}. The stable and unstable phases obtained from the free energy turn out to belong to different topological classes, which can be verified from the bulk~\cite{Wei:2022dzw} as well as from the boundary computations~\cite{Yerra:2022coh}.  \\

\noindent 
The aim of this note is to report further progress in addressing the issues posed above and extend the computations 
of~\cite{Yerra:2022coh} to more general situations in the gauge/gravity duality. The set up used in~\cite{Yerra:2022coh} 
involved computation of topological charges from an effective potential of the gauge theory which relied on inputs from the bulk 
free energy. It is desirable to have an independent computation of topological charges in the gauge 
theory and, to this effect, in this work, we use the $(a,b)$ matrix model on the boundary. 
We find the topological charges that follow directly from this model or it's 
suitable generalisation when the bulk is represented by either the Schwarzschild or the $R$-charged black 
holes. \\

\noindent 
We should mention that the matrix model studied in this work, is defined in the non-extended thermodynamic set up, where the cosmological constant is fixed (we in fact set it to unity). Of course, the comparison between the
thermodynamic behaviour of charged black holes in AdS in the canonical
ensemble and that of the van der Waal's fluid is cleanest in the extended phase
space approach, where one introduces a pressure-volume term in its thermodynamic description. This is emphasised in~\cite{Kubiznak:2012wp}, from which it is evident that the analogy of phase structure of the above class of black holes to the van der Waals fluid is at most qualitative in the non-extended thermodynamic description~\cite{Chamblin:1999tk,Chamblin:1999hg}, particularly because one is comparing different thermodynamic quantities on both sides (see the appendix of~\cite{Kubiznak:2012wp} for details). It should thus be interesting to study the topological charges of critical points in the strongly coupled gauge theories using the extended thermodynamic framework. The reason for employing the non-extended thermodynamic framework in this work, is two fold. {\it Firstly}, the aim here is only to compare the 
topological charges associated with the first order transition involving AdS and AdS black hole with that of  the
deconfining transition in the gauge theory living on the boundary of the AdS. This
leads to a classification of critical points of certain class of strongly coupled gauge theories dual to uncharged and charged black holes in five dimensions. This can be carried out in the non-extended phase space and was initiated in ~\cite{Yerra:2022coh}. {\it Secondly}, to make an attempt to understand the complete phase diagram of the gauge theory and subsequently  make a comparison with that of the van der Waal's fluid on the boundary (originating from the bulk analysis of ~\cite{Kubiznak:2012wp}) and carry out the topological analysis, is likely to be premature at this stage. 
This is primarily because the gauge/gravity duality in the extended phase space is not fully developed for the classes of black holes considered here (see~\cite{Frassino:2022zaz,Ahmed:2023snm} for important developments). Moreover, since the gauge theory under consideration is
strongly coupled, it is not immediately obvious how to generalise the 
conjectured matrix model set up proposed in ~\cite{Alvarez-Gaume:2005dvb,Basu:2005pj}
to accommodate the extended thermodynamic phase space of the bulk~\cite{Kubiznak:2012wp}.\\

\noindent
Structure of the rest of the paper is as follows. In section-(\ref{one}), we start by considering gauge theories with zero chemical potential. In subsection-(\ref{section:sch}), we briefly review the Bragg-Williams off-shell free energy in the bulk, and summarise the topological method used in the rest of the paper, for computation of charge of the Hawking-Page transition point, as well charges of various stable and unstable black hole phases. In subsection-(\ref{section:boundary for sch}), after a quick outline of the boundary matrix models~\cite{Alvarez-Gaume:2005dvb}, we compute the topological charges of the de-confining transition 
as well as equilibrium phases of the effective potential. Section-(\ref{two}) addresses similar issues alluded to above, but now in the presence of a chemical potential, where the system can be studied in the fixed charge (canonical) and fixed chemical potential (grand canonical) ensembles. Introducing a charge in the boundary corresponds to switching on the R-charge of ${\cal N} = 4$, $SU(N)$  SYM theory. Using a grand canonical ensemble, various qualitative features in the presence of a chemical potential can be incorporated in the boundary matrix model with suitable minimal modifications. The topological charges of the HP transition in the bulk and de-confining transition on the boundary are computed in subsections-(\ref{bulkgrand}) and (\ref{boundarygrand}), respectively. The computation of topological charges in the canonical ensemble are discussed in subsection-(\ref{canonical}). In the fixed charge ensemble, as elaborate in the sequel, there exists a second order critical point whose topological charge comes out to be opposite to the computations done earlier in the extended thermodynamic set up. The computation done using the boundary effective potential are given in subsection-(\ref{canonicalboundary}), which matches with the result obtained in the bulk. Remarks and conclusions are given in section-(\ref{conclusions}).
We end with two Appendices. Appendix-(\ref{A}) contains the construction of Bragg-Williams free energy of black holes studied in the bulk. In Appendix-(\ref{B}), we explain the reason why the topological charges of critical points in the extended and non-extended thermodynamic treatment come out to be opposite to each other.

\section{Black hole and its matrix dual: zero chemical potential} \label{one}
In what follows, we first consider the case of  Schwarzschild-AdS$_5$ black holes on the gravity side. We 
briefly review the computation of the topological charge carried by the HP transition point and also the 
topological charges of the equilibrium phases of the system.   We then consider  its matrix dual, namely the 
$(a,b)$ model in the 
boundary side and carry out the similar computation of topological charges to compare  the value of topological charges obtained in
the bulk.
\subsection{Bulk}\label{section:sch}
\begin{figure}[h!]
	
	{\centering
		
	\includegraphics[width=3in]{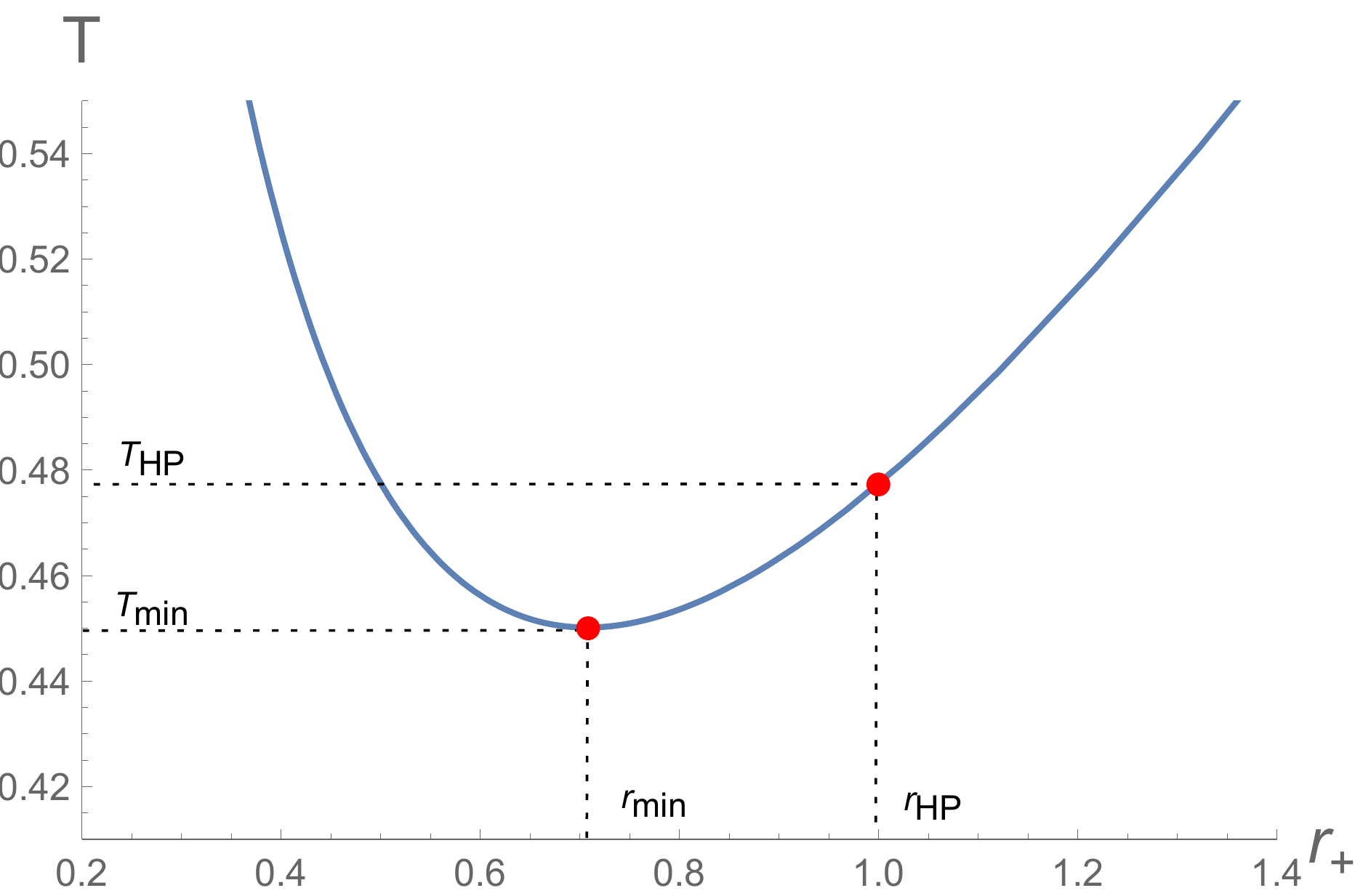}		
		\caption{\footnotesize For Schwarzschild black holes in $\text{AdS}_5$: Temperature $T$  as a function of horizon radius $r_+$, indicating the existence of small $(r_+ < r_{\rm min})$ and large $(r_+ > r_{\rm min})$ black holes for $T > T_{\rm min}$, and the  Hawking-Page (HP) transition for $ T = T_{\rm HP}$. Here, $(r_{\rm min}, T_{\rm min}) = (\frac{1}{\sqrt{2}}, \frac{\sqrt{2}}{\pi})$ and  $(r_{\rm HP}, T_{\rm HP}) = (1, \frac{3}{2\pi})$.}
	\label{fig:sch_bulk_eos}}	
\end{figure}

\begin{figure}[h!]
	
	{\centering
		
		\subfloat[]{\includegraphics[width=3in]{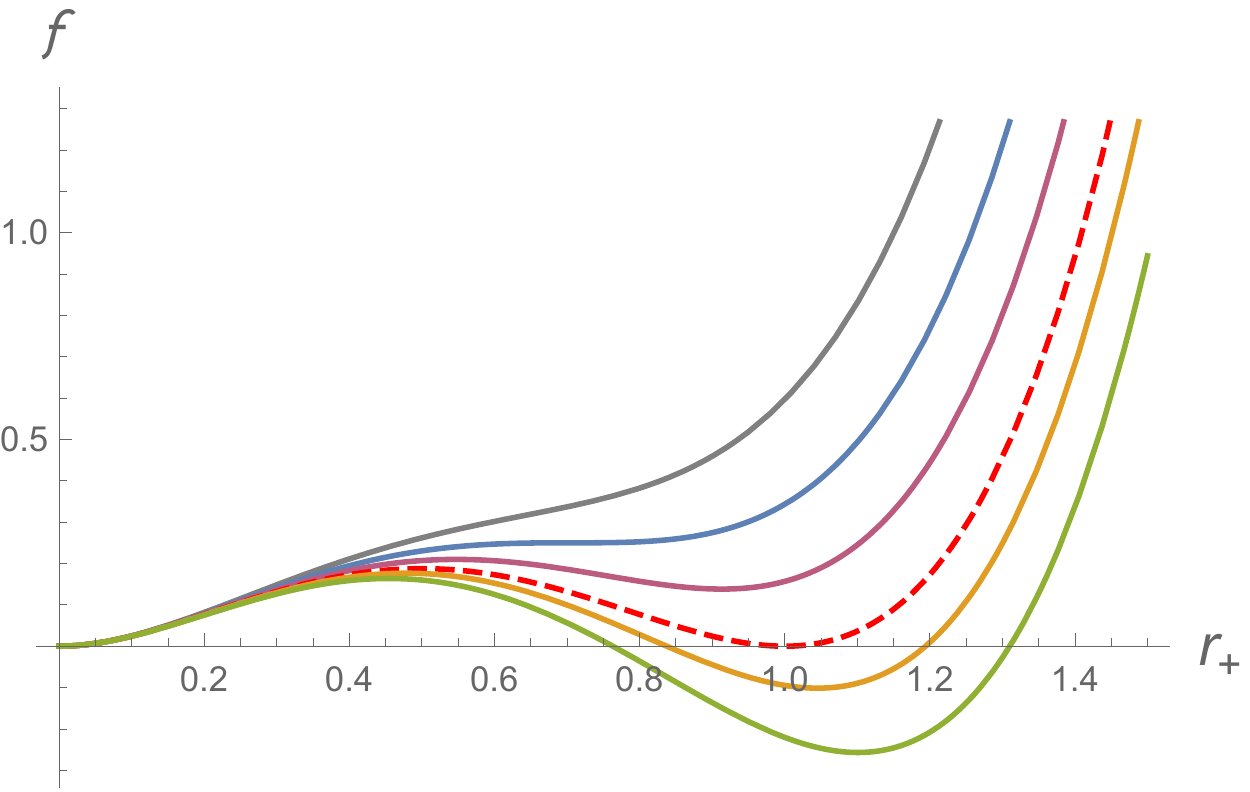}\label{fig:sch_bulk_free_energy_plot}}\hspace{0.5cm}	
		\subfloat[]{\includegraphics[width=2.8in]{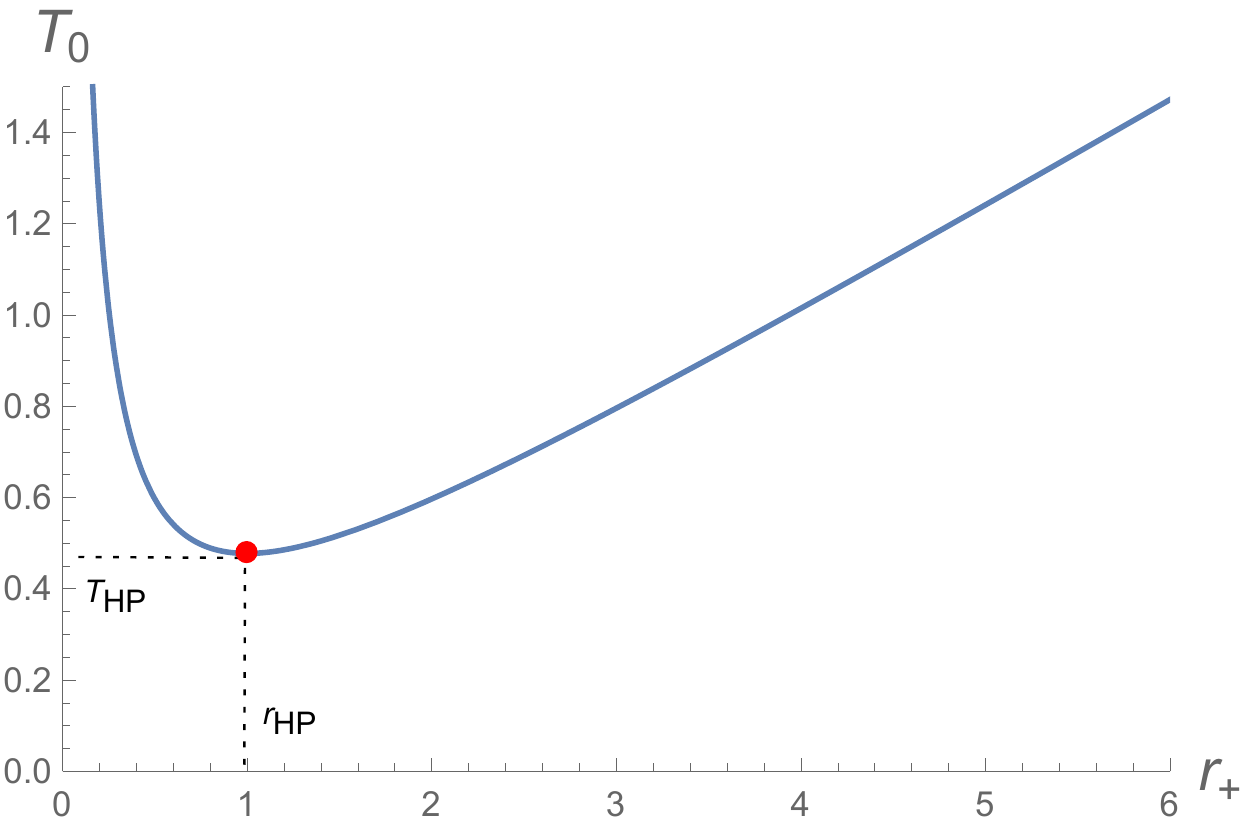}\label{fig:sch_bulk_T0_plot}}	
		
		\caption{\footnotesize For Schwarzschild black holes in $\text{AdS}_5$:  (a) Behaviour of the Bragg-Williams free energy $f$, as a function of horizon radius $r_+$ at different temperatures $T$. Temperature of the curves increases from top to bottom. Blue curve is for $T_{\rm min}$ and dashed red curve is for $T_{\rm HP}$. (b) The curve $T_{\rm 0}$  shows the HP transition point at its minima (red dot).}
	}
	
\end{figure}

We start with the thermodynamics of  Schwarzschild black holes in AdS$_5$~\cite{Hawking:1982dh}. There exists a minimum temperature $T_{\rm min}$, above which there is a nucleation of two black holes, small $(r_+ < r_{\rm min})$  and large $(r_+ > r_{\rm min})$, as shown in the Fig.~\ref{fig:sch_bulk_eos}. The smaller black hole is locally unstable due to negative specific heat, where as the larger black hole has positive specific heat and is locally stable.
There exists another temperature $T_{\rm HP} > T_{\rm min}$, where the larger black hole undergoes a phase transition called Hawking-Page (HP) transition at which the preferable phase switches from AdS to black hole. 
The expression for the black hole temperature $T$, in terms of horizon radius $r_+$, is given by~\footnote{We work with the units $\hbar = c = 1$, and we set AdS length $l$, volume of the three sphere $\omega_3$, and $16\pi G$ to 1. $G$ is the Newton's constant.}:
 \begin{equation}
 T = \frac{1+2r_+^2}{2\pi r_+}.
 \end{equation} 
\vskip 0.2cm 
\noindent  We, now, briefly review the method used for obtaining the topological charge (winding number) carried by the Hawking-Page (HP) transition point (details can be found in~\cite{Yerra:2022coh}). This  computation employs the Bragg-Williams (BW) construction of an off-shell free energy function $f$, given by~\cite{Banerjee:2010ve} (See the  Appendix -(\ref{A1}) for details):
\begin{equation}
f(r_+, T)= M-TS= 3r_+^2(1+r_+^2)-4\pi r_+^3T.
\end{equation}
Here $M$, $S$ are mass and entropy of the black hole and $T$ is an external parameter. The behavior of the free energy $f$ for various temperatures is shown in Fig.~\ref{fig:sch_bulk_free_energy_plot}. Using $f =0$, which is one of the two conditions $( f=0 \, \text{and} \,  \frac{\partial f}{\partial r_+} =0)$ useful in determining critical points, one obtains the temperature $T_0 = \frac{3(1+r_+^2)}{4\pi r_+}$. The HP transition point is identified from the minima of $T_0$ as seen from Fig.~\ref{fig:sch_bulk_T0_plot}.  
\begin{figure}[h!]
	{\centering
		\subfloat[]{\includegraphics[width=3in]{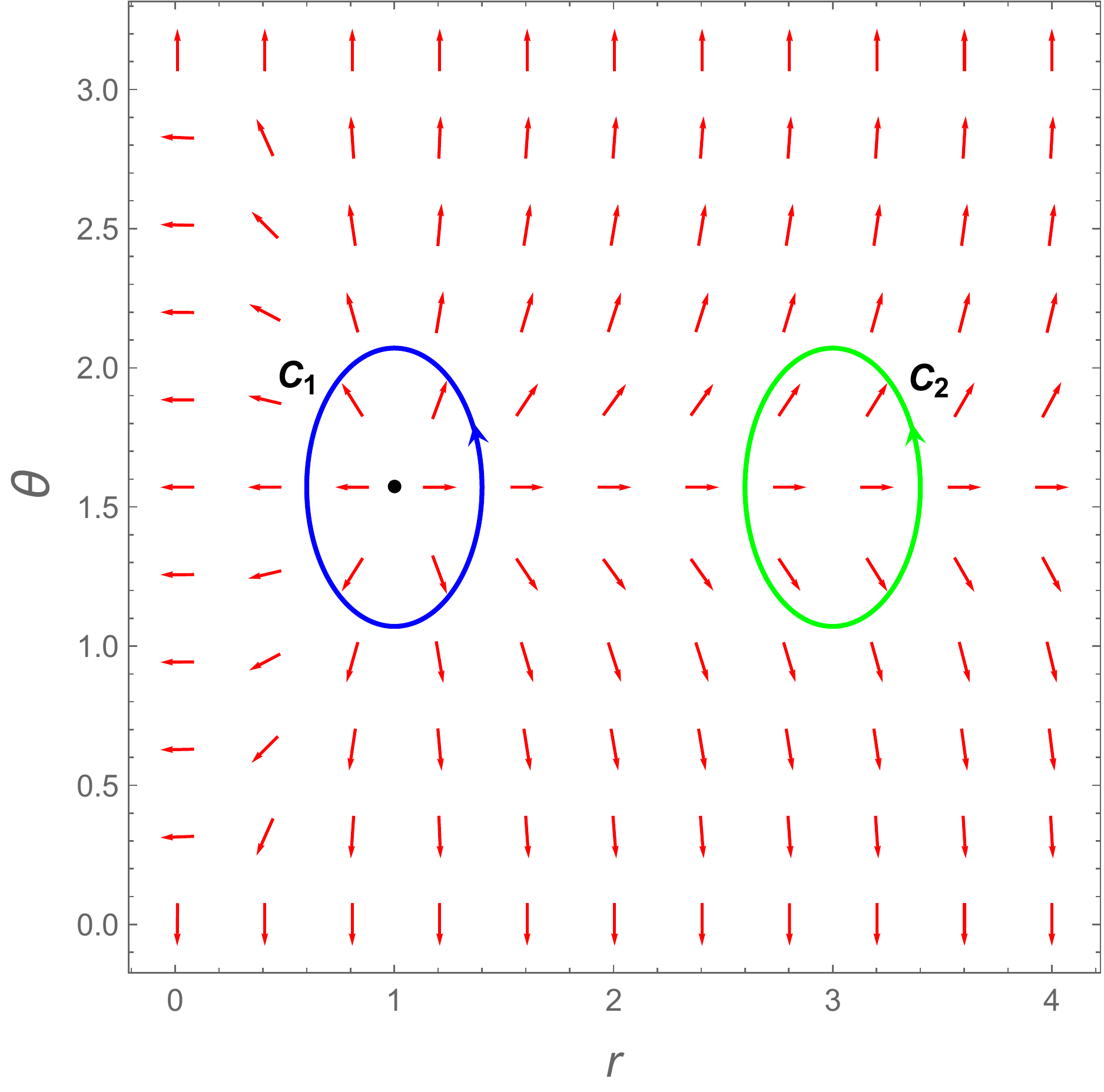}\label{Fig:sch_bulk_hp_vec_plot}}\hspace{0.5cm}	
		\subfloat[]{\includegraphics[width=3in]{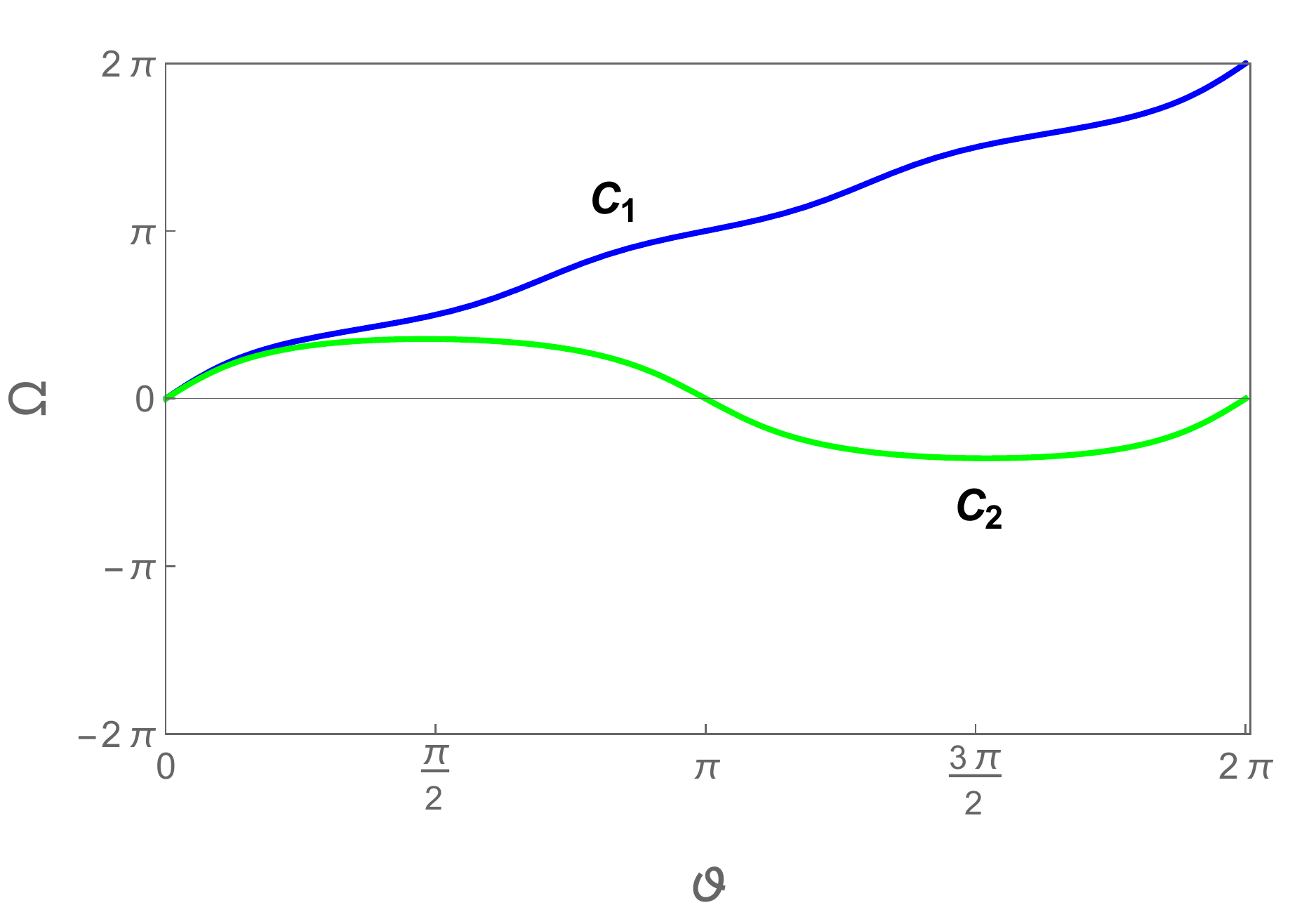}\label{Fig:sch_bulk_hp_omega_plot}}				
		
		\caption{\footnotesize For Schwarzschild black holes in $\text{AdS}_5$: (a) The  vector field $n$ in the $\theta-r$ plane, vanishes at Hawking-Page (HP) transition point (black dot, located at $r_{\rm HP} = 1$).  Contour $C_1$ contains the HP transition point, while the contour $C_2$ does not. 
			(b) $\Omega$ vs $\vartheta$ for contours $C_1$ (blue curve),   and $C_2$ (green curve). Here, parametric coefficients of the contours are $(A,B,r_0) =(0.4,0.5,1)$ for  $C_1$, and $(0.4,0.5,3)$ for $C_2$.} 
	}
\end{figure}
\vskip 0.2cm
\noindent
Now, in order to find the topological charge associated with the HP point, one defines a vector field $\phi (\phi^r, \,\phi^\theta)$, using $T_0 (r_+)$ as: 
\begin{eqnarray}
\phi^{r} &=& \partial_{r_+} \Phi, \\
\phi^{\theta} &=& \partial_{\theta} \Phi,
\end{eqnarray}
where, $\Phi=\frac{1}{\text{sin}\theta} T_{\rm 0}^{\phantom{0}} (r_+)$.
This vector field $\phi$ vanishes exactly at the Hawking-Page transition point, which can be seen clearly from the normalised vector field  $n=(\frac{\phi^{r}}{||\phi||},\frac{\phi^\theta}{||\phi||})$,  plotted in  Fig.~\ref{Fig:sch_bulk_hp_vec_plot}.
 The computation of topological charge can now be performed with the help of deflection angle $\Omega(\vartheta)$ (see Fig.~\ref{Fig:sch_bulk_hp_omega_plot}) and results in the assignment of a topological charge of $+1$ to the HP transition point~\cite{Yerra:2022coh}.
\vskip 0.2cm
\noindent
Further, in~\cite{Wei:2022dzw,Yerra:2022coh}, the topological charges carried by the black hole solutions i.e., the extremal points (equilibrium phases) of the Bragg-Williams off-shell free energy $f$, have been computed by defining the vector field $\phi$, such that: 
\begin{equation}\label{eq:vec field with f}
\phi(\phi^r, \phi^\theta) = \phi\Big(\frac{\partial f}{\partial r_+}, -\text{cot}\theta \text{csc}\theta \Big). 
\end{equation}
\begin{figure}[h!]
	{\centering
		\subfloat[]{\includegraphics[width=2in]{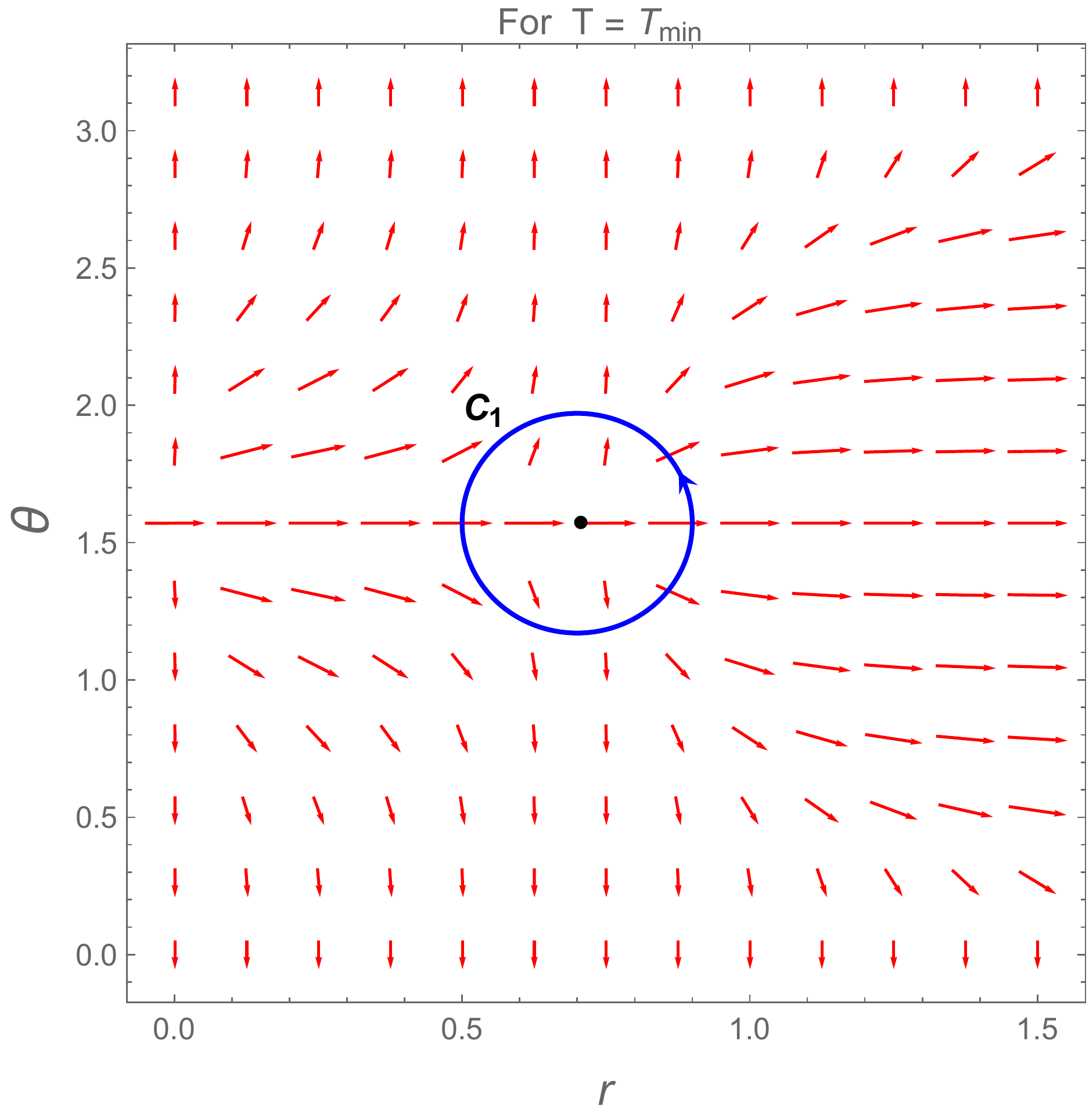}}\hspace{1.5cm}
		\subfloat[]{\includegraphics[width=2.5in]{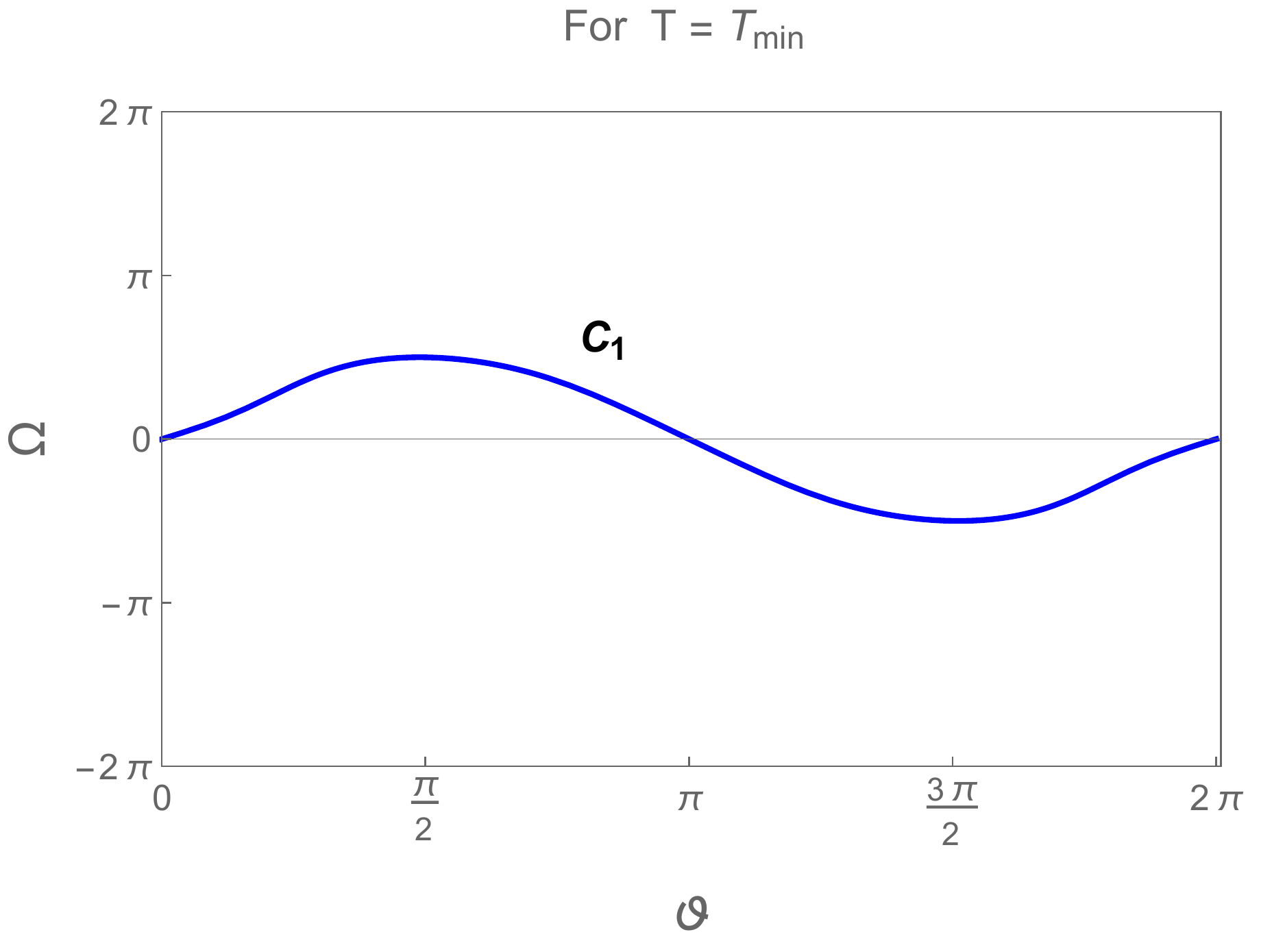}}\hspace{1cm}
		\subfloat[]{\includegraphics[width=2in]{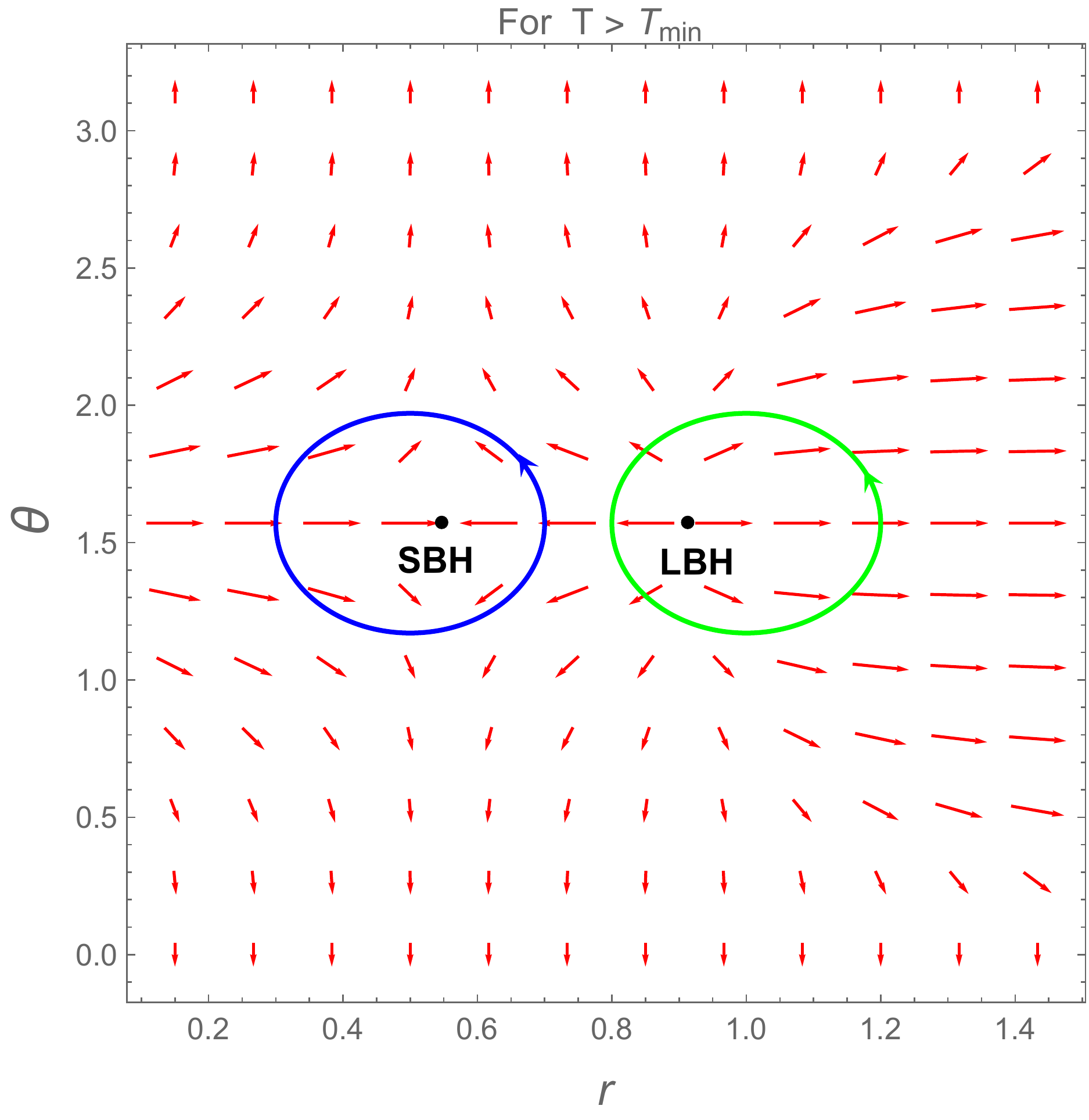}}\hspace{1.5cm}
		\subfloat[]{\includegraphics[width=2.5in]{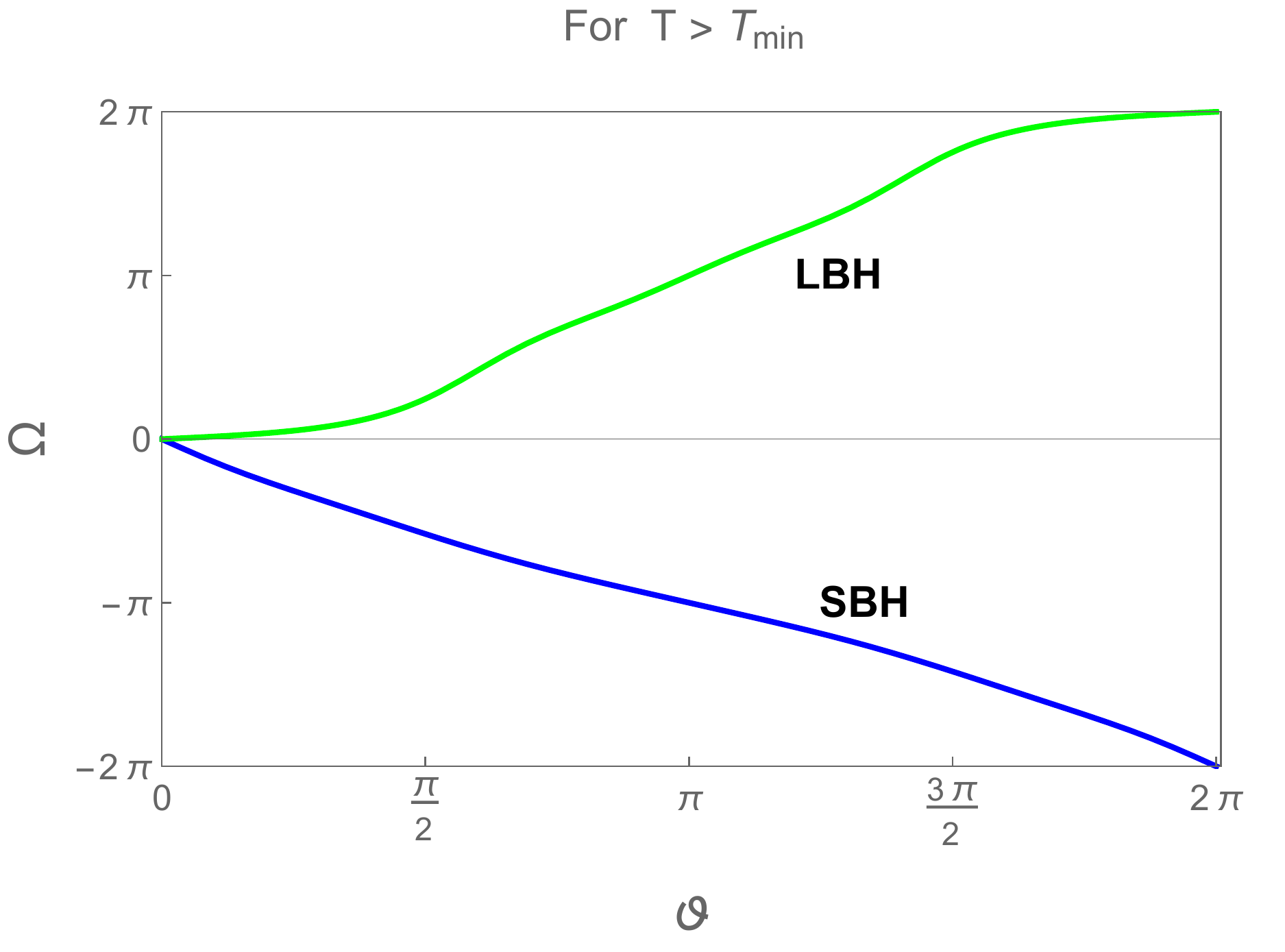}}
			
		\caption{\footnotesize For Schwarzschild black holes in $\text{AdS}_5$: Left panel shows the vanishing of vector field $\phi$ at extremal points of $f$ (black dots) i.e., at small black holes (SBH) and large black holes (LBH) and also at the black hole possessing lowest temperature $T_{\rm min}$. Right panel shows the behaviour of deflection angle $\Omega (\vartheta)$ for corresponding black holes in left panel.} 
		\label{fig:sch_bulk_equi_4plots}	}
\end{figure}
\noindent
A similar computation shows that the small black holes carry the topological charge $-1$, and all the large
black holes carry  the topological charge $+1$, while the black hole with minimum temperature $T_{\rm min}$, carries no topological charge (the corresponding plots of normalised vector field $n$ and deflection angle $\Omega(\vartheta)$ are shown in Fig.~\ref{fig:sch_bulk_equi_4plots}).
\vskip 0.2cm
\noindent
We now check the results obtained above by considering a phenomenological boundary matrix dual that reproduces the known qualitative features of the bulk theory.

\subsection{Boundary}\label{section:boundary for sch} 
To carry out a similar computation on the boundary, we use the $(a,b)$ model 
which captures, among others, the deconfining transition of $\mathcal{N}=4$, $SU(N)$ guage theory 
at large $N$. In what follows, we give a brief description of the model, more details can be found in
~\cite{Alvarez-Gaume:2005dvb,Aharony:2003sx}. We will closely follow the discussions that appeared 
in~\cite{Dey:2006ds,Dey:2008bw,Chandrasekhar:2012vh}.    
Here one starts by  writing down the partition function for SYM theory  as a matrix integral over the 
effective action  
involving Wilson-Polyakov loop operator $(\text{tr}U)/N$
\begin{equation}
Z(\lambda, T)=  \int dU e^{S_\text{eff}(U)},
\end{equation}
where, $U = P\, \text{exp}(i \int_{0}^{\beta} A d\tau)$ is the unitary $U(N)$ matrix, and $A(\tau)$ is the zero mode of the time component of the
gauge field in $S^3$. In general $S_\text{eff}(U)$ is a polynomial in the  traces of $U$ and its powers that 
are allowed by the $Z_N$ symmetry. The phenomenological $(a,b)$ model is a truncated version of the above where only 
a couple of 
terms are retained in $S_\text{eff}(U)$. It has the form
\begin{equation}\label{eq: a_b_matrixmodel_for_sch}
Z(a, b) = \int dU \, \text{exp}[ a (\text{tr}U \, \text{tr}U^{\dag}) + \frac{b}{N^2} (\text{tr}U \, \text{tr}U^\dag)^2].
\end{equation}
Here, $a$ and $b$ are the parameters of the model which have nontrivial dependence on the  temperature $T$ and 
the ’t Hooft coupling $\lambda$. The temperature dependence, in particular can be found as
follows~\cite{Alvarez-Gaume:2005dvb,Dey:2006ds}. 
The effective 
potential
arising from this model can be written in terms of the order parameter $\rho$ ( which is the expectation value of the Polyakov loop $ \frac{1}{N} \langle\text{tr} \, U\rangle$) that characterises the deconfined phase of the gauge theory as:
\begin{eqnarray}\label{eq:Veff_sch_boundary}
V(\rho) &=& \frac{1-a}{2}\rho^2 -\frac{b}{2} \rho^4      \hspace{4cm} \text{for} \quad 0 \leq \rho \leq \frac{1}{2}   \nonumber \\
&=& -\frac{a}{2} \rho^2 -\frac{b}{2} \rho^4 -\frac{1}{4} \text{log}[2(1-\rho)] +\frac{1}{8}  \hspace{0.8cm}  \text{for} \quad \frac{1}{2} \leq \rho \leq 1, 
\end{eqnarray}  
where, $\rho^2 = (\frac{1}{N^2}) \text{tr}\, U \text{tr}\, U^\dag$. The saddle point equation is given by
\begin{eqnarray}
a\rho +2b\rho^3 &=& \rho   \hspace{2cm} \text{for} \quad 0 \leq \rho \leq \frac{1}{2}   \nonumber \\
&=& \frac{1}{4(1-\rho)}  \hspace{0.8cm}  \text{for} \quad \frac{1}{2} \leq \rho \leq 1. 
\end{eqnarray} 
 The parameters are bounded as $a <1$ and $b>0$.  The temperature dependence on the parameters are then fixed 
using the bulk data.
Let us consider a temperature $T > T_\text{\rm min}$, with $T_\text{\rm min}$ being the black 
hole 
pair-nucleation temperature. At this temperature, we can write:
\begin{eqnarray}
2a \rho_{\rm 1,2}^2 + 2 b \rho_{\rm 1,2}^4 +\text{log}(1-\rho_{\rm 1,2})+f &=&-I_{\rm 1,2}, \label{eq_sch_boundary_V=I}\\
a \rho_{\rm 1,2} + 2 b \rho_{\rm 1,2}^3 &=& \frac{1}{4(1-\rho_{\rm 1,2})}. \label{eq_sch_boundary_saddle_pt}
\end{eqnarray}
\begin{figure}[h!]
	
	{\centering
		
		\subfloat{\includegraphics[width=2.8in]{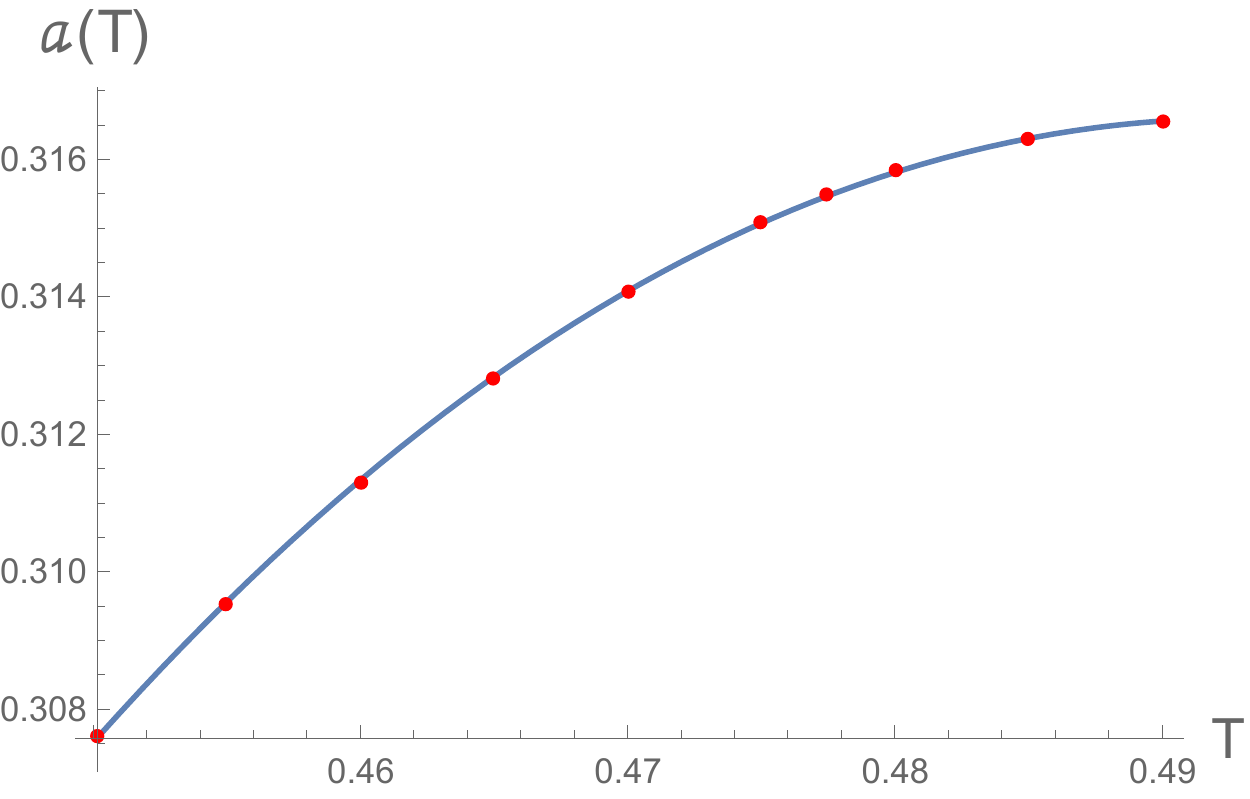}}\hspace{1cm}	
		\subfloat{\includegraphics[width=2.8in]{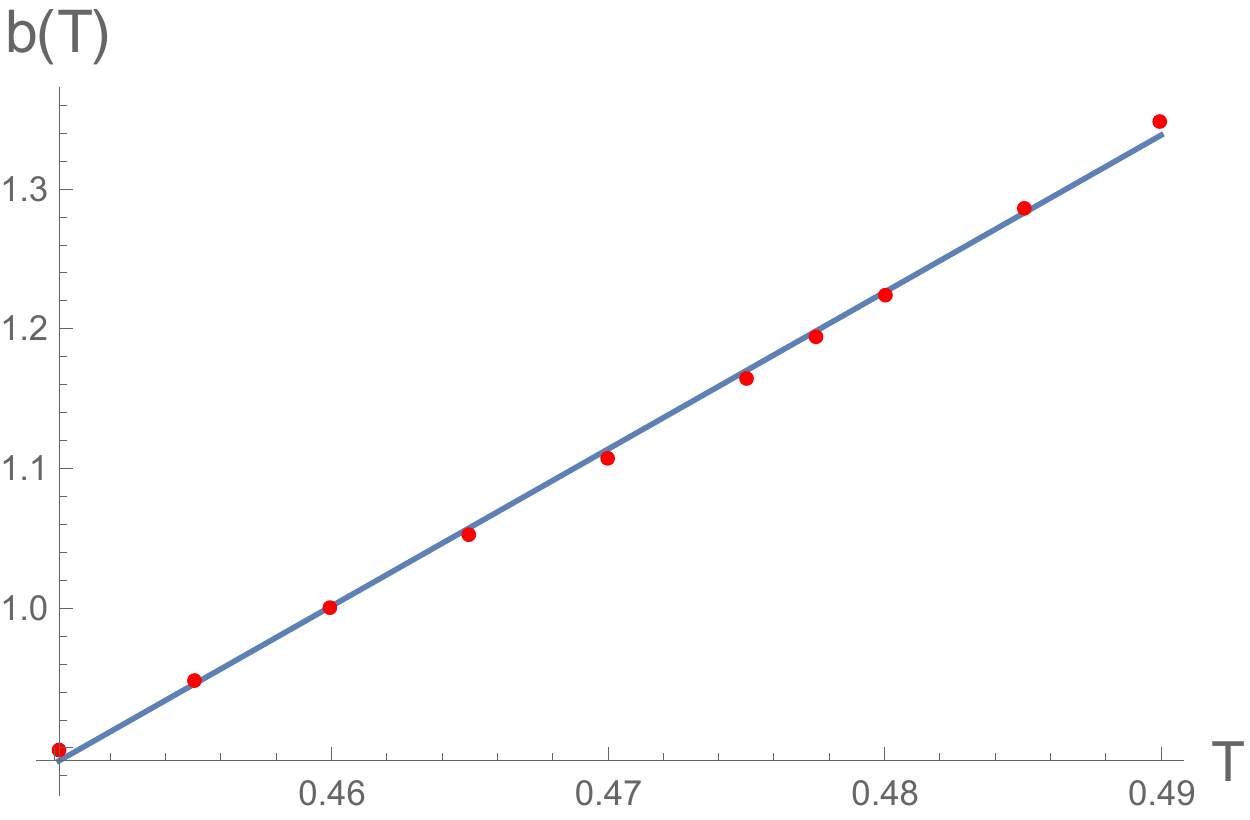}}	
		
		\caption{\footnotesize For $(a,b)$ matrix model with zero chemical potential: The temperature $T$ dependence of the parameters $a(T)$ and $b(T)$. Red dots indicate the data points, while blue colored curves are their fitting curves . For $a(T)$, the fitting curve is $a(T) =c_1 \text{log}(T) +c_2 T +c_3$, and for  $b(T)$, it is $b(T) =c_4 T +c_5$. $(\text{Here,}\, c_1=2.257, c_2=-4.579, c_3=4.171, c_4=11.239, c_5=-4.169.)$.}
		\label{fig:sch_boundary_a_b_plots}	}
	
\end{figure}
\noindent
Here, $I_{\rm 1,2}$~\footnote{ $I = \pi r_+^3\Big(\frac{1-r_+^2}{1+2r_+^2}\Big)$~\cite{Alvarez-Gaume:2005dvb}. The comparison between
the matrix model potential and the bulk action is valid only when one neglects the string loop effects.} are the actions for the small and the large black holes. $\rho_{\rm 1,2}$ are the corresponding
solutions in the matrix model. The constant $f = \text{log}(2)-1/2 $   is introduced to make the potential continuous at $\rho = 1/2$.
On solving the four equations \eqref{eq_sch_boundary_V=I} and \eqref{eq_sch_boundary_saddle_pt}, we obtain $a$, $b$ and $\rho_{1,2}$ at a given temperature. Thus, the dependence of the parameters $a(T)$ and $b(T)$ on the temperature $T$ can be obtained numerically and the result
is shown in the Fig.~\ref{fig:sch_boundary_a_b_plots}.
\begin{figure}[h!]
	
	{\centering
		
		\subfloat[]{\includegraphics[width=3in]{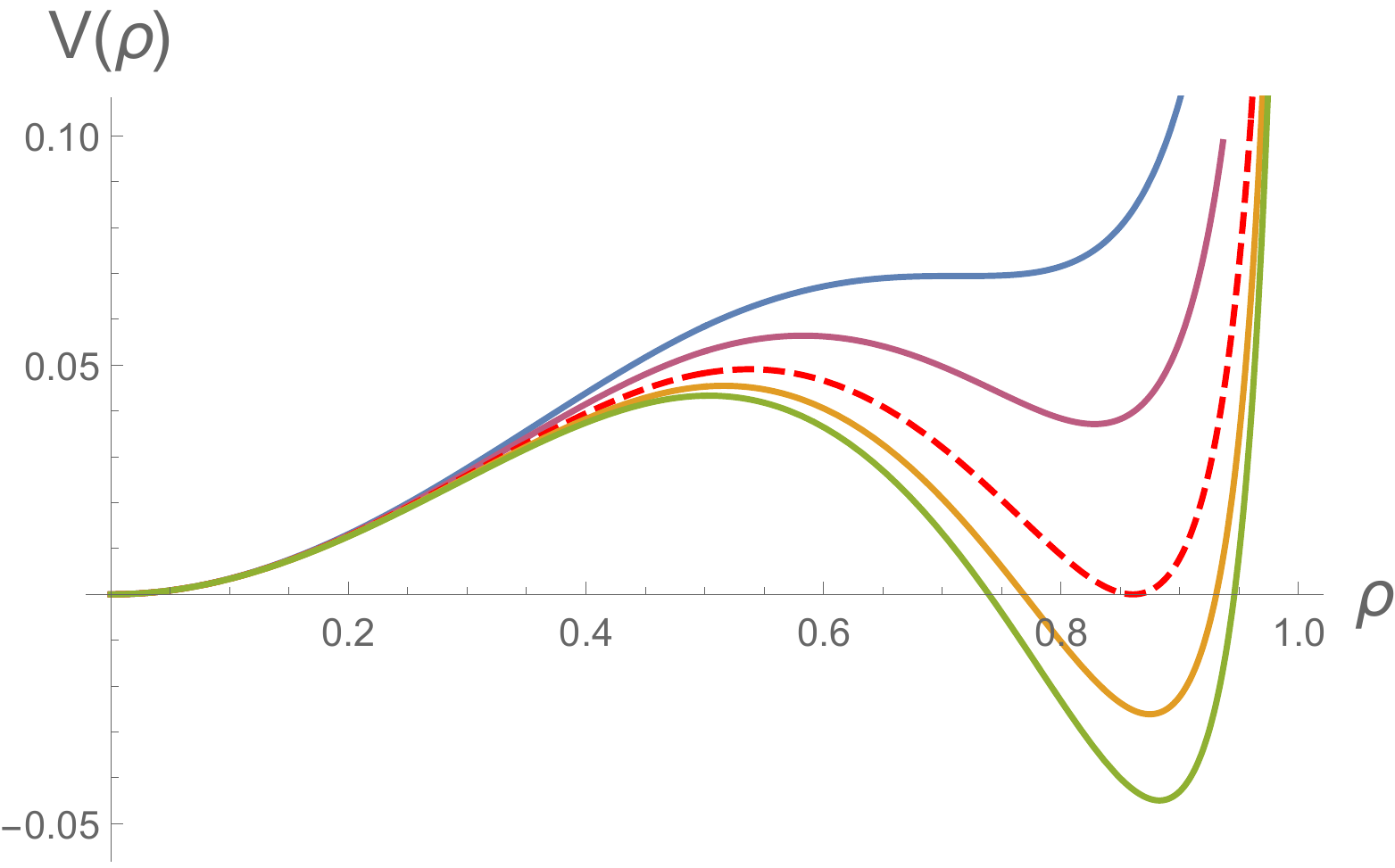}	\label{fig:sch_boundary_effective_potential_plot}}\hspace{0.5cm}	
		\subfloat[]{\includegraphics[width=2.8in]{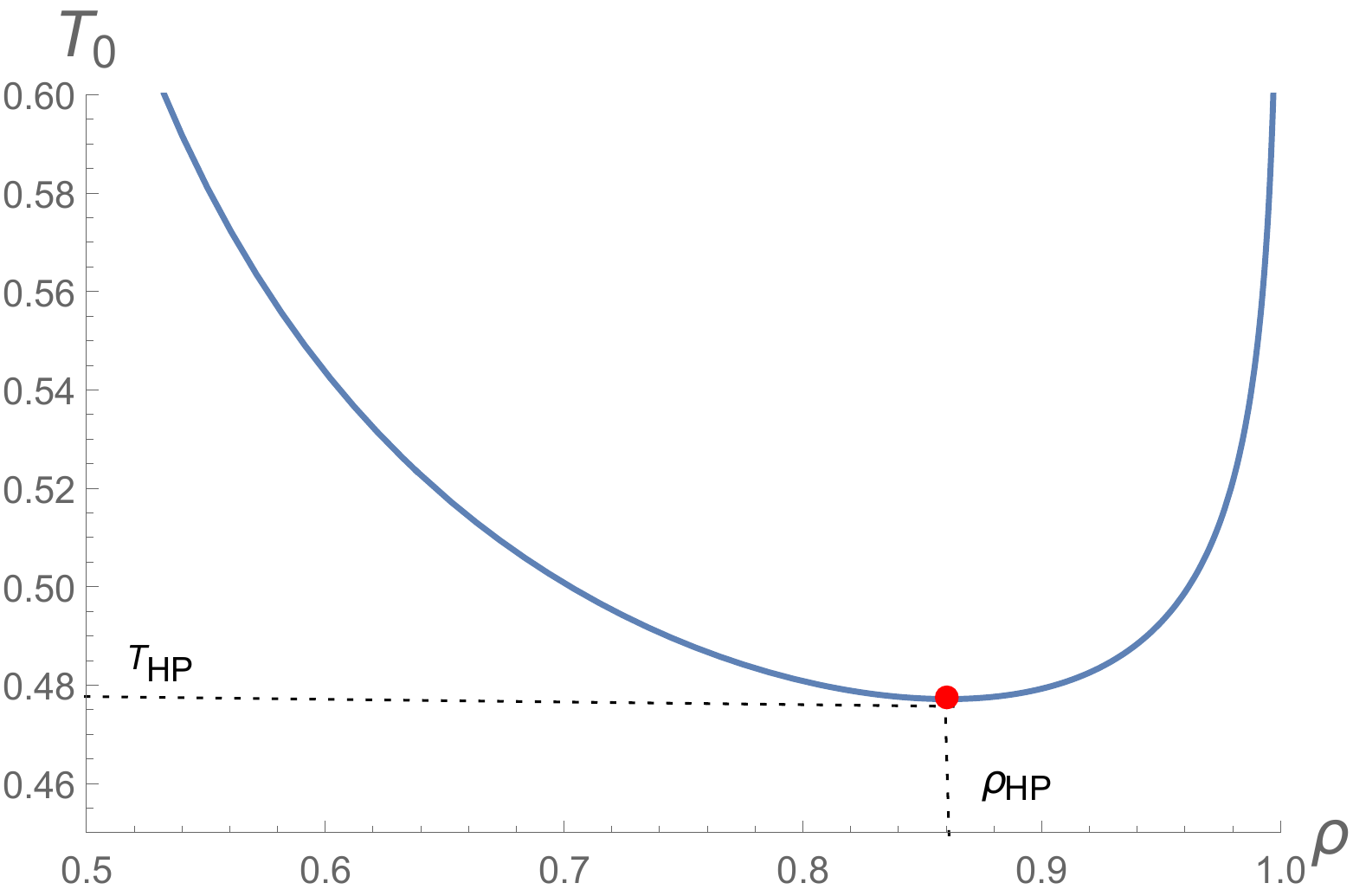}	\label{fig:sch_boundary_T0_plot}}	
		
		\caption{\footnotesize For $(a,b)$ matrix model with zero chemical potential: (a) Behaviour of the effective potential $V(\rho)$ at various temperatures $T$. Temperature of the curves increases from top to bottom. Blue curve is at $T_{\rm min}$, and dashed red curve,  showing the HP transition at $\rho_{\rm HP} = 0.861$, is at $T_{\rm HP}$ for which $a=0.3155$ and $b=1.194$. (b) The curve $T_0$ shows the phase transition point at its minima (red dot).}
		}	
\end{figure}
\vskip 0.2cm
\noindent
Now, the behaviour of the effective potential $V(\rho)$ at various temperatures can be seen from the Fig.~\ref{fig:sch_boundary_effective_potential_plot}.
In  $\rho > 1/2$ region, the saddle points of $V(\rho)$  reproduce the phase structure of bulk 
theory~\cite{Alvarez-Gaume:2005dvb}. For a given temperature $T > T_{\rm min}$, there exist two saddle points, among which, the minimum represents the stable large black hole (LBH) phase, while the maximum represents the unstable small black hole (SBH) phase on gravity side.  For $T=T_{\rm min}$, the saddle point, which is 
neither a minimum nor a maximum, represents the black hole pair-nucleation point. 
$T=T_{\rm HP}$, for which degenerate minima appear at $\rho = 0$ and $\rho=0.861$   
represents the HP transition temperature in the bulk and the deconfining temperature at the boundary.
We note here that, $\rho = 0$ is always a solution and represents the thermal $\text{AdS}_5$ phase.  
We also note here that, the supergravity description breaks down in the $\rho \leq 1/2 $ region. However, at $\rho=1/2$, the matrix model shows a third-order phase transition called Gross-Witten transition~\cite{Gross:1980he}, though it is not visible in supergravity, it has a natural interpretation in string theory as a Horowitz-Polchinski correspondence point~\cite{Horowitz:1996nw}.   
\begin{figure}[h!]
	{\centering
		\subfloat[]{\includegraphics[width=2.8in]{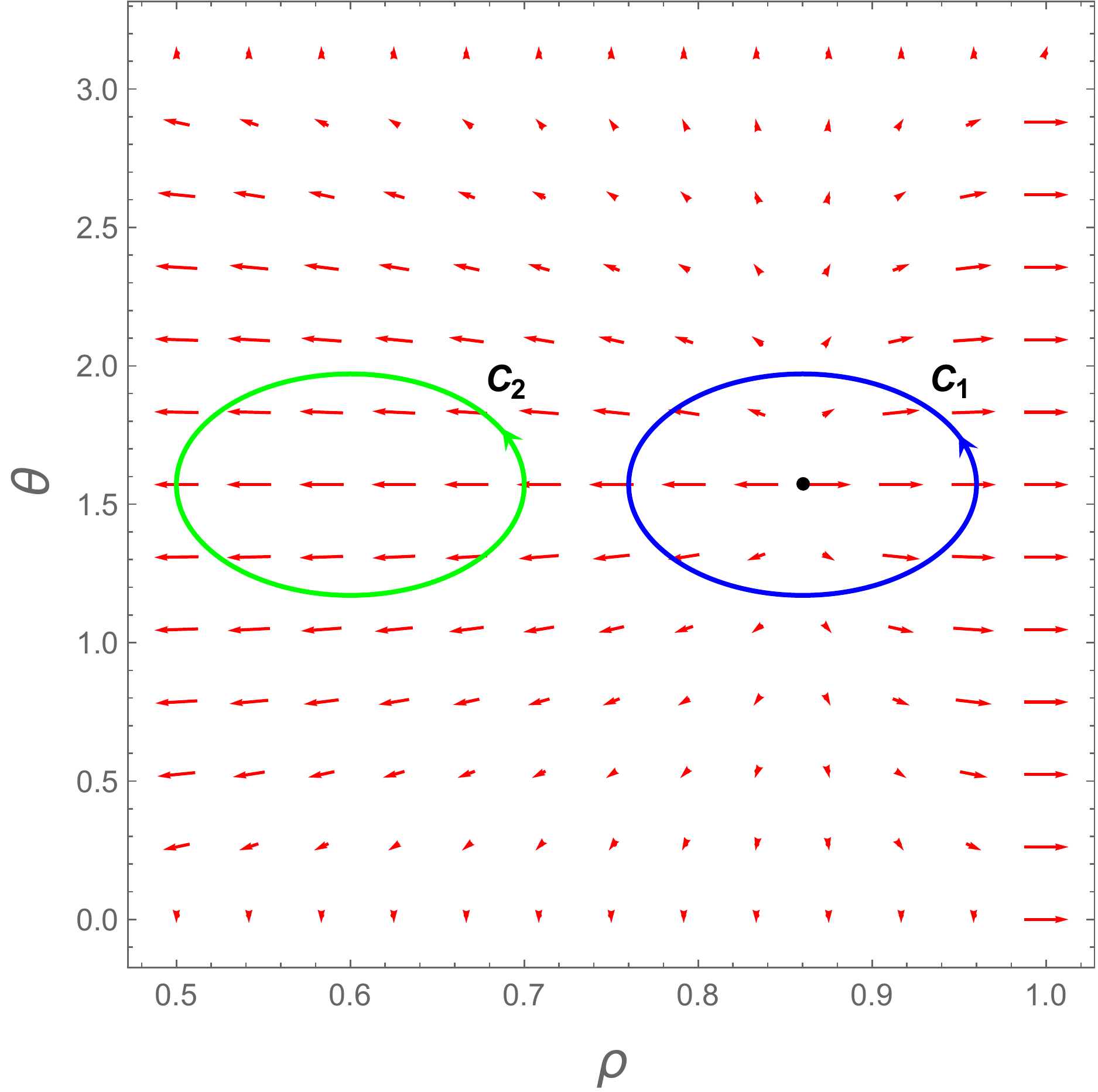}\label{Fig:sch_boundary_hp_vec_plot}}\hspace{0.7cm}	
		\subfloat[]{\includegraphics[width=3in]{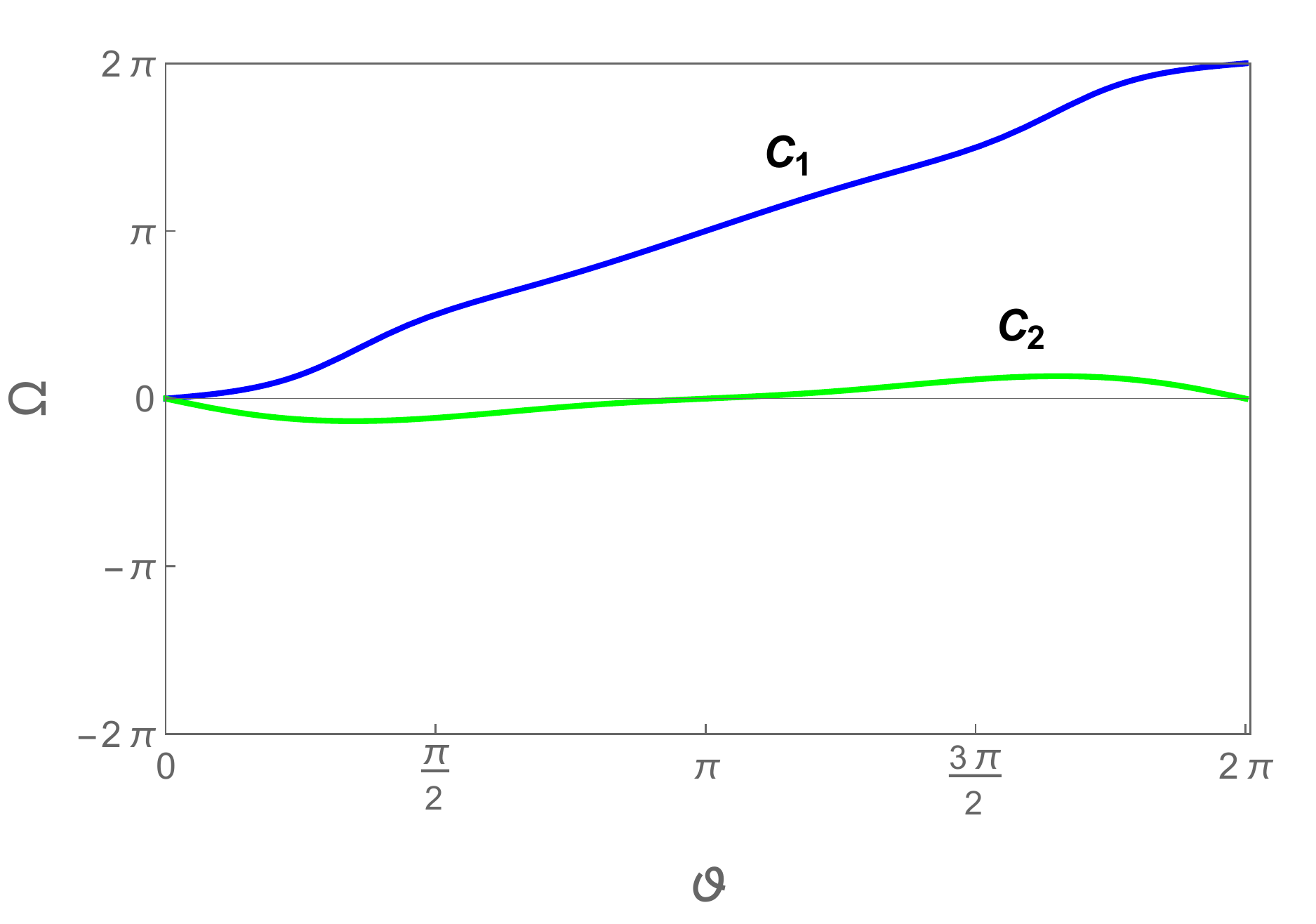}\label{Fig:sch_boundary_hp_omega_plot}}				
		
		\caption{\footnotesize For $(a,b)$ matrix model with zero chemical potential: (a) The normalised vector field of $\phi$ in the $\theta-\rho$ plane, vanishes at the phase transition point (black dot, located at $\rho_{\rm HP} = 0.861$) that represents HP transition in the bulk.  Contour $C_1$ contains the transition point, while the contour $C_2$ does not. 
			(b) $\Omega$ vs $\vartheta$ for contours $C_1$ (blue curve),   and $C_2$ (green curve). Here, parametric coefficients of the contours are $(A,B,\rho_0) =(0.1,0.4,0.86)$ for  $C_1$, and $(0.1,0.4,0.6)$ for $C_2$.} 
	}
\end{figure}
\vskip 0.2cm
\noindent
Now, we turn to compute the topological charge (winding number) associated with the phase transition point of $V(\rho)$, that represents HP transition in the bulk.
This can be done by defining a temperature $T_0(\rho)$, where $V(\rho) = 0$ in $\rho > 1/2 $ region, and using the fitting curves for $a(T)$ and $b(T)$, then one obtains:
\begin{equation}
	T_0(\rho) = \frac{c_1}{(c_2 + c_4 \rho^2)}\, \text{ProductLog} \Big[ \frac{e^{\frac{1-4c_3 \rho^2 -4 c_5 \rho^4}{4c_1 \rho^2}} (2-2\rho)^{-\frac{1}{2c_1 \rho^2}} (c_2 + c_4\rho^2)}{c_1} \Big],
\end{equation}
where, its minima, shown in Fig.~\ref{fig:sch_boundary_T0_plot}, gives the HP transition of the bulk. Then, using $T_0(\rho)$, we define the vector field $\phi (\phi^\rho, \phi^\theta)$, as:
\begin{eqnarray}
\phi^\rho &=& \partial_\rho \big(T_0/\text{sin}\theta \big),\\
\phi^\theta &=& \partial_\theta \big(T_0/\text{sin}\theta \big).
\end{eqnarray} 
As can be seen in Fig.~\ref{Fig:sch_boundary_hp_vec_plot}, this vector filed $\phi$ vanishes exactly at the phase transition point of $V(\rho)$, that represents  the HP transition in the bulk. For this transition point, the computation of the topological charge (using the deflection angle $\Omega(\vartheta)$ from the Fig.~\ref{Fig:sch_boundary_hp_omega_plot}), turns out to have the topological charge $+1$, which matches exactly with the topological charge of the HP transition  in the bulk. 
\begin{figure}[h!]
	{\centering
		\subfloat[]{\includegraphics[width=2in]{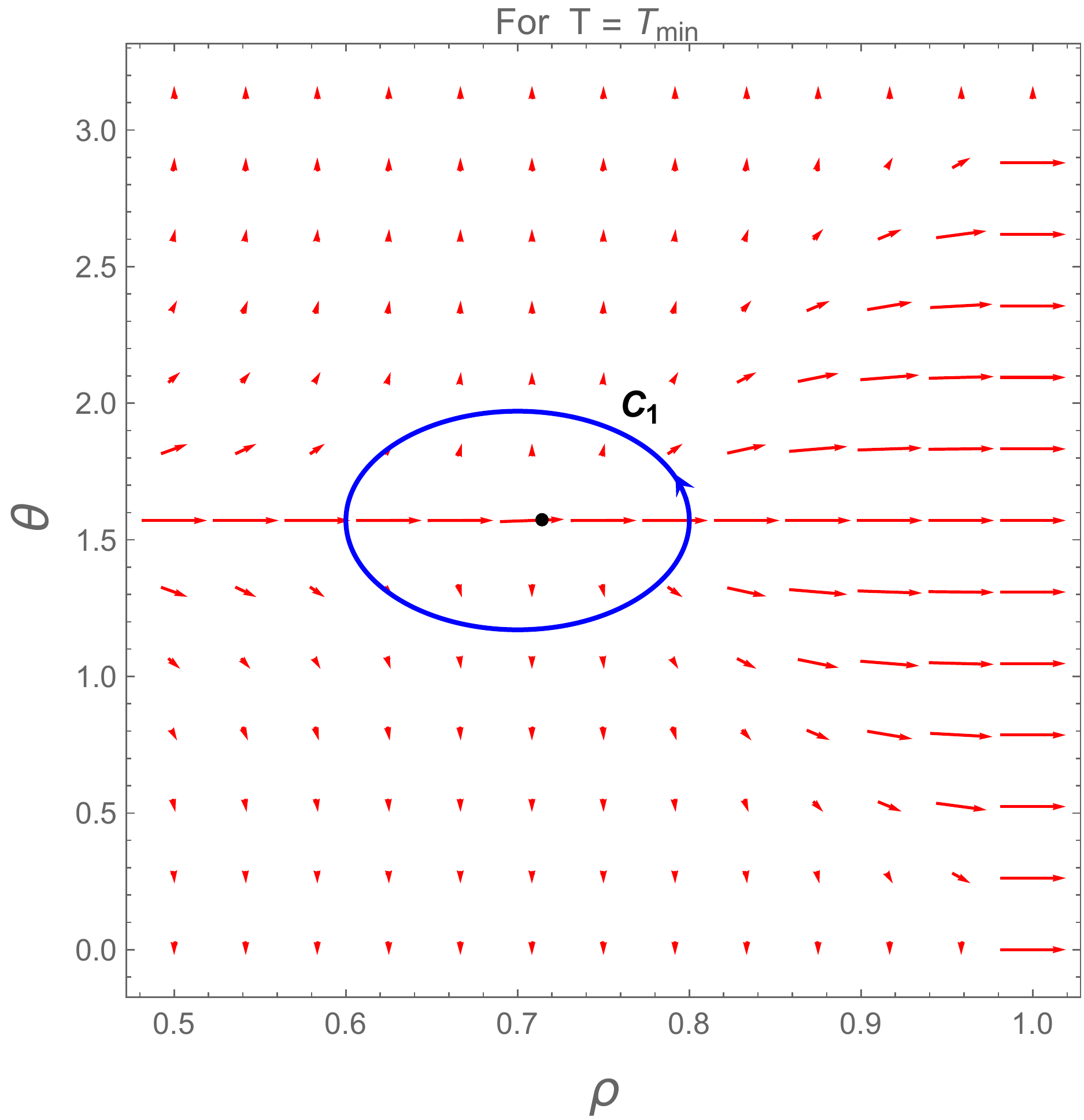}}\hspace{1.5cm}
		\subfloat[]{\includegraphics[width=2.5in]{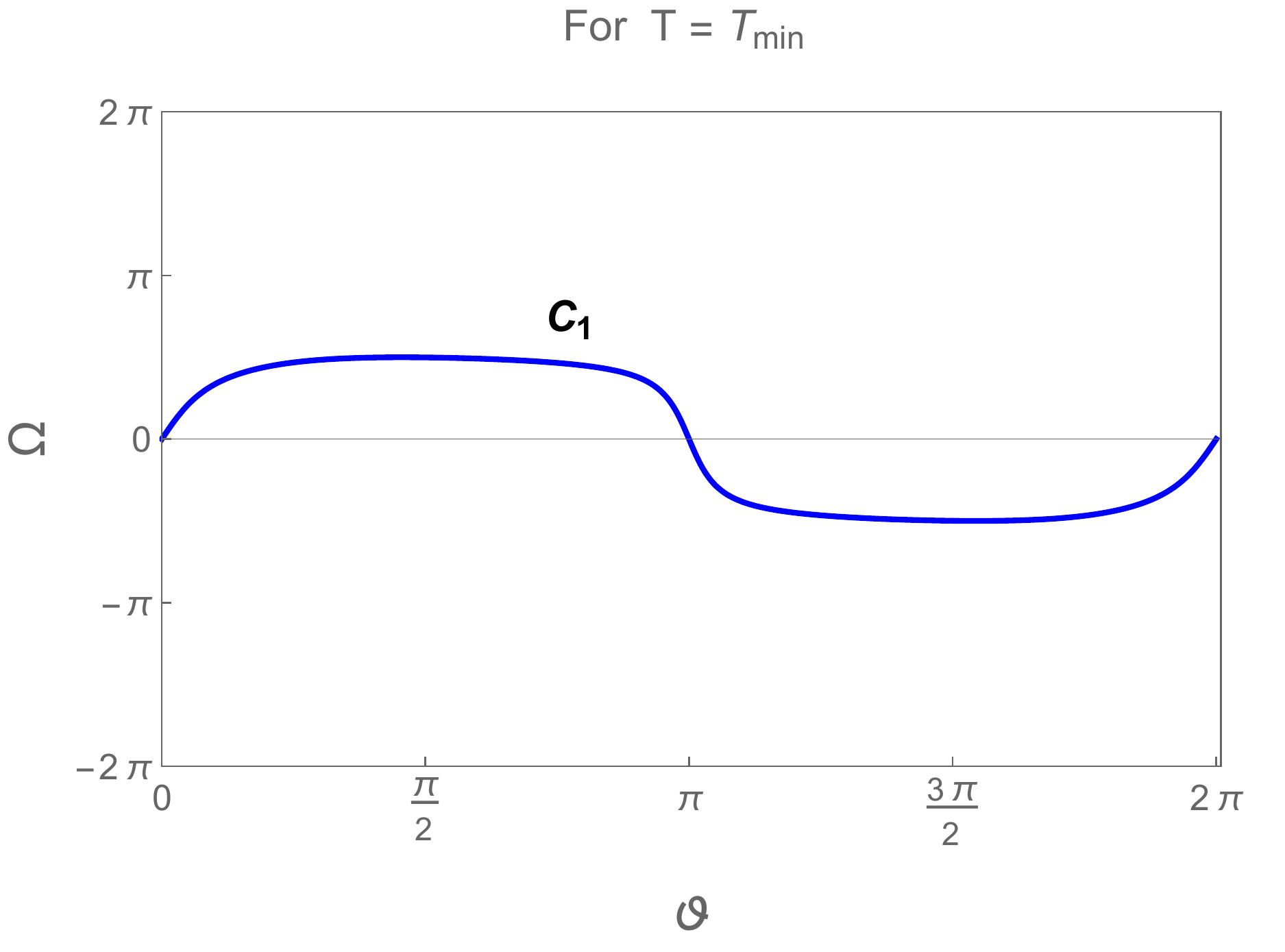}}\hspace{1cm}
		\subfloat[]{\includegraphics[width=2in]{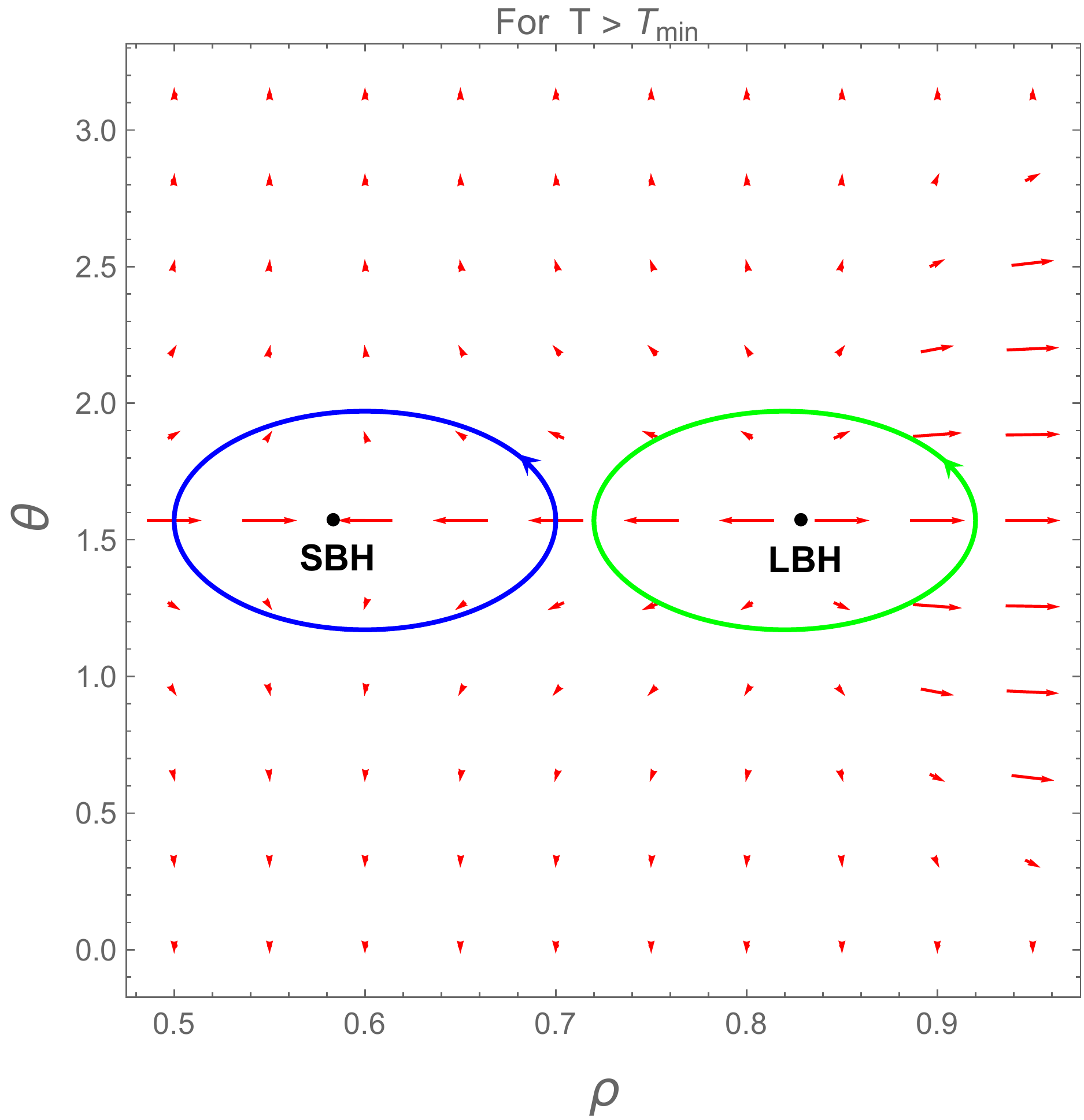}}\hspace{1.5cm}
		\subfloat[]{\includegraphics[width=2.5in]{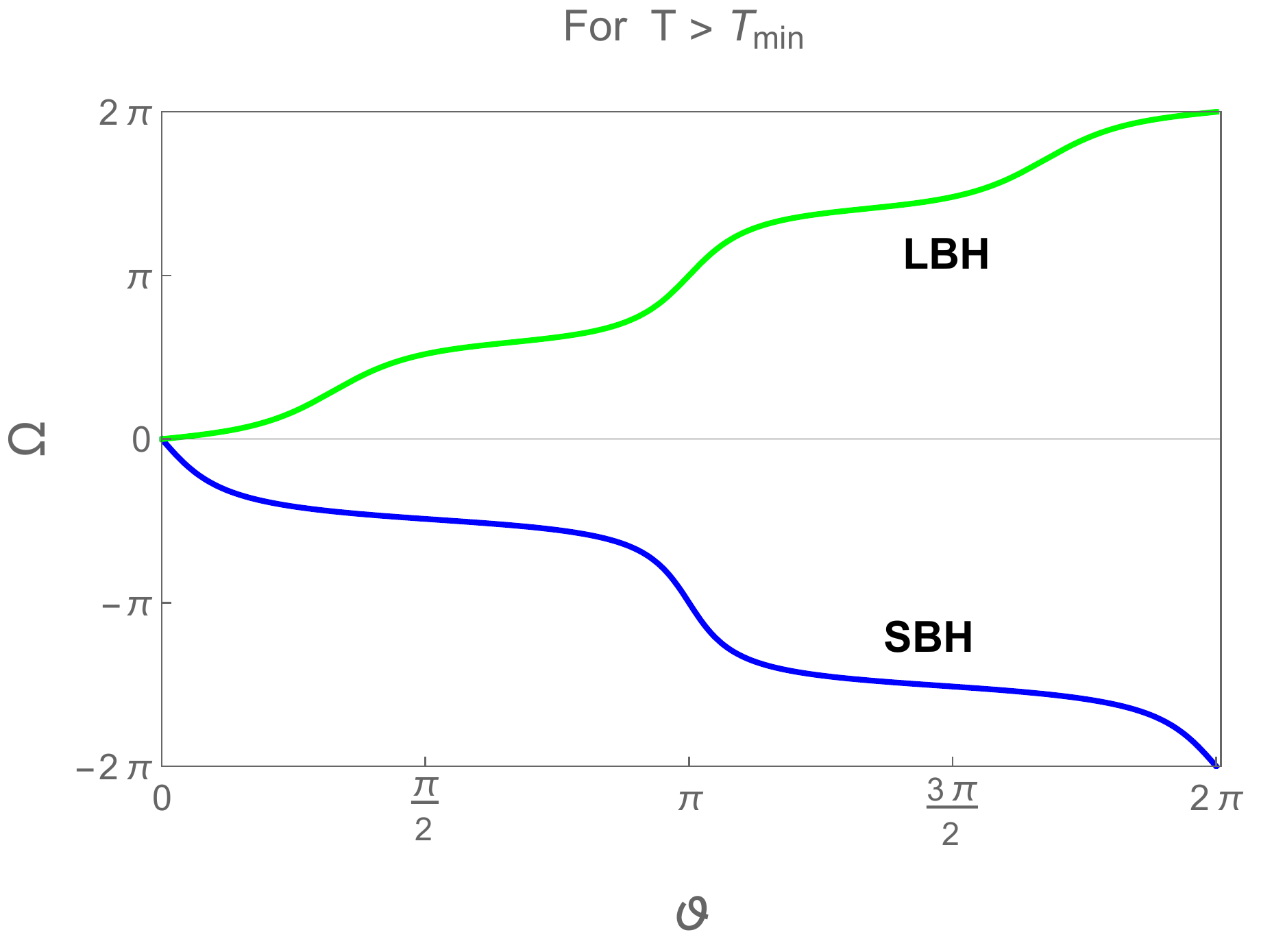}}
		
		\caption{\footnotesize For $(a,b)$ matrix model with zero chemical potential: Left panel shows the vanishing of vector field $\phi$ at saddle points of $V(\rho)$ (black dots). Right panel shows the behavior of deflection angle $\Omega (\vartheta)$ for corresponding saddle points in left panel.} 
		\label{fig:sch_boundary_equi_4plots}	}
\end{figure}
\vskip 0.2cm
\noindent
We now move to compute the topological charges carried by the saddle points (equilibrium phases) of $V(\rho)$, that represent the black hole solutions in the bulk.  We use the effective potential $V(\rho)$ in $\rho > 1/2$ region, to define the vector field, as:
\begin{equation}
\phi(\phi^\rho, \phi^\theta) = \phi\Big( \frac{\partial V}{\partial \rho}, -\text{cot}\theta \, \text{csc}\theta \Big).
\end{equation} 
This vector field $\phi$ vanishes exactly at the  saddle points of $V(\rho)$. The computation of topological charges for these saddle points reveals that  all the stable (minima) saddle points (which represent the large black holes in the bulk) carry the topological charge $+1$, whereas,  all the unstable (maxima) saddle points (which represent the small black holes in the bulk) carry the topological charge $-1$,  while,  the saddle point which is neither maxima nor minima (which represents the black hole with minimum temperature in the bulk) carries no topological charge (See Fig.~\ref{fig:sch_boundary_equi_4plots} for the plots of corresponding vector field and deflection angle.). Therefore, these topological charges are in conformity with the corresponding ones in the bulk. 

\section{Black hole and its matrix dual: non-zero chemical potential} \label{two}

In this section, we  consider the R-charged black holes in AdS$_5$ and compute 
the topological charges of the phase transition point as well as of various equilibrium 
phases using an off-shell free energy, both in the canonical and grand canonical ensemble. 
We only briefly mention the results for the computation in the grand canonical ensemble, 
as the details are already available in~\cite{Yerra:2022coh}. For the canonical ensemble, 
the topological charges of the critical points were computed in~\cite{Wei:2021vdx}, 
but our result differs from theirs and hence the details are presented here. 
One reason for the difference may possibly be because the phase transition in~\cite{Wei:2021vdx} was studied 
in the space of extended thermodynamics with different parameters. We then perform the computation in the 
$(a,b)$ matrix model, now generalised to allow $a$ and $b$ to depend on the chemical potential, 
in addition to the temperature. 

\subsection{Grand canonical ensemble: fixed potential} \label{bulkgrand}
\subsubsection{Bulk}

In the grand canonical ensemble, the phase structure  of  Reissner-Nordstrom-AdS$_5$  black holes at  fixed potential $\mu < \sqrt{3}/2 $~\cite{Chamblin:1999tk,Chamblin:1999hg}, is same as that of the Schwarzschild-AdS$_5$ black holes that we studied in the section~\eqref{section:sch}, i.e., there exist a minimum temperature $T_{\rm min}$, above which the small (unstable) and large (stable) black holes  are formed. The HP transition happens again between the larger  black hole and its thermal AdS$_5$ background.  
The expression for black hole temperature now depends on the chemical potential $\mu$, and is given by~\cite{Chamblin:1999tk,Chamblin:1999hg}:
\begin{equation}
T(r_+, \mu) = \frac{(1-\frac{4}{3}\mu^2)+2r_+^2}{2\pi r_+},
\end{equation}
where its minimum temperature is $T_{\rm min} = \frac{\sqrt{2-\frac{8\mu^2}{3}}}{\pi}$. The HP transition happens at:
\begin{equation}
 T_{\rm HP} = \frac{3}{2\pi}\sqrt{1-\frac{4\mu^2}{3}}, \quad \text{and}\quad r_{\rm HP} = \sqrt{1-\frac{4\mu^2}{3}}.
\end{equation}
The BW free energy  can be computed now as~\cite{Banerjee:2010ve} (See the Appendix-(\ref{A2})  for details):
\begin{equation}
f(r_+, T, \mu)= M-TS-\mu Q= 3r_+^2\big(1-\frac{4\mu^2}{3}\big)-4\pi r_+^3T + 3r_+^4,
\end{equation}
where $Q$ is the electric charge of the black hole with conjugate chemical potential $\mu$.
One can see that for a fixed potential $\mu$, the plot of BW free energy at various temperatures is similar to Fig.~\ref{fig:sch_bulk_free_energy_plot}.  
\vskip 0.2cm
\noindent 
Performing a computation similar to the one discussed in section~\eqref{section:sch}, the topological charges for HP transition and  equilibrium phases of these black holes at a fixed potential $\mu$, computed earlier in~\cite{Yerra:2022coh}, can be reproduced. The results are found to be same as the of Schwarzschild-AdS$_5$ black hole case, i.e., the topological charge for HP point is $+1$, and for SBH/LBH/black hole with lowest temperature, it is -1/+1/0, respectively.
\vskip 0.2cm
\noindent 
We now move to compute the corresponding topological charges  for its boundary matrix dual to facilitate a
comparison.
\subsubsection{Boundary} \label{boundarygrand}
\begin{figure}[h!]
	
	{\centering
		
		\subfloat{\includegraphics[width=2.8in]{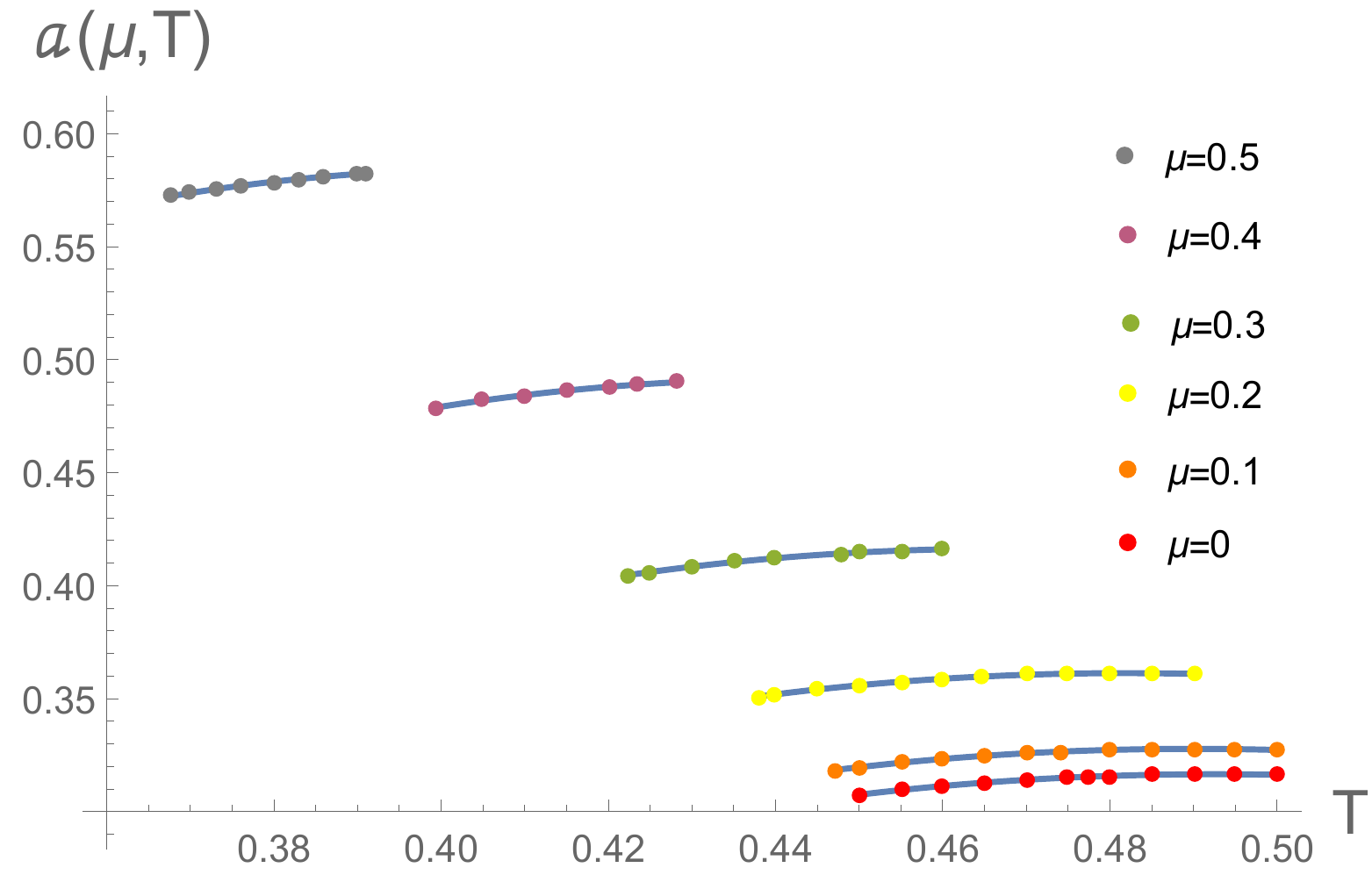}}\hspace{1cm}	
		\subfloat{\includegraphics[width=2.8in]{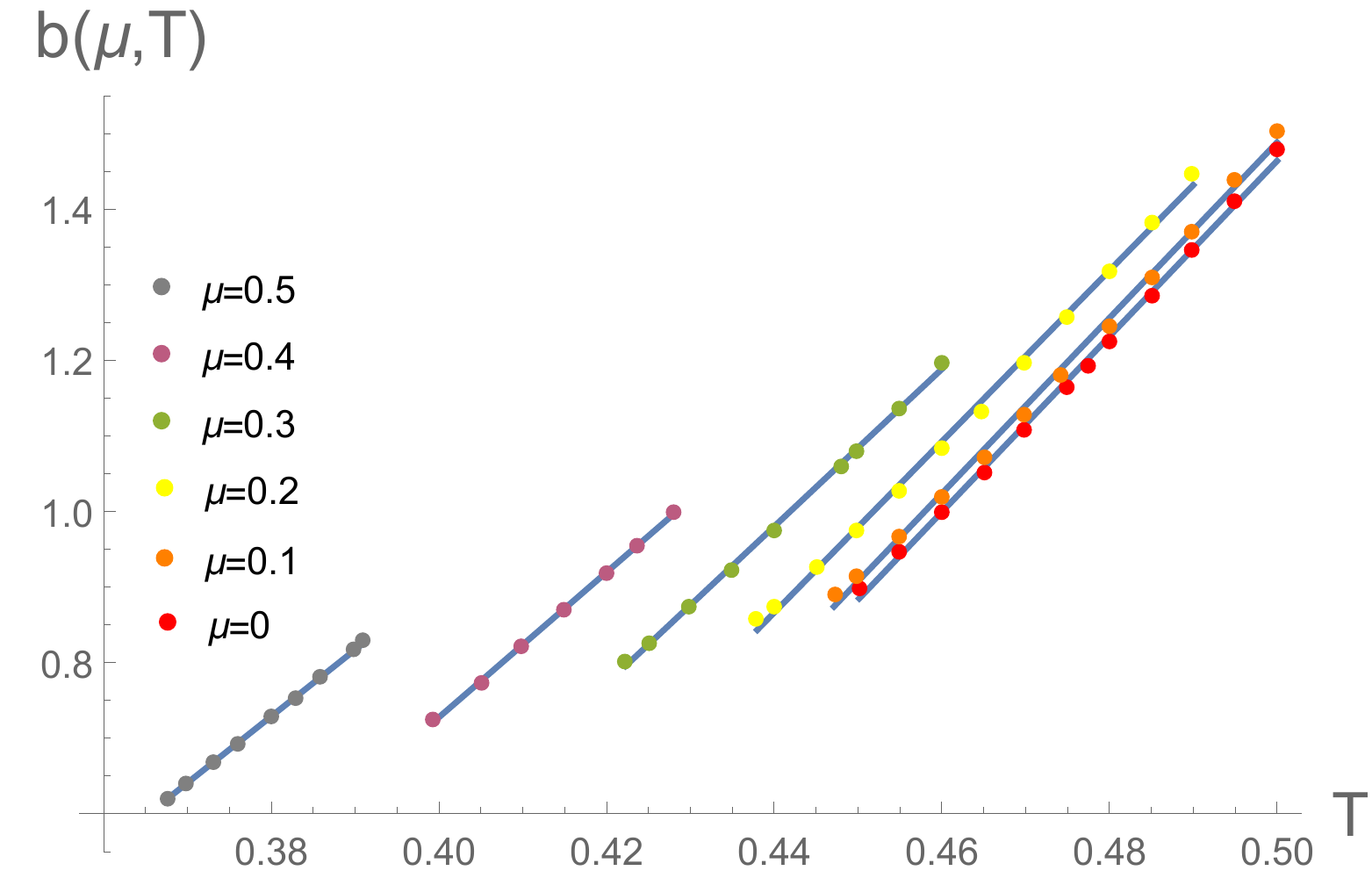}}	
		
		\caption{\footnotesize For $(a,b)$ matrix model with non-zero chemical potential (grand canonical ensemble): The temperature $T$ dependence of the parameters $a(\mu,T)$ and $b(\mu,T)$, at fixed potential $\mu$. Dots indicate the data points, while the solid curves are their fitting curves. For $a(\mu,T)$, the fitting curve is of the form $a(\mu, T) =c_1 \text{log}(T) +c_2 T +c_3$, and for  $b(\mu, T)$, it is  $b(\mu, T) =c_4 T +c_5$. (\text{Here, the coefficients}\, $c_1, c_2, c_3, c_4$, and $c_5$ are $\mu$ dependent).}
		\label{fig:RN_grand_canon_boundary_a_b_plots}	}
	
\end{figure}

We start with the previous $(a, b)$ matrix model~\eqref{eq: a_b_matrixmodel_for_sch}, as a boundary dual  
for the Reissner-Nordstrom-AdS$_5$  black holes in the grand canonical 
ensemble (fixed potential)~\cite{Chandrasekhar:2012vh}. However, we allow the parameters $a$ and $b$ now 
to depend on the chemical potential $\mu$ as well as the temperature.
One can follow the same strategy as before to figure out the dependence of $a$, $b$ on $\mu$ and $T$.
The only difference is that now in~\eqref{eq_sch_boundary_V=I}, we need to substitute $I_{\rm 1,2}$ for the Reissner-Nordstrom-AdS$_5$  black hole.
Carrying out the computation numerically~\cite{Chandrasekhar:2012vh}, we get $a$, $b$ as shown in the Fig.~\ref{fig:RN_grand_canon_boundary_a_b_plots}. 
We see that, for a fixed $\mu$, the behavior of the effective potential $V(\rho)$ (eqn.~\ref{eq:Veff_sch_boundary}) is same as  in the Fig.~\ref{fig:sch_boundary_effective_potential_plot}. 
Thus, this $(a,b)$ matix model, at fixed potential $\mu$, mimics the phase structure of the bulk.
\begin{figure}[h!]
	{\centering
		
		\subfloat[]{\includegraphics[width=2.5in]{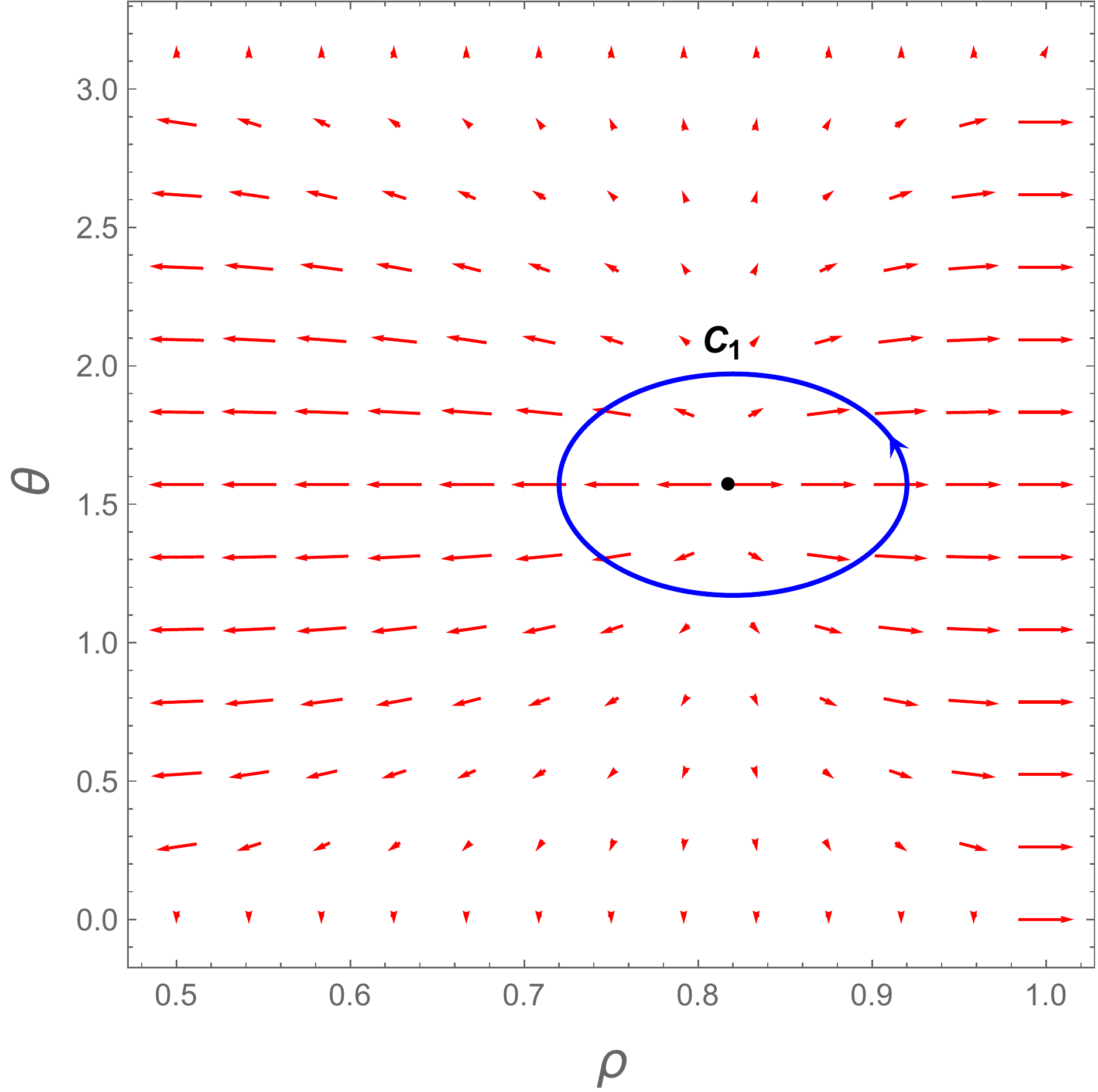}}\hspace{1.2cm}	
		\subfloat[]{\includegraphics[width=3in]{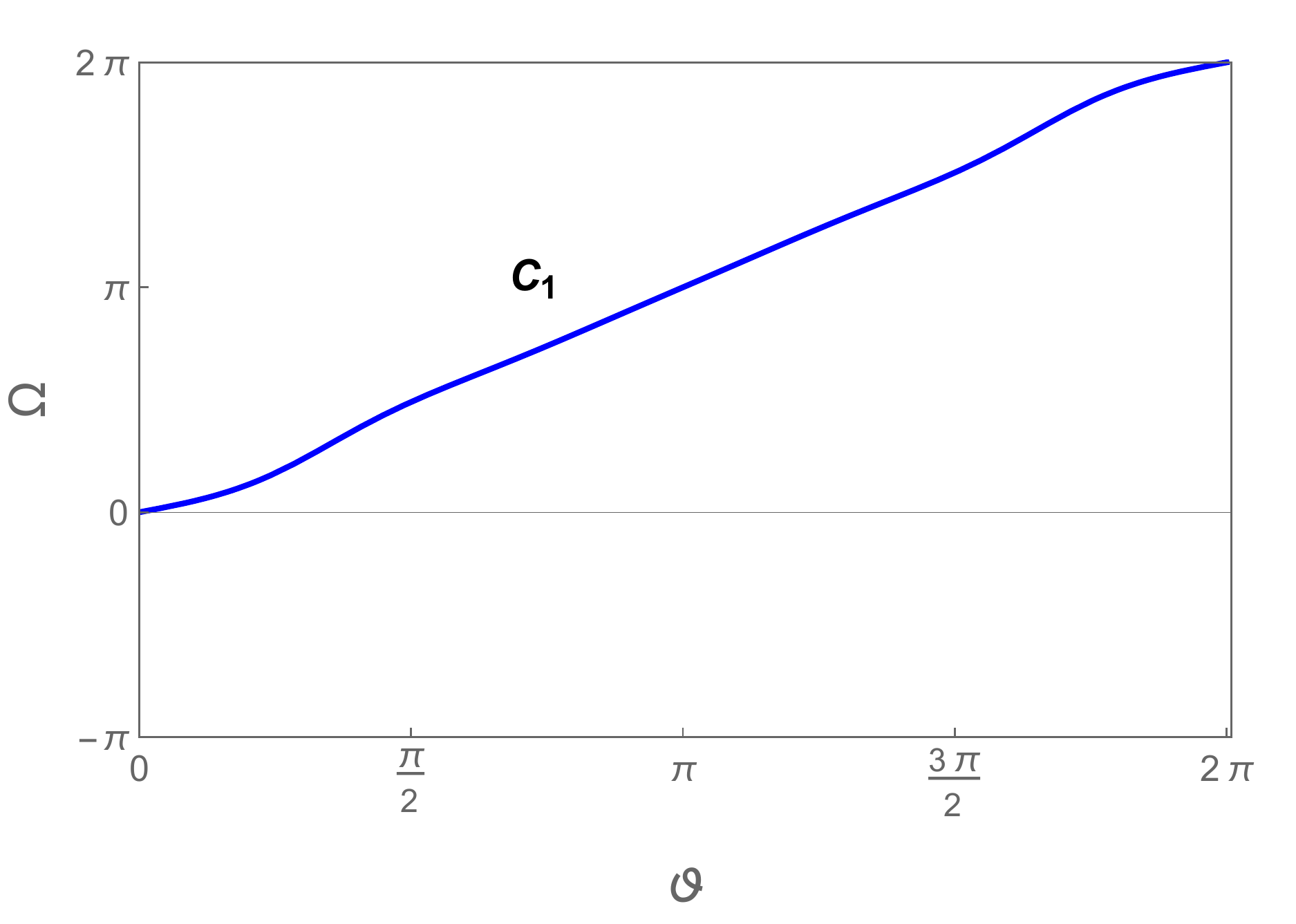}}				
		
		\caption{\footnotesize For $(a,b)$ matrix model with non-zero chemical potential (grand canonical ensemble): (a) The normalised vector field of $\phi$ in the $\theta-\rho$ plane, vanishes at the phase transition point (black dot, located at $\rho_{\rm HP} = 0.817$ for a fixed $\mu=0.5$) that represents 
HP transition in the bulk.   
			(b) $\Omega$ vs $\vartheta$ for contour $C_1$, giving the topological charge $+1$ for phase transition point of $V(\rho)$, that represents HP transition in the bulk. Here, parametric coefficients of the contour are $(A,B,\rho_0) =(0.1,0.4,0.82)$.} 
\label{Fig:RN_grand_canon_boundary_hp_vec_omega_plots}	}
\end{figure}
\vskip 0.2cm
\noindent
Now, for a fixed $\mu$, one can proceed to compute the topological charges for the phase transition point and equilibrium phases of $V(\rho)$ in the $\rho > 1/2 $ region, as done in the section~\ref{section:boundary for sch}. We avoid the details for brevity and mention that the results match the ones obtained in the bulk (see Fig.~\ref{Fig:RN_grand_canon_boundary_hp_vec_omega_plots}).
\subsection{Canonical ensemble: fixed charge}  \label{canonical}
\subsubsection{Bulk}
In the canonical ensemble, Reissner-Nordstrom-AdS$_5$  black holes are known to exhibit a rich phase structure,  resembling that of van der Waals fluid. This can also be inferred from the behavior of the equation of state $T(r_+,q)$, given by~\cite{Chamblin:1999tk,Chamblin:1999hg}:
\begin{equation}\label{eq:RN_canon_bulk_eos}
T = \frac{2r_+^6+r_+^4-q^2}{2\pi r_+^5}.
\end{equation}
\begin{figure}[h!]
	
	{\centering
		
		\subfloat[]{\includegraphics[width=2.3in]{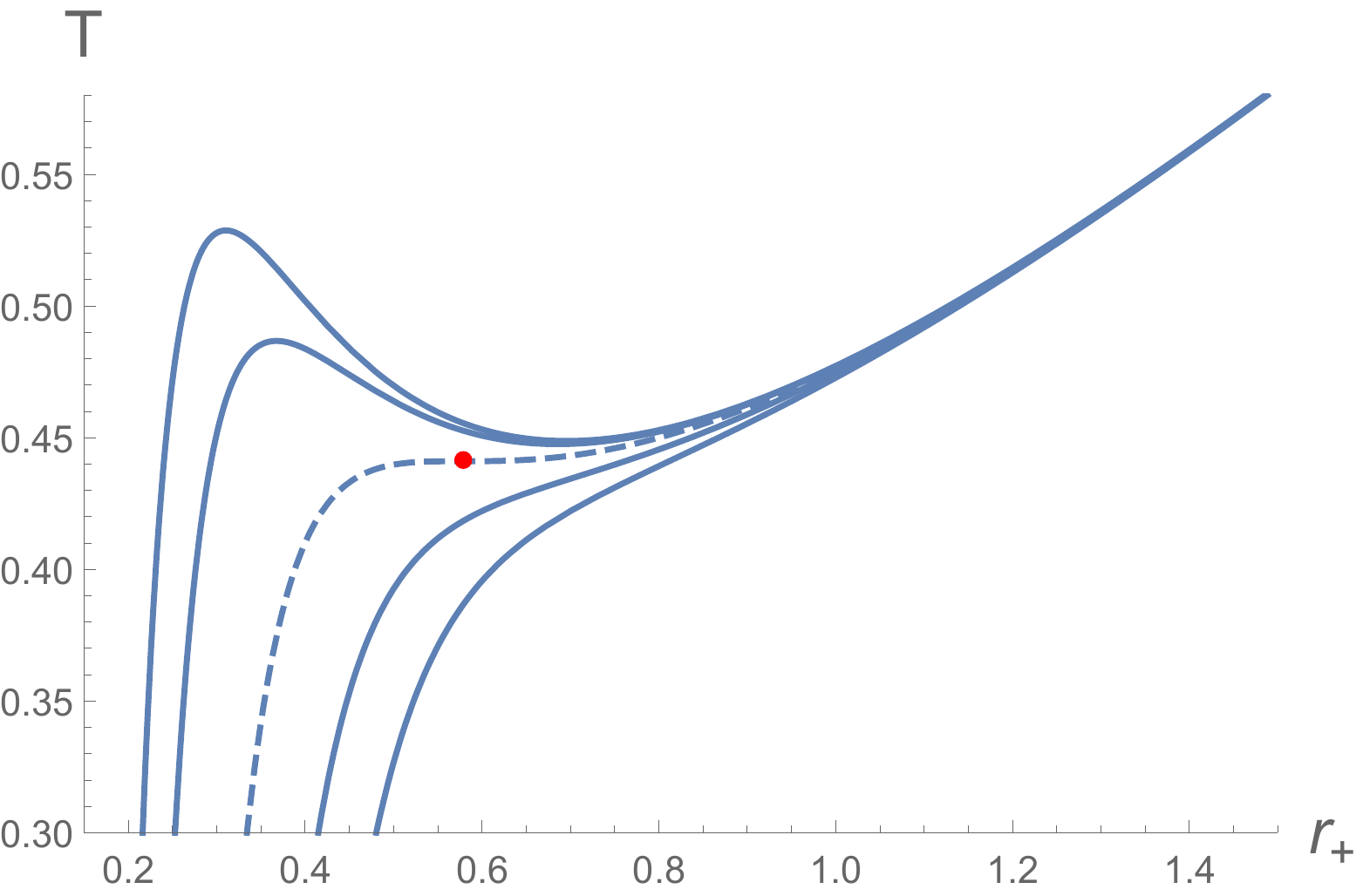}\label{fig:RN_canon_bulk_eos}}\hspace{0.2cm}
		\subfloat[]{\includegraphics[width=1.7in]{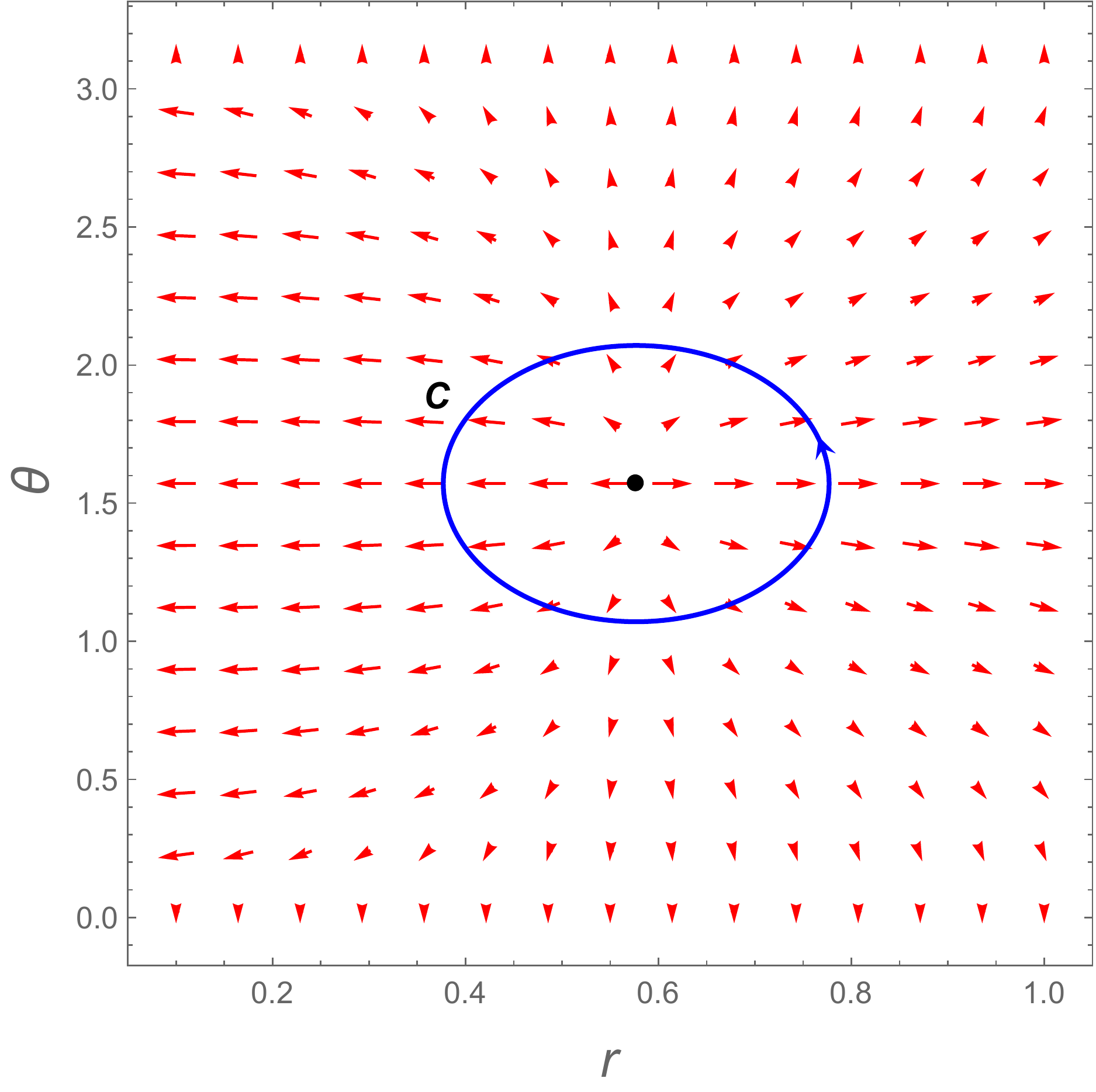}\label{fig:RN_canon_bulk_vdw_vec_plot}}\hspace{0.3cm}	
		\subfloat[]{\includegraphics[width=1.8in]{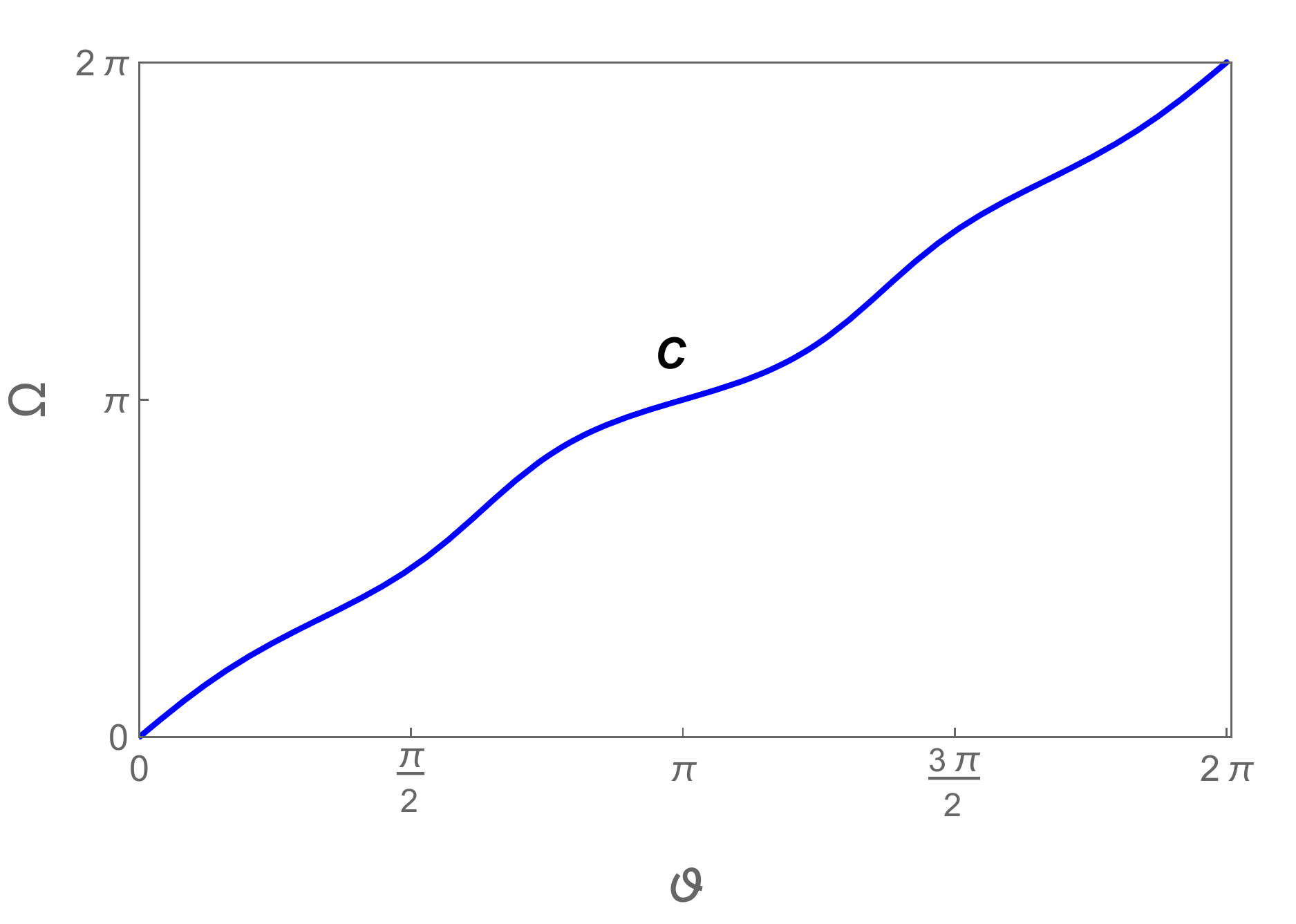}\label{fig:RN_canon_bulk_vdw_omega_plot}}			
		\caption{\footnotesize For Reissner-Nordstrom black holes in $\text{AdS}_5$ (canonical ensemble): (a) Behaviour of the equation of state $T(r_+,q)$ at various charges $q$, indicating the existence of  three branches of black hole
			solutions (small, intermediate, and large black holes) for $ q <q_{\rm cr}$, and one branch of black hole solutions for $ q > q_{\rm cr}$. Charge of the curves increases from top to bottom. Dashed curve is for $q=q_{\rm cr}$. Red dot is the critical point. (b) Vector field $\phi$, vanishes at the critical point (black dot, at $r_{\rm cr} = 0.577$). (c) $\Omega$ vs $\vartheta$ for contour $C$. }
	}	
\end{figure}
\noindent
As shown in  Fig.~\ref{fig:RN_canon_bulk_eos}, there exists a critical charge $q_{\rm cr}$, above which the black holes are in a unique phase, while for the charge $q< q_{\rm cr}$,  black holes can exist in three phases (called as, small, intermediate, and large black hole phases).  
The small and large black hole phases are locally stable due to positive specific heat, while, the intermediate black hole phase is unstable due to negative specific heat. There exists a first-order phase transition between the small and large black holes, that terminates in a second-order critical point. The critical point can be obtained using the conditions for a stationary point of inflection, i.e.,
\begin{equation}
\frac{\partial T}{\partial r_+} = 0, \quad \text{and}, \quad \frac{\partial^2 T}{\partial r_+^2} = 0.
\end{equation}
This gives the critical point as $(T_{\rm cr}, \, r_{\rm cr},\, q_{\rm cr}) = (\frac{4\sqrt{3}}{5\pi}, \frac{1}{\sqrt{3}}, \frac{1}{3\sqrt{15}})$. 
The topological charge carried by the critical point can be computed, following~\cite{Wei:2021vdx},  by defining the vector field $\phi(\phi^r,\, \phi^\theta)$ as:
\begin{eqnarray}
\phi^r &=& \partial_{r_+} \big(T(r_+)/\text{sin}\theta \big),\\
\phi^\theta &=& \partial_\theta \big(T(r_+)/\text{sin}\theta \big). \label{phitheta1}
\end{eqnarray}
Here, we write $T(r_+)$ as in~\eqref{eq:RN_canon_bulk_eos} with $q$ eliminated using the condition $\frac{\partial T}{\partial r_+} = 0$, at the critical point. This vector field $\phi$ vanishes exactly at the critical point and one can find that the topological charge carried by the critical point would be $+1$ (see figures~\ref{fig:RN_canon_bulk_vdw_vec_plot} and ~\ref{fig:RN_canon_bulk_vdw_omega_plot}). This topological charge is in fact opposite to the value computed in extended-phase space, where it was found to be $-1$~\cite{Wei:2021vdx}.  This opposite behaviour of the topological charges is an expected result, as the topological charge of the critical point depends on the behaviour of the phase structure (i.e., on the behaviour of the vector field) around that critical point. This is explained in Appendix-(\ref{B}).
\begin{figure}[h!]
	
	{\centering
		
		\subfloat[]{\includegraphics[width=2.7in]{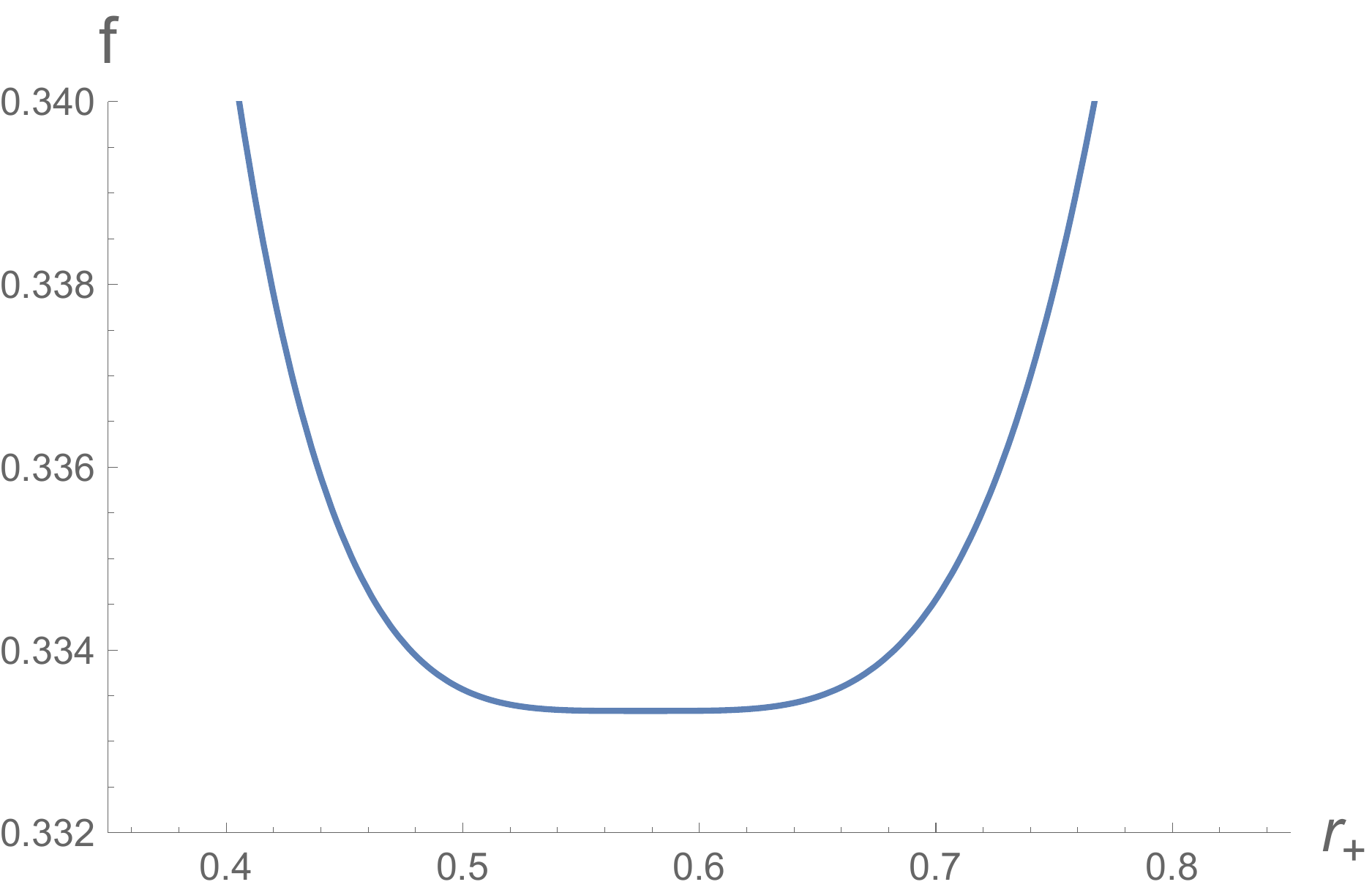}}\hspace{0.5cm}	
		\subfloat[]{\includegraphics[width=3.2in]{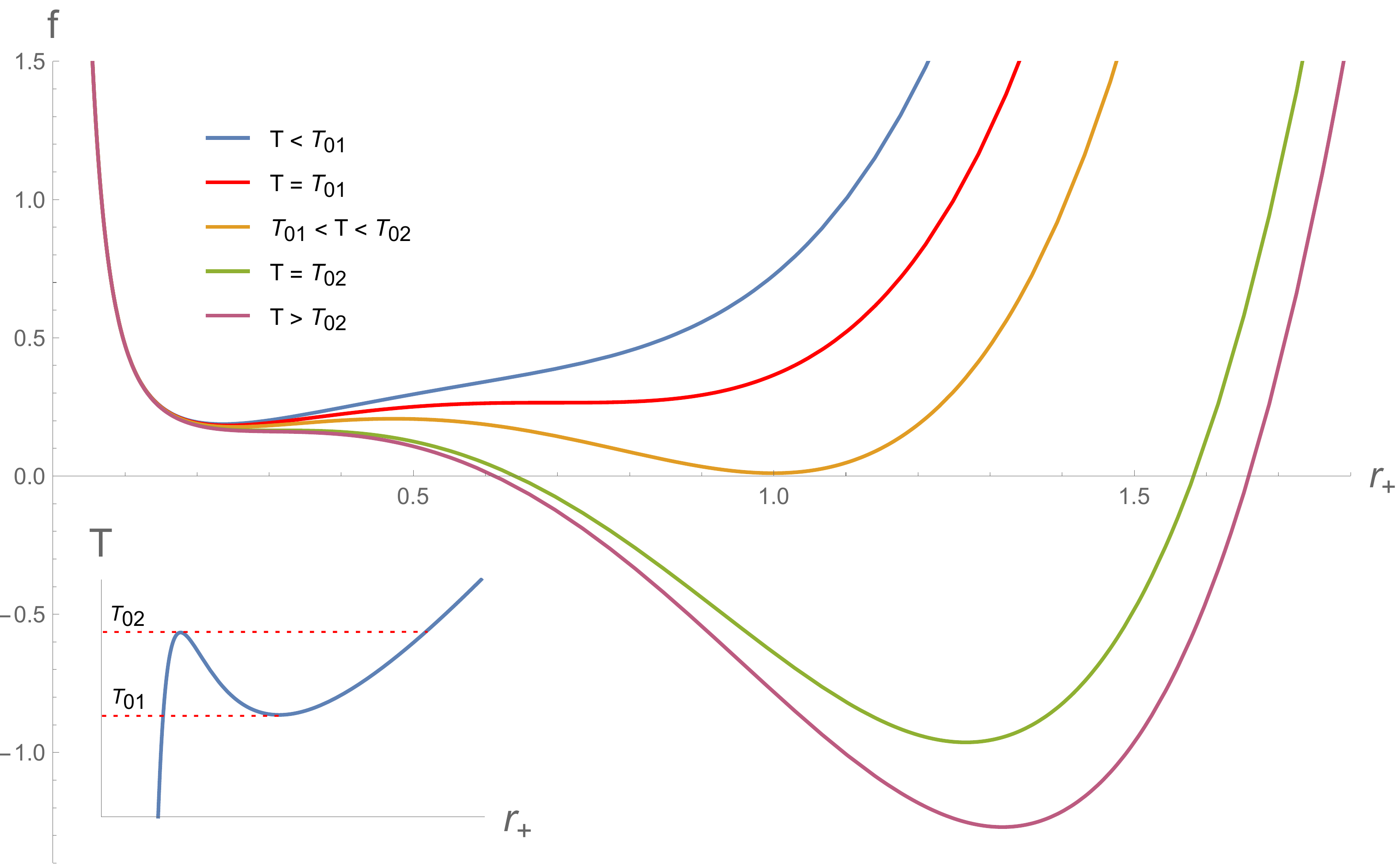}}	
		
		\caption{\footnotesize For Reissner-Nordstrom black holes in $\text{AdS}_5$ (canonical ensemble): Behavior of off-shell free energy $f$ (a) for $q=q_{\rm cr}$,  and  $T=T_{\rm cr}$. (b ) for $q < q_{\rm cr}$, and  at corresponding  temperatures of the inset plot. The extremal points of $f$, represent the black hole solutions. }
\label{fig:RN_canon_bulk_f1f2_plots}	}	
\end{figure}
\begin{figure}[h!]
	{\centering
		\subfloat[]{\includegraphics[width=2.4in]{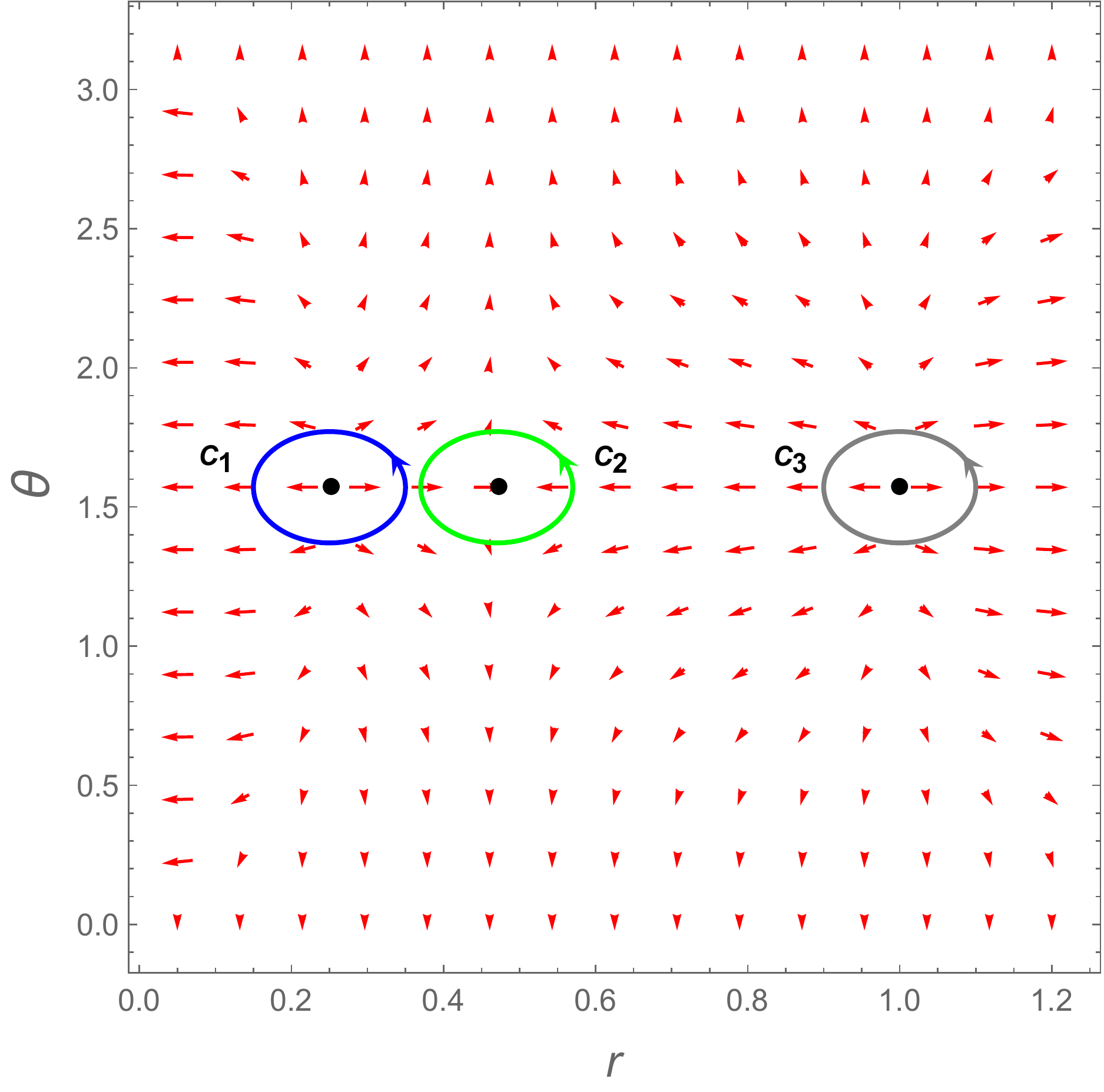}}\hspace{2cm}	
		\subfloat[]{\includegraphics[width=2.7in]{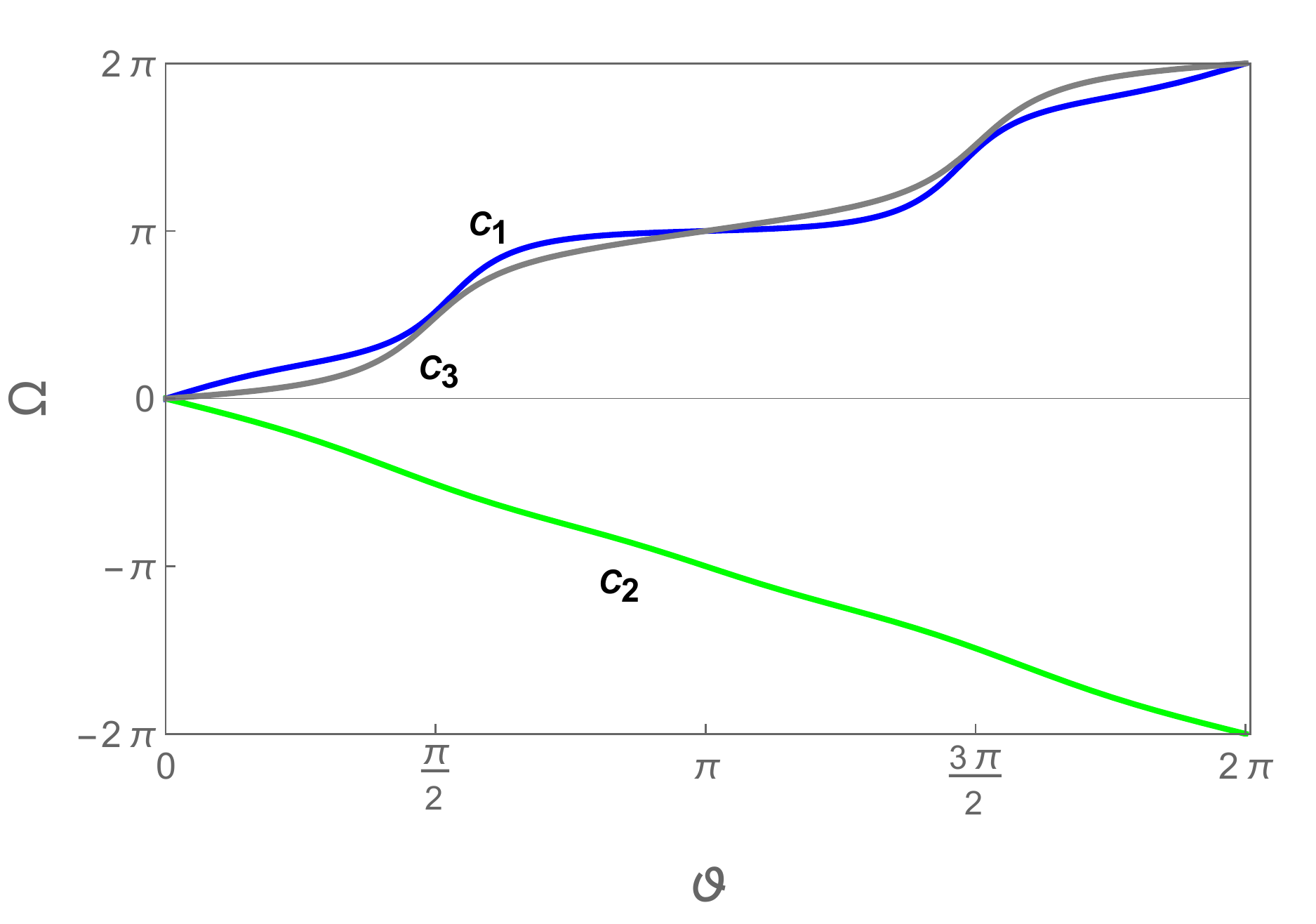}}				
		
		\caption{\footnotesize For Reissner-Nordstrom  black holes in $\text{AdS}_5$ (canonical ensemble): (a) The  vector field $\phi$, vanishes at small, intermediate, and large black hole solutions (black dots, located at $r_+ = 0.25, 0.47, 1$). (b) $\Omega$ vs $\vartheta$ for contours $C_1$, $C_2$, and $C_3$. Here, we used $q < q_{\rm cr}$, and $T_{01} < T < T_{02}$.} 
\label{Fig:RN_canon_bulk_equi_vec_omega_plots}	}
\end{figure}
\vskip 0.2cm
\noindent
We can now proceed to compute the topological charges associated with the small/intermediate/large black holes using the off-shell free energy f, given by~\cite{Chamblin:1999hg,Li:2020nsy} (See the Appendix-(\ref{A2})  for details):
\begin{equation}
f = M-TS=\frac{3}{r_+^2} (r_+^6+r_+^4+q^2) - 4\pi r_+^3T,
\end{equation}
whose its extremal points represent the black hole solutions, as shown in fig.~\ref{fig:RN_canon_bulk_f1f2_plots}. The vector field  $\phi$, defined as in~\eqref{eq:vec field with f}, vanishes at these extremal points.  A similar analysis gives the topological charge for small/intermediate/large black hole to be +1/-1/+1, respectively (see fig.~\ref{Fig:RN_canon_bulk_equi_vec_omega_plots} for details). These results perfectly match the values obtained in extended phase space~\cite{Wei:2022dzw}.
We note here that, at the  merging point (where, the small and intermediate black holes merge)/nucleation point  (where, the  intermediate and large black hole pair nucleates)/ critical point (where, the small,  intermediate, and large black holes merge),  the topological charges of the corresponding black holes will be added, thus they carry 0/0/+1 topological charges, respectively.   
\subsubsection{Boundary}\label{canonicalboundary}
\begin{figure}[h!]
	
	{\centering
		
		\subfloat{\includegraphics[width=2.7in]{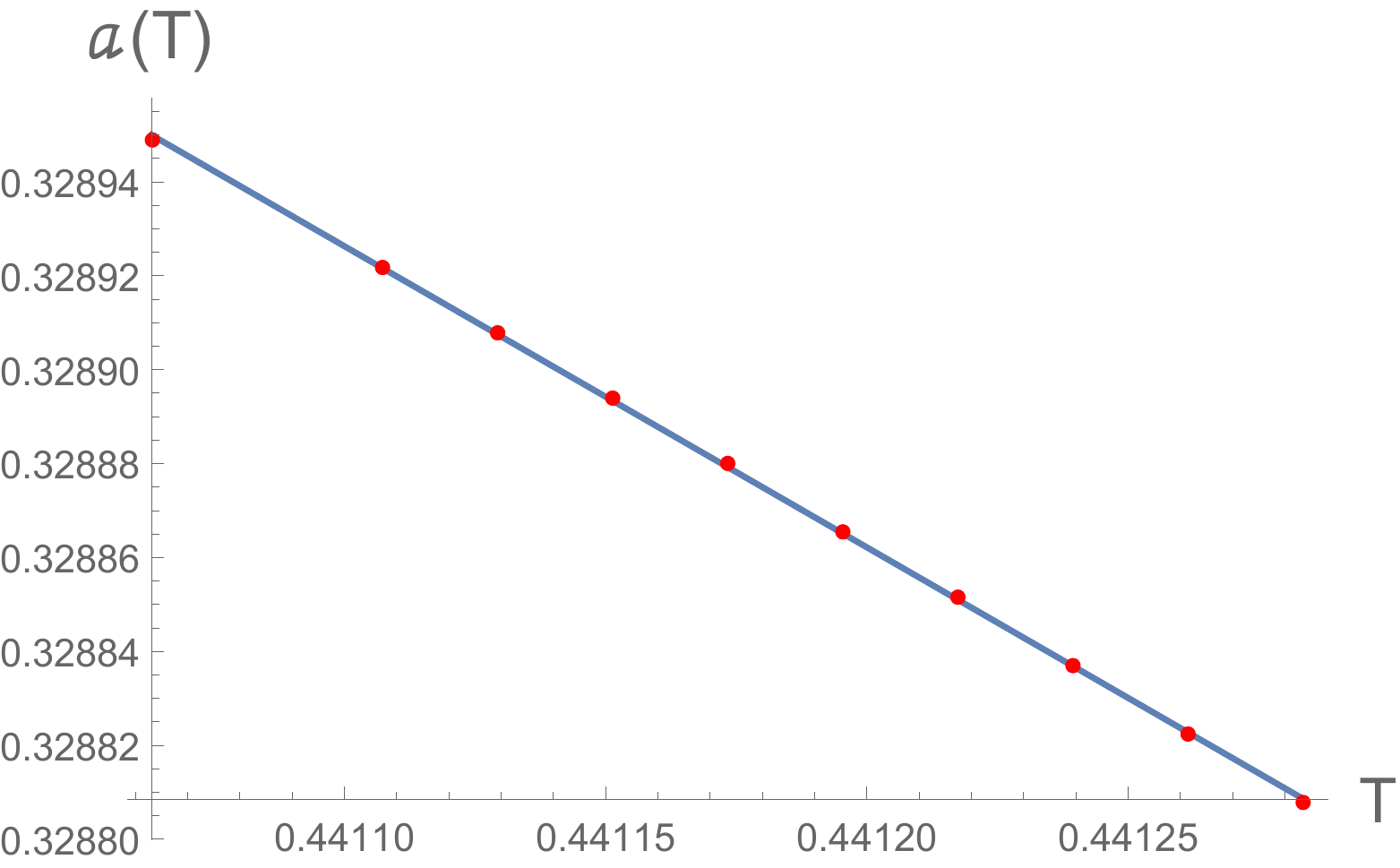}}\hspace{1.5cm}	
		\subfloat{\includegraphics[width=2.7in]{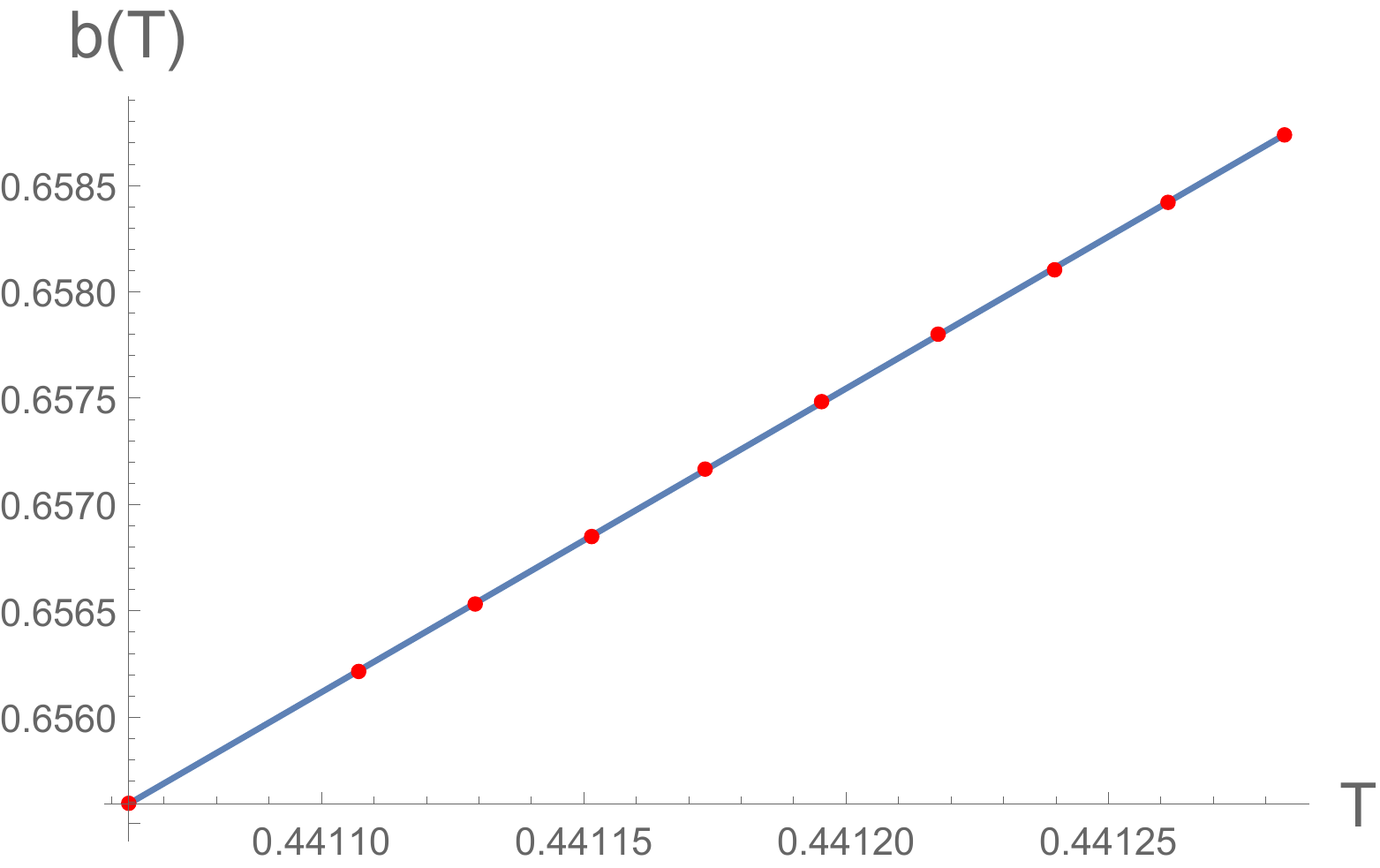}}	
		
		\caption{\footnotesize For $(a,b)$ matrix model with non-zero chemical potential (canonical ensemble): The temperature  dependence of the parameters $a(T)$ and $b(T)$ for $ T \geq T_{\rm cr}$. Red dots indicate the data points, while blue coloured curves are their fitting curves . For $a(T)$, the fitting curve is $a(T) =c_1 T +c_2$, and for  $b(T)$, it is $b(T) =c_3 T +c_4$. $(\text{Here,}\, c_1=-0.641573, c_2=0.611924, c_3=14.2782, c_4=-5.64198)$.}
		\label{fig:RN_canon_boundary_a_b_plots}	}
	
\end{figure}

\begin{figure}[h!]
	{\centering
		\subfloat[]{\includegraphics[width=2.7in]{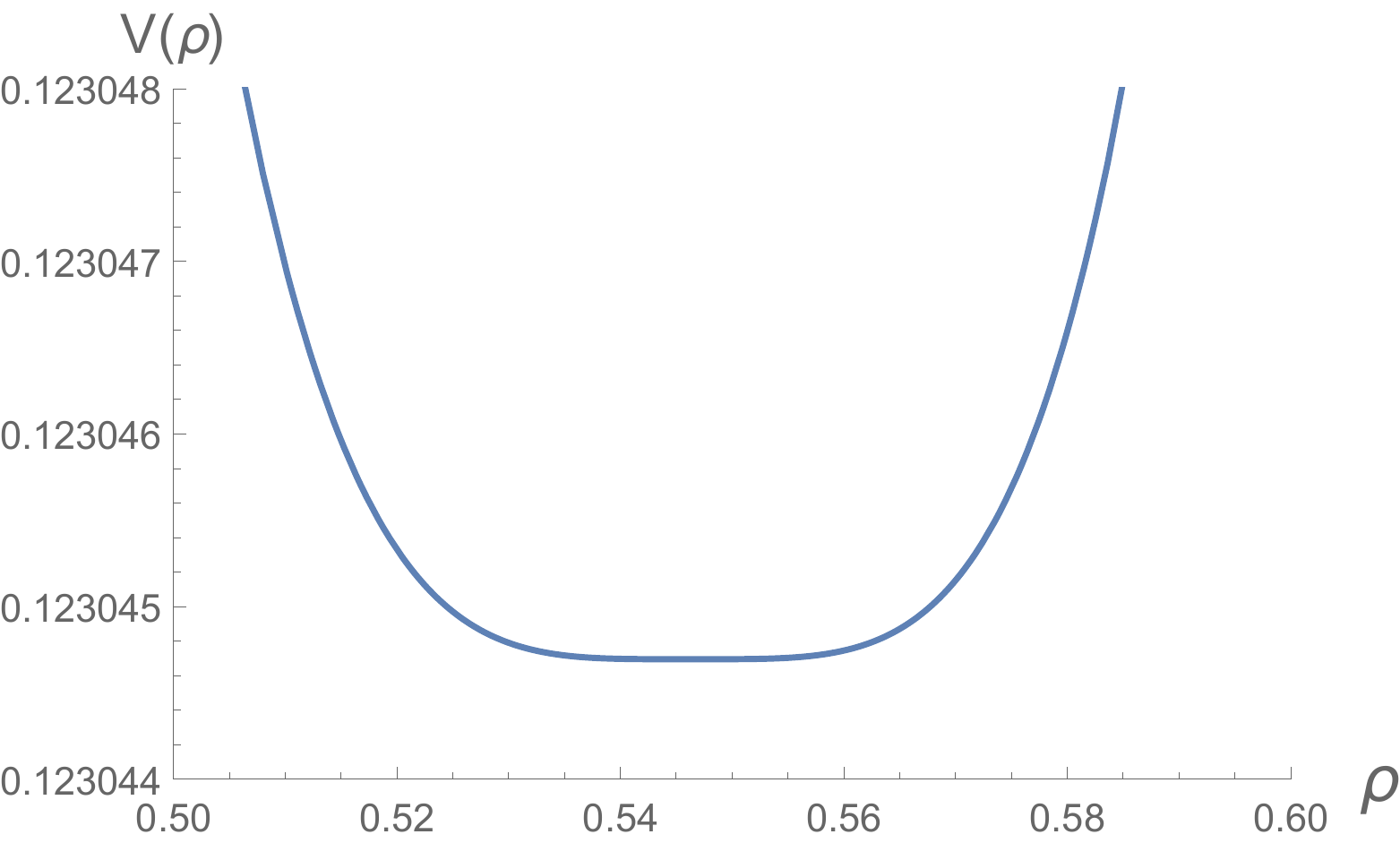}}\hspace{1.5cm}	
		\subfloat[]{\includegraphics[width=2.7in]{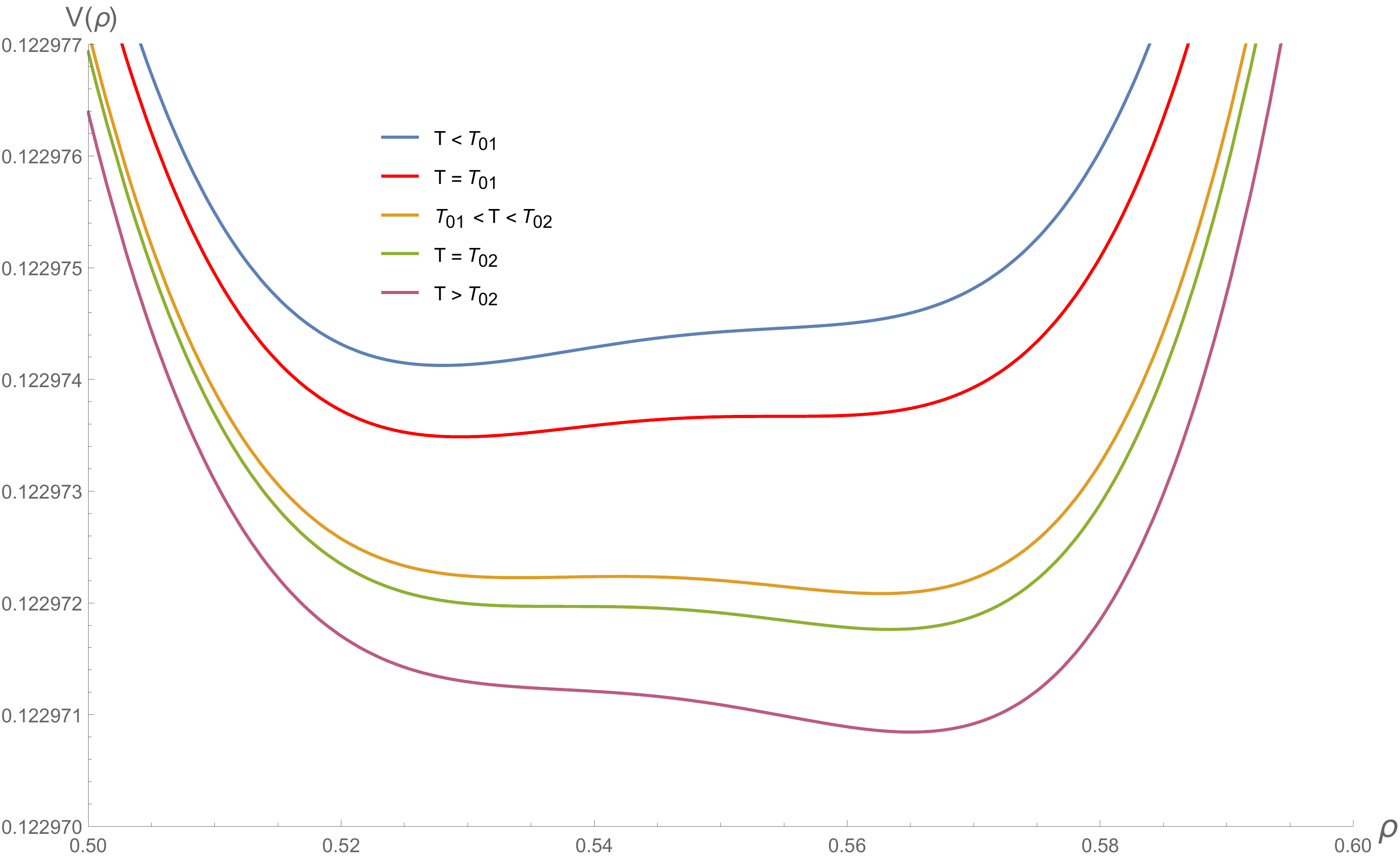}}				
		
		\caption{\footnotesize For $(a,b)$ matrix model with non-zero chemical potential (canonical ensemble): behaviour of the  potential $V(\rho)$  for a fixed charge $q$ and at various temperatures $T$.  (a) for  $ q= q_{\rm cr}$, and  $T=T_{\rm cr}$. (b) for  $ q < q_{\rm cr}$.} 
		\label{Fig:RN_canon_boundary_v1v2_plots}	}
\end{figure}

\begin{figure}[h!]
	
	{\centering
		
		\subfloat[]{\includegraphics[width=2.5in]{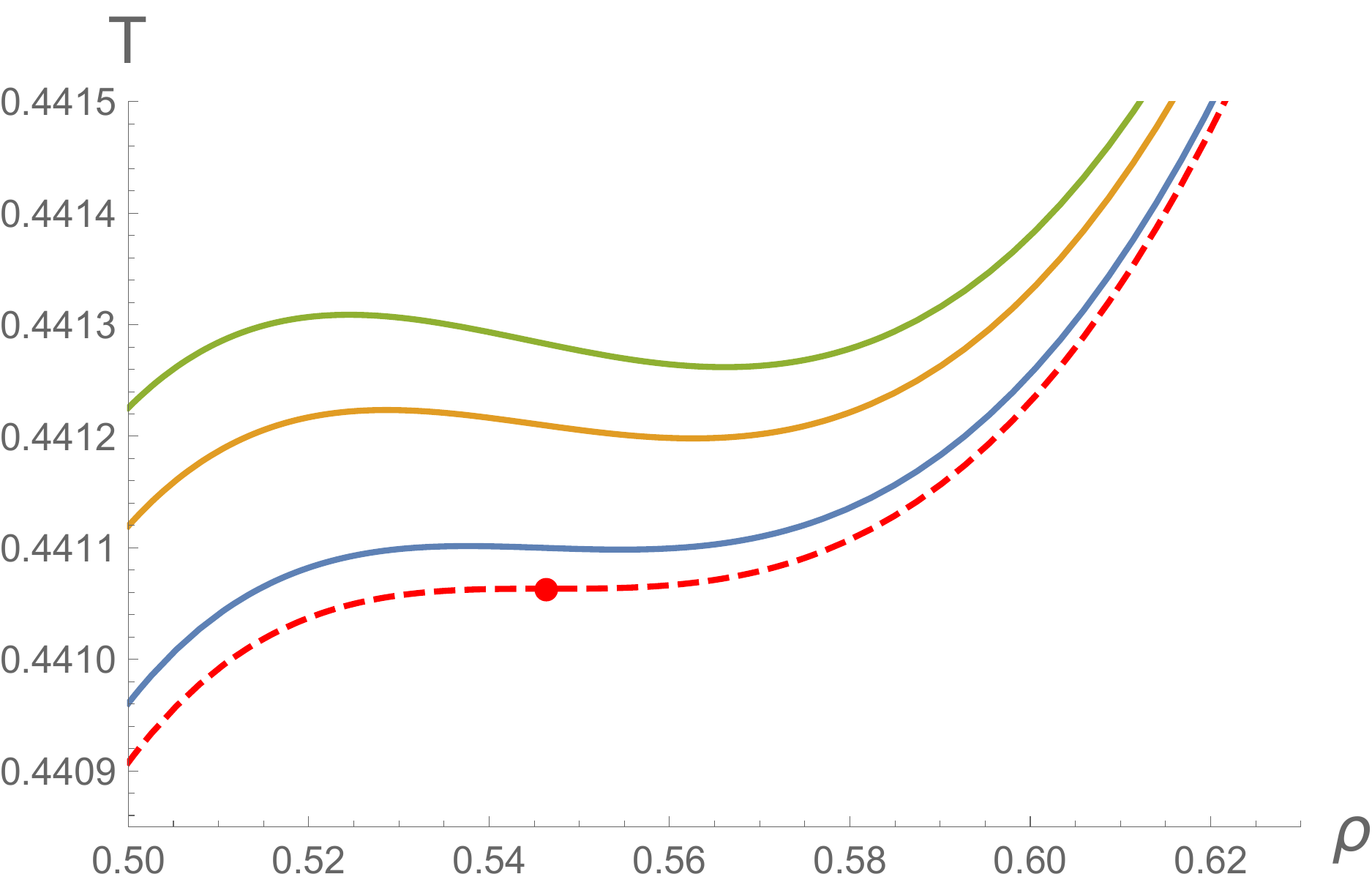}\label{fig:RN_canon_boundary_eos}}\hspace{0.2cm}
		\subfloat[]{\includegraphics[width=1.5in]{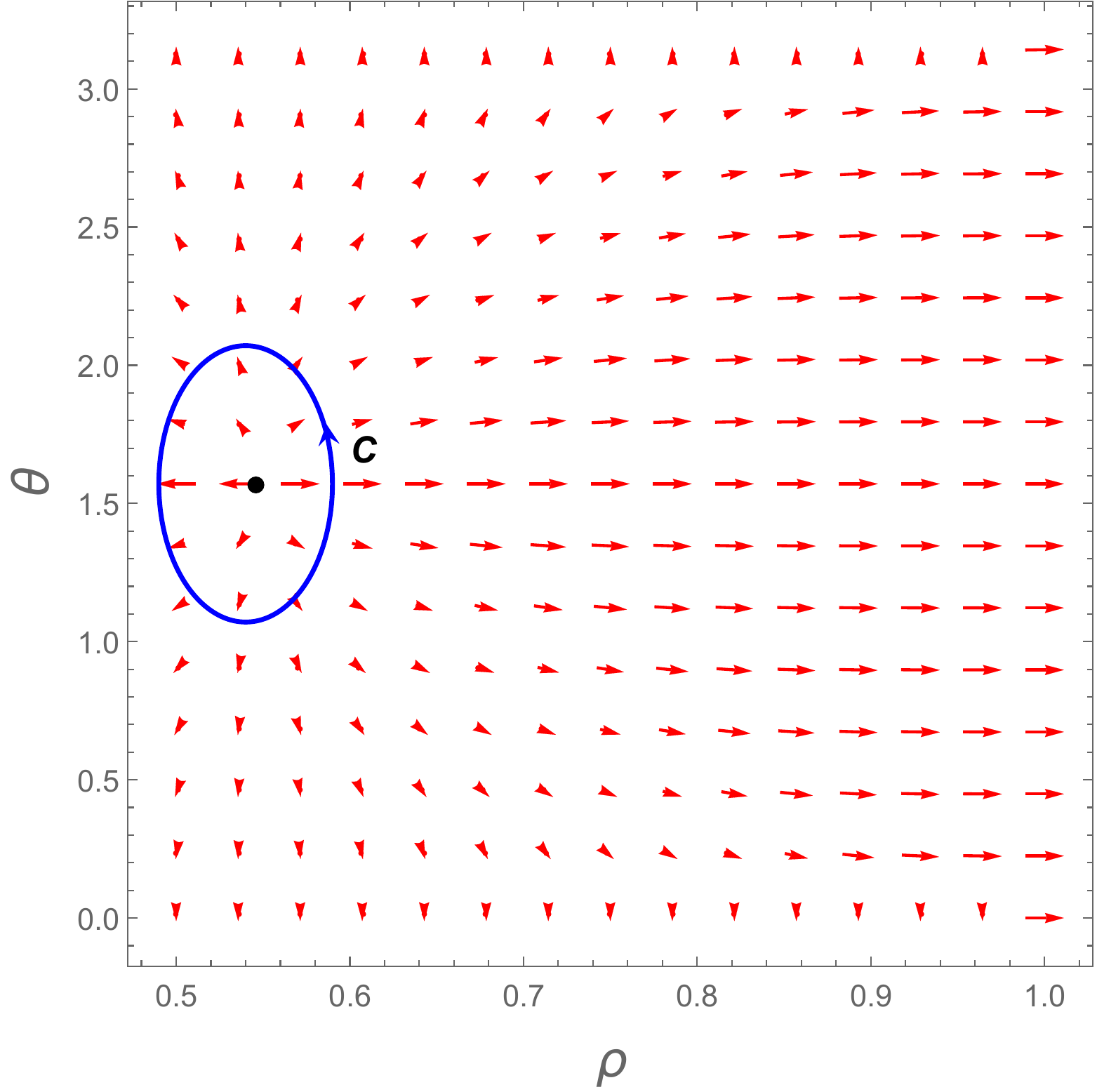}\label{fig:RN_canon_boundary_vdw_vec_plot}}\hspace{0.2cm}	
		\subfloat[]{\includegraphics[width=2in]{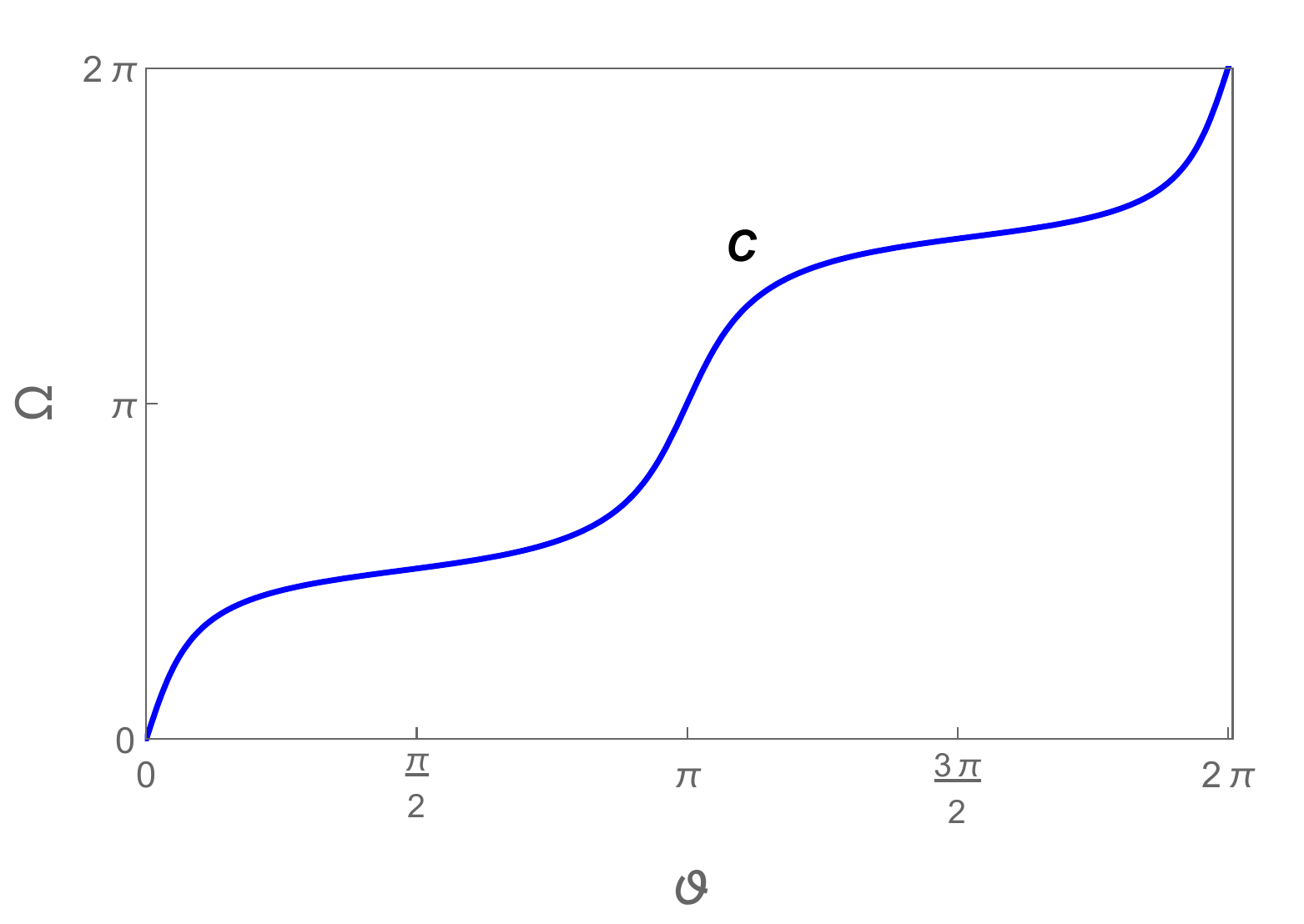}\label{fig:RN_canon_boundary_vdw_omega_plot}}			
		\caption{\footnotesize  (a) Behavior of the equation of state $T(\rho, q)$ at fixed charges $q \leq q_{\rm cr}$. Charge of the curves increases from top to bottom. Dashed curve is for $q=q_{\rm cr}$. Red dot is the critical point. (b) Vector field $\phi$, vanishes at the critical point (black dot, at $\rho_{\rm cr} = 0.5463$). (c) $\Omega$ vs $\vartheta$ for contour $C$, giving the topological charge $+1$. }
	}	
\end{figure}

One could use the $(a, b)$ matrix model given in eqn.~(\ref{eq: a_b_matrixmodel_for_sch}), with an additional logarithmic term in the effective potential for a
fixed nonzero charge, as a boundary dual for Reissner-Nordstrom $\text{AdS}_5$  black holes in  canonical ensemble. The corresponding effective action is given by~\cite{Basu:2005pj}: 
\begin{equation}
S_q = S\big(a(T), \, b(T), \, \rho \big) + q \, \text{log}(\rho),
\end{equation} 
where the saddle point equations are:
\begin{eqnarray}
	\rho F +q &=& \rho^2   \hspace{2cm} \text{for} \quad 0 \leq \rho \leq \frac{1}{2}    \\
	&=& \frac{\rho}{4(1-\rho)}  \hspace{0.8cm}  \text{for} \quad \frac{1}{2} \leq \rho \leq 1, \label{eq:saddle pt eq for rho >1/2 for RN boundary}
\end{eqnarray} 
with $ F(\rho) = a\rho +2b\rho^3$. The effective potential becomes:
\begin{eqnarray}
V(\rho) &=& \frac{1-a}{2}\rho^2 -\frac{b}{2} \rho^4  -q\, \text{log}(\rho)     \hspace{4cm} \text{for} \quad 0 \leq \rho \leq \frac{1}{2}   \nonumber \\
&=& -\frac{a}{2} \rho^2 -\frac{b}{2} \rho^4 -q\, \text{log}(\rho)  -\frac{1}{4} \text{log}[2(1-\rho)] +\frac{1}{8}  \hspace{0.8cm}  \text{for} \quad \frac{1}{2} \leq \rho \leq 1. 
\end{eqnarray}
The parameters are bounded as $a(T) < 1/2$ and $b(T) >1/4$ and can be computed as in~\cite{Basu:2005pj}.
At places our way to arrive at the phase structure is somewhat different from~\cite{Basu:2005pj}. 
So, in what follows,  we provide some details.
We assume that the saddle point equation $q(\rho, T)$ for $\rho > 1/2$ 
(i.e., eqn.~\ref{eq:saddle pt eq for rho >1/2 for RN boundary}), mimics the bulk 
equation of state $q(r_+,T)$. Thus, at the critical point, we have:
\begin{eqnarray}
\frac{\partial q}{\partial \rho} \Big\vert_{cr} & = & 0, \\
\frac{\partial^2 q}{\partial \rho^2} \Big\vert_{cr} & = & 0. 
\end{eqnarray}
The critical point also satisfies the saddle point equation,
\begin{equation}
q_{\rm cr} = \frac{\rho_{\rm cr}}{4(1-\rho_{\rm cr})} - \rho_{\rm cr}(a\rho_{\rm cr}+2b\rho_{\rm cr}^3).
\end{equation}
Solving the above three equations, we get $a(T_{\rm cr})$, $b(T_{\rm cr})$ and $\rho_{\rm cr}$.
Since, for a fixed $T  > T_{\rm cr}$, the curve $q(\rho, T)$  contains the  nucleation and merging points  (extremal points),   then we have:
\begin{eqnarray}
\frac{\partial q}{\partial \rho} \Big\vert_{\rm n,m} & = & 0,\\
q_{\rm n,m} &=& \frac{\rho_{\rm n,m}}{4(1-\rho_{\rm n,m})} - \rho_{\rm n,m}(a\rho_{\rm n,m} + 2b\rho_{\rm n,m}^3), 
\end{eqnarray}
where, the suffix n/m stands for nucleation/merging point, respectively.
Solving the above four equations, we get $a(T)$, $b(T)$, $\rho_m$ and $\rho_n$, for a fixed $T  > T_{\rm cr}$. The temperature dependence of the parameters $a(T)$ and $b(T)$, is shown in the Fig.~\ref{fig:RN_canon_boundary_a_b_plots}.
\vskip 0.2cm
\noindent
Now, one can see  the behaviour of  effective potential $V(\rho)$ at a fixed charge $q$ for various  temperatures,  from the Fig.~\ref{Fig:RN_canon_boundary_v1v2_plots}. The saddle points of $V(\rho)$ for $\rho > 1/2$, represent the corresponding black hole solutions in bulk. Stable/unstable saddle points correspond to stable/unstable black holes, respectively. Let us summarise the phase structure we obtain. For a fixed $q < q_{\rm cr}$, at low temperatures $T < T_{01}$, there exist only one  saddle point for $V(\rho)$, corresponds to small stable black hole in the bulk. However, when the temperature is raised to  $T_{01}$, there is a nucleation of two new saddle points, corresponding to unstable intermediate and stable large black holes in the bulk.  On the other hand, there are three saddle points (corresponding to small-intermediate-large black holes) when $ T_{01} <T < T_{02}$. Here though, there is a merging of two saddle points (corresponding to small-intermediate black holes) when temperature raised to $T_{02}$.  Finally, there exists only one saddle point (corresponding to large black hole) when $T>T_{02}$. 
For $q=q_{\rm cr}$, we have a critical point, where all the three saddle points merge to one and this happens at the critical temperature $T_{cr}$.  
With this identification, the phase structure of the effective potential $V(\rho)$,  matches with that of the bulk, for a fixed charge $q \leq q_{\rm cr}$~\footnote{ One can refer~\cite{Basu:2005pj}, for further details of the phase structure.}. We note here that,  $\rho = 0 $ is never a solution in the nonzero charge sector.
\vskip 0.2cm
\noindent
Now, using the fitting curves for $a(T)$ and $b(T)$, we solve the saddle point eqn.~\ref{eq:saddle pt eq for rho >1/2 for RN boundary}, for temperature, which gives:
\begin{equation}
T(\rho, q) = \frac{4 q (1-\rho) + 8c_4(1-\rho)\rho^4 + 4c_2(1-\rho)\rho^2 - \rho \,}{4(\rho - 1)\rho^2 (c_1 + 2c_3 \rho^2)}.
\end{equation}
The qualitative behaviour of the above expression obtained from curve fitting matches with the known bulk formula, as can be seen from fig.~\ref{fig:RN_canon_boundary_eos}, and its critical point can also be located at $(T_{\rm cr}, \, q_{\rm cr}, \, \rho_{\rm cr}) = (\frac{4\sqrt{3}}{5\pi}, \, \frac{1}{3\sqrt{15}}, \, 0.546293)$. 
\begin{figure}[h!]
	{\centering
		\subfloat[]{\includegraphics[width=2.4in]{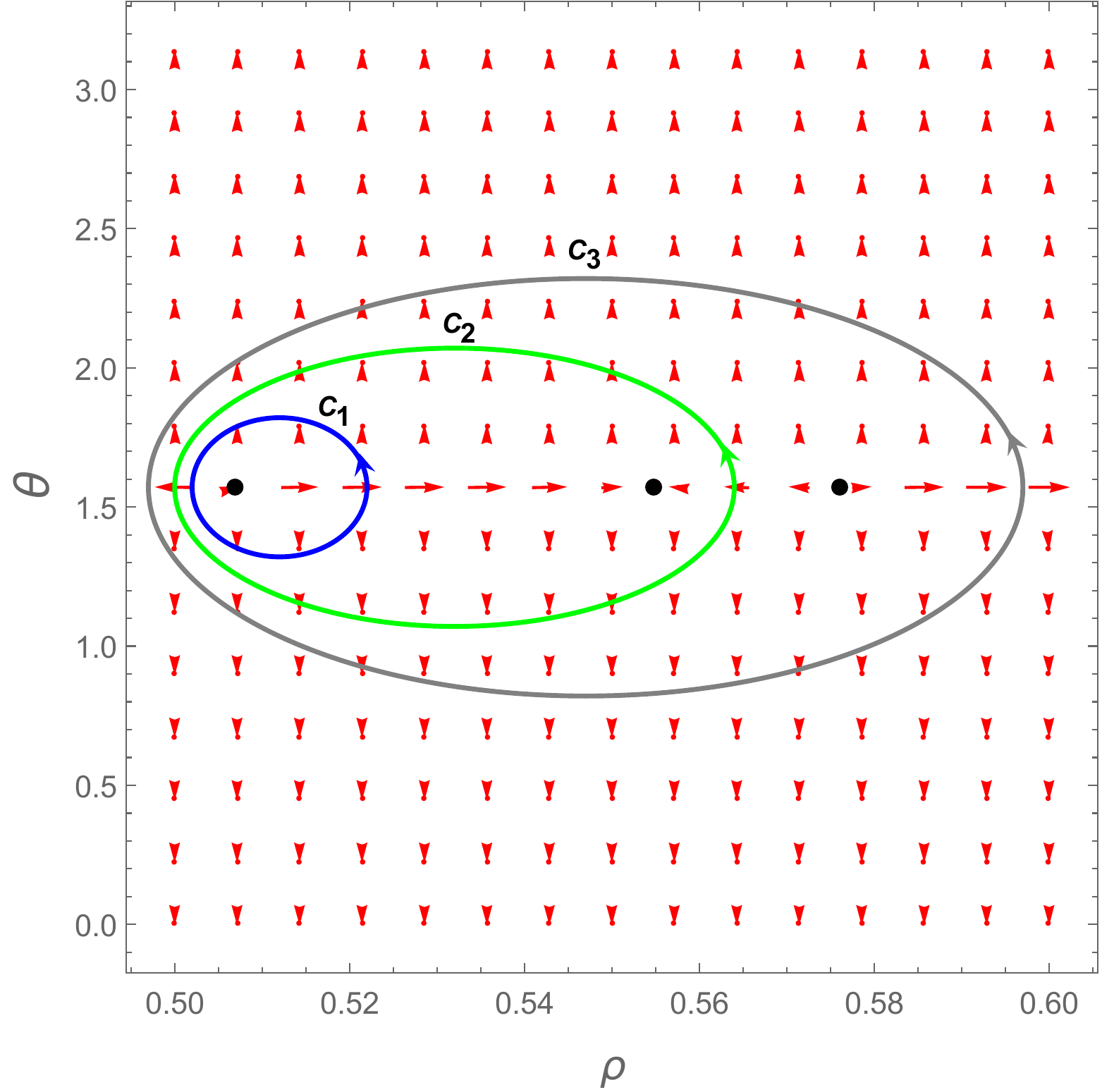}}\hspace{2cm}	
		\subfloat[]{\includegraphics[width=2.7in]{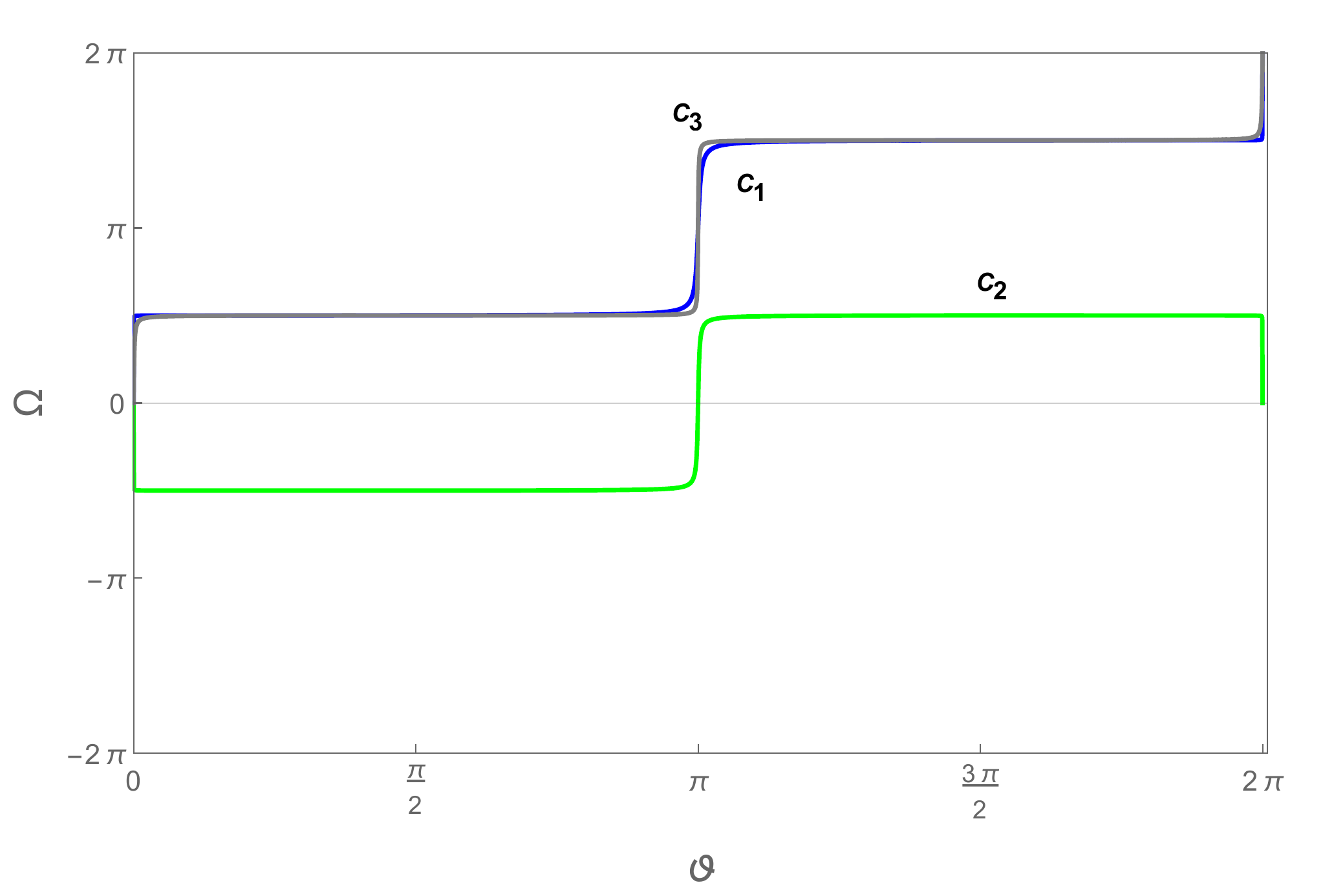}}				
		
		\caption{\footnotesize  (a) The  vector field $\phi( \frac{\partial V}{\partial \rho}, \, -\text{cot}\,\theta \,\text{csc}\,\theta)$ vanishes exactly  at the saddle points (black dots, located at $\rho = 0.507, 0.555, 0.576$) of $V(\rho)$, correspond to small, intermediate, and large black hole solutions in the bulk. (b) $\Omega$ vs $\vartheta$ for contours $C_1$, $C_2$, and $C_3$. The  saddle points, correspond to small/intermediate/large black holes, carry the topological charges +1/-1/+1, respectively.} 
		\label{Fig:RN_canon_boundary_equi_vec_omega_plots}	}
\end{figure}
\vskip 0.2cm
\noindent
As done in the bulk, one can now compute the topological charge carried by the critical point,  by defining the vector field $\phi(\phi^\rho,\, \phi^\theta)$.  
It turns out that the critical point carries the same topological charge as in bulk, i.e., $+1$, (See, Figures~\ref{fig:RN_canon_boundary_vdw_vec_plot} and ~\ref{fig:RN_canon_boundary_vdw_omega_plot}).
Further, one can see, from the Fig.~\ref{Fig:RN_canon_boundary_equi_vec_omega_plots},  that the  topological charges associated with the saddle points of $V(\rho)$, match with that of the corresponding black holes in the bulk, as well.
\section{Conclusions} \label{conclusions}

In this paper, we pursued a set up where the thermodynamics and phase 
transition points of Schwarzschild and charged black holes in AdS could 
be studied via an off-shell Bragg-Williams free energy. The boundary 
dual corresponds to ${\cal {N}} =4$, $SU(N)$ SYM theory on $S^3$, which can be studied via 
a phenomenological matrix model. In this holographic set up, 
we computed the topological charges of various equilibrium phases and critical points, 
both for black holes in the bulk and phase transition points in the dual matrix model. 
In the bulk, we studied HP transition points of Schwarzschild and charged black holes in AdS and found the topological charge to be $+1$, which is a novel charge as per the classification scheme proposed in~\cite{Wei:2021vdx}. The effect of chemical potential on the 
matrix model on the boundary is incorporated by allowing parameters $(a,b)$
to depend on the chemical potential as well as the temperature. 
We then showed that the same value of topological charge ensues from the effective potential of the boundary matrix model at the confinement-deconfinement transition point. 
 We also studied the charges associated with various equilibrium phase of the free energy in the bulk and the matrix model effective potential on the boundary, which both gave the same value of +1 (-1) for stable (unstable) configurations. The results in the bulk are in accord with the conjecture in~\cite{Wei:2022dzw}, that the total topological charge for both Schwarzschild and charged black holes (in grand canonical ensemble) is zero and hence they belong to the same topological class. \\

\noindent
In the canonical ensemble, we examined the second order critical points of charged black holes in AdS, within the original spirit of black hole thermodynamics~\cite{Chamblin:1999tk,Chamblin:1999hg}, and found the topological charge to be $+1$. 
This value is opposite to the one found for the same point in~\cite{Wei:2021vdx}. This opposite behaviour of the topological charges is an expected result, as the topological charge of the critical point depends on the behaviour of the phase structure (i.e., on the behaviour of the vector field) around that critical point (see Appendix-(\ref{B}). Thus, charged black holes in AdS in the canonical ensemble, where there exists a second order critical point (analogous to the van der Waals system) and the grand canonical ensemble group into different topological classes~\cite{Wei:2021vdx}. The boundary computation was set up in the matrix model studied in section-(\ref{one}), but now with an additional logarithmic term in the effective action~\cite{Basu:2005pj}. The values of topological charges computed from the effective potential match the results in bulk.\\

\noindent
It would be interesting to check the topological charges of the reverse HP transitions~\cite{Mbarek:2018bau} and also novel reentrant HP transitions~\cite{Cui:2021qpu}. Of course, it is also interesting to pursue whether and how the Hawking-Page point sets off a topological transition where a black hole emerges from the background space time as a topological defect. Some of these questions become more interesting to pursue when there are higher derivative terms in the action, such as Gauss-Bonnet terms. Such terms give rise to interesting phase structure and topological classification in the bulk has be done recently in the extended phase space framework~\cite{Yerra:2022alz}. Setting up of boundary matrix models which take into account the variations of cosmological constant in the bulk are quite interesting and should be pursued in future, following recent developments in novel holographic set ups in the brane world scenarios~\cite{Ahmed:2023snm,Frassino:2022zaz,Bueno:2022log,Chen:2020uac,Emparan:2020znc,Meessen:2022hcg}. It might also be useful to study the topological classification of various phases for black holes in higher derivative theories can possibly be addressed in the boundary matrix models, with appropriate assumptions on the model parameters $(a,b)$\cite{Dey:2008bw}. 
On another front, it is an interesting exercise is to look for boundary duals of threshold points corresponding to fault-tolerance~\cite{Bao:2022agm} and compute their topological charges, which might give inputs on the connection of quantum error correction and holography.

\section*{Acknowledgements}
One of us (C.B.) thanks the DST (SERB), Government of India, for financial support through the Mathematical Research Impact Centric Support (MATRICS) grant no. MTR/2020/000135. 
We thank the referee for helpful suggestions which improved the manuscript.

\renewcommand{\thesection}{\Alph{section}}
\appendix
\section*{Appendices}

\section{Construction of Bragg-Williams Free Energy for black holes in AdS}\label{A}
	\subsection{Schwarzschild-AdS black holes} \label{A1}
	We start with the $(n+1)$-dimensional Schwarzschild-AdS black holes, whose action and the line element are given by~\cite{Witten:1998zw}:
	\begin{equation}
		I = -\frac{1}{16{\pi}G} \int d^{n+1}x \sqrt{-g} \left[R  +
		\frac{n(n-1)}{l^2}\right],	
	\end{equation}
	\begin{equation}
		ds^2 = -V(r)dt^2 + \frac{dr^2}{V(r)} +r^{2}d{\Omega}^2_{n-1},
	\end{equation}
	\noindent where, $G$ is the Newton's constant, $d{\Omega}^2_{n-1}$ is the metric on the round
	unit $(n{-}1)$--sphere having volume $\omega_{n-1}$, and the metric function $V(r)$ takes the form
	\begin{equation}
		V(r) = 1 - \frac{m}{r^{n-2}} +  \frac{r^2}{l^2}.
	\end{equation}
	
	\noindent Here, $l$ is the AdS length and $m$ is related to the ADM mass $M$ of the black hole.
	The energy $E$, temperature $T$, entropy $S$, and the free energy $F$ of the black hole are given, in terms of horizon radius $r_+$, by:
	\begin{eqnarray}
		E&=&M =\frac{(n-1)\omega_{n-1}}{16\pi G}m  \nonumber \\
		&=&\frac{(n-1)\omega_{n-1}}{16\pi G}\big(r_+^{n-2}+\frac{r_+^n}{l^2}\big),\\
		T&=& \frac{(n-2)l^2+nr_+^2}{4\pi l^2r_+}, \\
		S&=&\frac{\omega_{n-1}}{4G}r_+^{n-1},\\
		F&=& M-TS.
	\end{eqnarray}
	They satisfy the first law of thermodynamics: $dE = TdS$. Now, by treating the horizon radius $r_+$ as the order parameter and the temperature $T$ as the external parameter, 
	one can construct the Bragg-Williams off-shell free energy $f$, as~\cite{Banerjee:2010ve}:
	\begin{eqnarray}
		f(r_+, T) &=& M-TS \nonumber \\
		&=&\frac{(n-1)\omega_{n-1}}{16\pi G}\big(r_+^{n-2}+\frac{r_+^n}{l^2}\big) - T\frac{\omega_{n-1}}{4G}r_+^{n-1}. \end{eqnarray} 
	\noindent In five dimensions (i.e., $n=4$), and setting $l= \omega_3= 16\pi G =1$, the Bragg-Williams free energy becomes
	\begin{equation}
		f(r_+, T)=3\big(r_+^{2}+r_+^4\big) - 4\pi r_+^{3}T.
	\end{equation} 
	\noindent We note here that our free energy $f$ differs from the free energy given in~\cite{Banerjee:2010ve,Yerra:2022coh} by a factor of $16\pi$, as we have set here $16\pi G =1$,  instead of setting $G=1$ there.
	 
\subsection{Reissner–Nordstrom-AdS black holes}\label{A2}

	We consider the  Reissner–Nordstrom-AdS black holes in $(n+1)$-dimensions, whose action and the line element are now turn out to be~\cite{Chamblin:1999tk,Chamblin:1999hg}:
	\begin{equation}
		I = -\frac{1}{16{\pi}G} \int d^{n+1}x \sqrt{-g} \left[R - F^2 +
		\frac{n(n-1)}{l^2}\right],	
	\end{equation}
	\begin{equation}
		ds^2 = -V(r)dt^2 + \frac{dr^2}{V(r)} +r^{2}d{\Omega}^2_{n-1},
	\end{equation} 
	\noindent where, the metric function $V(r)$ and gauge potential are:
	\begin{eqnarray}
		V(r) &=& 1 - \frac{m}{r^{n-2}} + \frac{q^2}{r^{2n-4}}+  \frac{r^2}{l^2}, \\
		A &= & \left(-{1\over c}{q\over r^{n-2}}+\mu\right)dt.
	\end{eqnarray}
	\noindent Here  $ c=\sqrt{2(n-2)\over n-1} $, and the parameter
	$q$ yields the black hole's charge
	\begin{equation}
		Q=\sqrt{2(n-1)(n-2)}\left({\omega_{n-1}\over 8\pi G}\right)q.
	\end{equation}
	\noindent The gauge potential is chosen to vanish on the horizon at $r = r_+$, that fixes the $\mu$ to be 
	\begin{equation}
		\mu={1\over c}{q\over r_+^{n-2}},
	\end{equation}
	which represents the electrostatic potential difference between the horizon and infinity.
	\subsubsection*{Grand canonical ensemble:} In grand canonical ensemble (i.e., in fixed potential $\mu$ ensemble),   
	the energy $E$, temperature $T$,   entropy $S$, and the free energy $F$ of the black hole are given by:
	\begin{eqnarray}
		E&=&M =\frac{(n-1)\omega_{n-1}}{16\pi G}m  \nonumber \\
		&=&\frac{(n-1)\omega_{n-1}}{16\pi G}\big(r_+^{n-2}+ \mu^2c^2 r_+^{n-2} +\frac{r_+^n}{l^2}\big),\\
		T&=& \frac{(n-2)l^2(1-c^2\mu^2)+nr_+^2}{4\pi l^2r_+}, \\
		S&=&\frac{\omega_{n-1}}{4G}r_+^{n-1},\\
		F&=& M-TS-\mu Q.
	\end{eqnarray}
	\noindent They satisfy the first law: $ dE = TdS +\mu dQ$.	Now,  the Bragg-Williams  free energy $f$, is given by~\cite{Banerjee:2010ve,Yerra:2022coh}:
	\begin{eqnarray}
		f(r_+, T, \mu) &=& M-TS-\mu Q \nonumber \\
		&=&\frac{(n-1)\omega_{n-1}}{16\pi G}\big(r_+^{n-2}- \mu^2c^2r_+^{n-2} +\frac{r_+^n}{l^2}\big) - T\frac{\omega_{n-1}}{4G}r_+^{n-1}. \end{eqnarray} 
	\noindent Here, the temperature $T$ and the potential $\mu$ are treated as the external parameters. For,  $n=4$, and  $l= \omega_3= 16\pi G =1$, the Bragg-Williams free energy becomes
	\begin{equation}
		f(r_+, T, \mu)=3\Big(r_+^{2}-\frac{4}{3}\mu^2r_+^2 +r_+^4\Big) - 4\pi r_+^{3}T.
	\end{equation}
	\subsubsection*{Canonical ensemble:} In  canonical ensemble (i.e., in fixed charge $Q$ ensemble),   
	the energy $E$, temperature $T$,   entropy $S$, and the free energy $F$ of the black hole are given by:
	\begin{eqnarray}
		E&=&M =\frac{(n-1)\omega_{n-1}}{16\pi G}m  \nonumber \\
		&=&\frac{(n-1)\omega_{n-1}}{16\pi G}\big(r_+^{n-2}+ \frac{q^2}{r_+^{n-2}} +\frac{r_+^n}{l^2}\big),\\
		T&=& \frac{nr_+^{2n-2}+(n-2)l^2 r_+^{2n-4}-(n-2)q^2l^2}{4\pi l^2r_+^{2n-3}}, \label{TRN1} \\
		S&=&\frac{\omega_{n-1}}{4G}r_+^{n-1},\\
		F&=& M-TS.
	\end{eqnarray}
	\noindent They satisfy the first law: $ dE = TdS +\mu dQ$.	Now,  the Bragg-Williams  free energy $f$, is given by~\cite{Li:2020nsy}:
	\begin{eqnarray}
		f(r_+, T, Q) &=& M-TS \nonumber \\
		&=&\frac{(n-1)\omega_{n-1}}{16\pi G}\big(r_+^{n-2}+ \frac{q^2}{r_+^{n-2}} +\frac{r_+^n}{l^2}\big) - T\frac{\omega_{n-1}}{4G}r_+^{n-1}. \end{eqnarray} 
	\noindent Here, the temperature $T$ and the charge $Q$ are treated as the external parameters. For,  $n=4$, and  $l= \omega_3= 16\pi G =1$, the Bragg-Williams free energy becomes
	\begin{equation}
		f(r_+, T, Q)=3\Big(r_+^{2}+\frac{q^2}{r_+^2} +r_+^4\Big) - 4\pi r_+^{3}T.
	\end{equation}

\section{Extremal points of the temperature}\label{B}

\begin{figure}[h!]
	{\centering
		\subfloat[]{\includegraphics[width=2.2in]{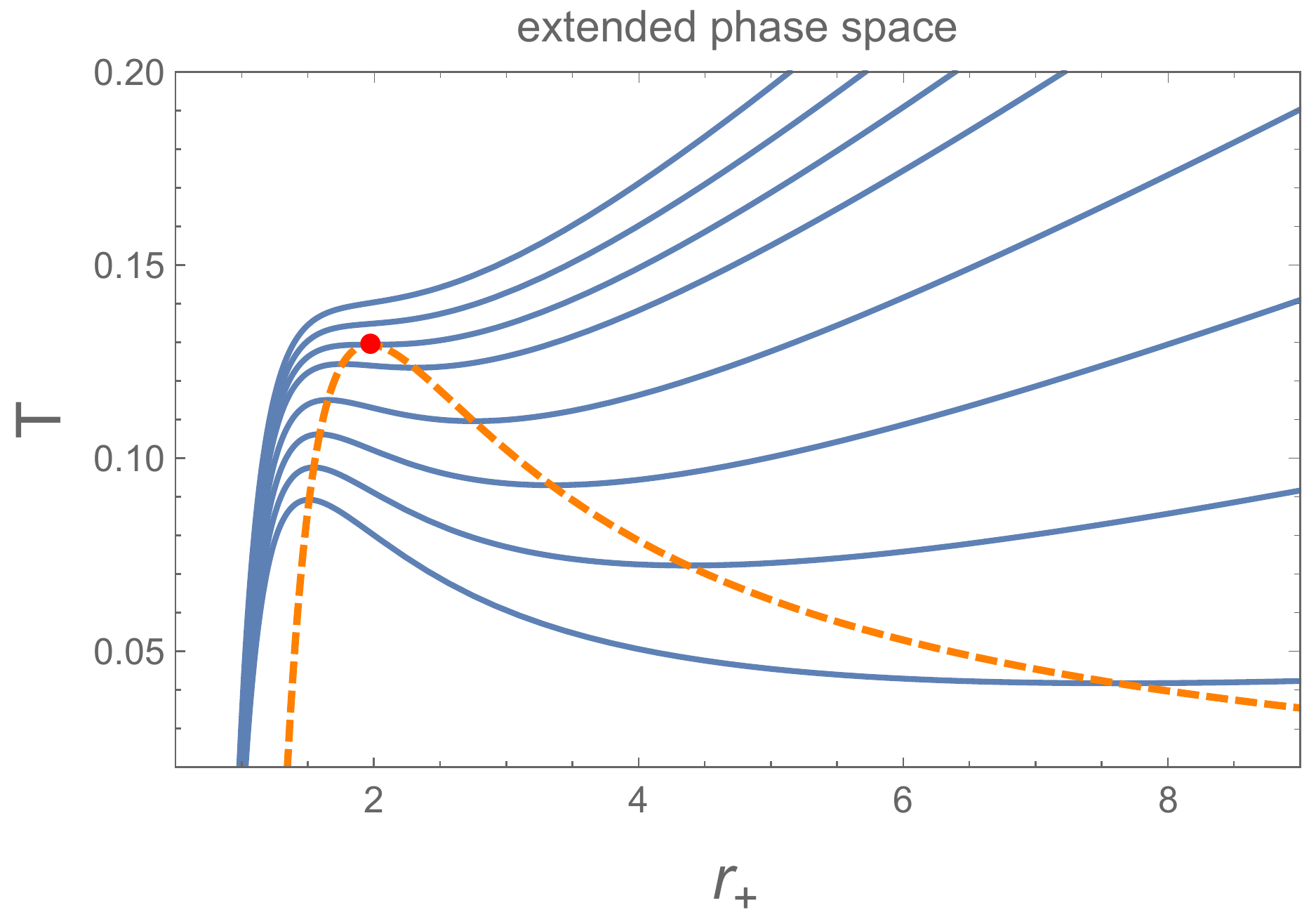}}\hspace{1.5cm}
		\subfloat[]{\includegraphics[width=2.2in]{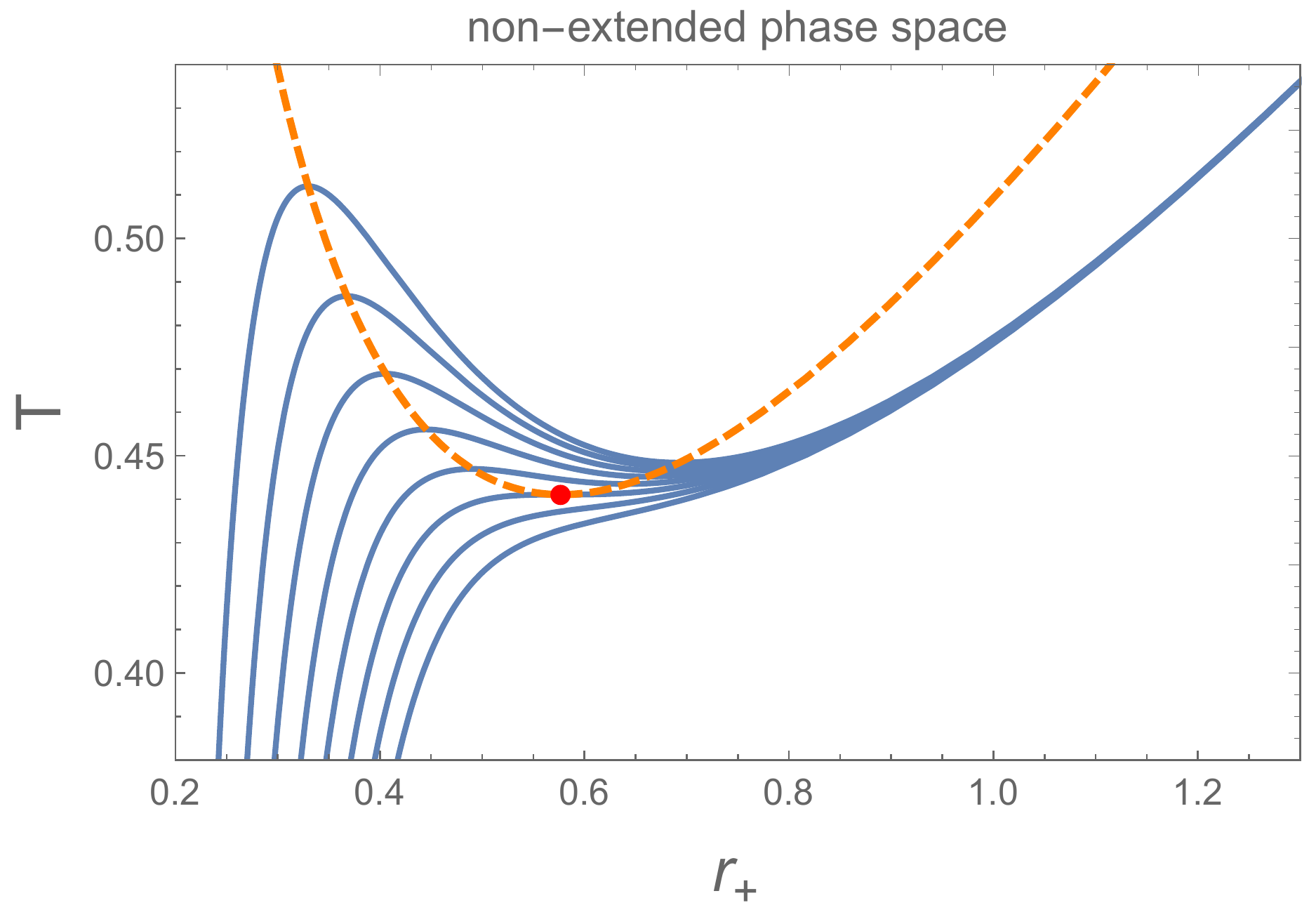}}
		
		\caption{\footnotesize  Behaviour of the curve corresponding to the extremal points of temperature (dashed orange coloured) for RN-AdS black holes in $T-r_+$ plane (a) in extended phase space. Blue curves are isobars with $q=1$. (b) in non-extended phase space. Blue curves are isocharge curves with $l=1$. Red dot is the critical point.} 
		\label{fig:spin_curves}	}
\end{figure}

\noindent
The discrepancy of the topological charges of the critical points in the extended and non-extended phase space of RN-AdS black holes can be understood, from the behaviour of  curves giving the extremal points of the temperature. 
After eliminating the charge $q$ (in non-extended phase space) or the pressure $P$ (in extended phase space~\cite{Wei:2021vdx}) from the equation of state, using one of the conditions for the critical point (i.e., $\frac{\partial T}{\partial r_+} = 0\, \text{and}\, \frac{\partial^2 T}{\partial r_+^2} = 0 $), we obtain the curve for extremal points of the temperature, given by: 
\begin{eqnarray}
T(r_+) &=& \frac{r_+^4 - 3q^2}{\pi r_+^5} \, \quad \quad \text{(In extended phase space}).  \label{TE1} \\
T(r_+) &=& \frac{2(l^2+3r_+^2)}{5\pi l^2 r_+ } \, \quad \text{(In non-extended phase space)}. \label{TNE1}
\end{eqnarray}
\noindent 
For the non-extended phase space, we obtained eqn.(\ref{TNE1}), by eliminating charge from eqn.(\ref{eq:RN_canon_bulk_eos}) (see discussion around eqn.(\ref{phitheta1})) and for the  extended phase space, a similar computation given in the appendix of~\cite{Wei:2021vdx}, yields eqn. (\ref{TE1}).
The nature of these  curves (dashed orange coloured) can be inferred from Fig.~\ref{fig:spin_curves}. 
The critical point can now be located to be at the maxima (minima) of the curve in extended phase space (non-extended phase space). In the computation of topological charge of the critical points, we have defined the vector field $\phi \big(\partial_{r_+}(T/\text{sin}\theta), \partial_\theta(T/\text{sin}\theta)\big)$, using the curve corresponding to the extremal points of temperature $T(r_+)$.
 Now, the vanishing of the vector field $\phi$ at the critical point happens at the maxima (minima) of the curve drawn in Fig.~\ref{fig:spin_curves} in the extended phase space (non-extended phase space), which yields the topological charge -1 (+1). Thus, the positive (negative) sign of the topological charge captures, the stable (unstable) nature of the critical point  on the curve corresponding to the extremal point of the temperature. That the stable and unstable points in general turn out to have oppose topological charges, is now known~\cite{Wei:2022dzw,Yerra:2022coh}.

\bibliographystyle{apsrev4-1}
\bibliography{topology_matrix}
\end{document}